# Ferroelectric Nematic and Smectic Liquid Crystals with Sub-Molecular Spatial Correlations

Parikshit Guragain[a,+], Arjun Ghimire[b,+] Manisha Badu[b], Netra Prasad Dhakal[c], Pawan Nepal[a], James T. Gleeson[b], Samuel Sprunt[b,c,*], Robert J. Twieg[a,*], Antal Jákli[b,c,*]

[a] Department of Chemistry and Biochemistry, Kent State University, Kent, OH 44242, USA

[b] Department of Physics, Kent State University, Kent, OH 44242, USA

[c] Materials Science Graduate Program and Advanced Materials and Liquid Crystal Institute, Kent State University, Kent, OH 44242, USA

*: Corresponding authors: rtwieg@kent.edu, ssprunt@kent.edu and ajakli@kent.edu

+: These authors contributed equally to this work

## Abstract

A number of highly polar three ring rod-shaped compounds with a terminal thiophene ring have been synthesized and the physical properties of a subset are reported in detail. On cooling from the isotropic fluid, they directly transition to a ferroelectric nematic liquid crystal ($N_F$) phase that shows the strongest spatial correlations corresponding to 1/3 of the molecular length ($L$/3). The set of thiophene compounds reported here have ferroelectric polarizations of $P_S \approx 7.8 \pm 0.3\ \mu C/cm^2$, which is about 20% larger than that of usual ferroelectric nematic liquid crystal materials. Such large polarization values are due to the ~20% larger mass densities of these thiophene compounds compared to most of the $N_F$ materials with short terminal chains. These unusual properties are consequences of tighter molecular packing due to the lack of flexible terminal chains.

Below the $N_F$ phase, compounds with a single nitro or two cyano polar groups on the terminal benzene ring exhibit a so far never observed smectic phase with periodicity ~1/3 the molecular length. Based on our experimental results, we propose a model of this phase featuring antipolar packing of the molecules within the layers.

## 1. Introduction

The long-sought ferroelectric nematic $N_F$ liquid crystal phase, wherein macroscopic polar order is combined with 3-dimensional fluidity, was recognized[1] after several groups independently synthesized materials 3(4) CN[2], DIO[3] and RM734[4] consisting of rod like molecules having permanent dipole moments of 10 Debye or greater. In the $N_F$ phase, these electric dipoles that spontaneously align in a common direction. This results in a macroscopic polarization

$$\vec{P}_S = \frac{\langle\vec{\mu}\rangle \rho N_A}{M} = \frac{S_P \vec{\mu} \rho N_A}{M}, \tag{1}$$

where $\rho$ is the mass density, $N_A \approx 6 \cdot 10^{23}$ is Avogadro's number, $M$ is the molar mass and $S_P \leq 1$ is the dipolar order parameter. Assuming typical values such as $\mu \approx 10$ D ($3.3 \cdot 10^{-29}$ Cm), $\rho \approx 1.3 \cdot 10^3 kg/m^3$,[5] $M \approx 0.4\, kg$, we obtain $P = 6.4 \cdot 10^{-2} C/m^2$ for perfect polar order ($S_P = 1$). Experimental observations on $N_F$ material[3,6,7] are consistent with $S_P \geq 0.95$. This spontaneous polarization is significantly larger than is typically observed in chiral SmC* smectic ferroelectrics[8–10] and polar bent-shape smectic liquid crystals[11]. Since their discovery, $N_F$ materials have attracted significant interest because of their extraordinary physical properties[12–25] as well as the promise of new technological applications. Since polar materials also lack inversion symmetry, $N_F$ materials also show remarkable linear electromechanical responses[26], large second harmonic generation (SHG) [22,27,28] due to large second-order non-linear optical (NLO) coefficients [29,30], exotic entangled photon-pair generation[31], and bulk photovoltaic effect[32].

Since 2017, hundreds of individual compounds exhibiting the $N_F$ phase have been reported, but most belong to a few discrete structural classes.[33–42] As work proceeds to explore the physical properties of existing materials, researchers are also striving to define structure-function relationships that would guide the synthesis of next-generation materials. The creation of diverse sets of $N_F$ materials is a prerequisite both to thoroughly understanding the physics of this novel phase and achieving real-world applications. Furthermore, the understanding of what governs not only the macroscopic properties but also the microscopic structure of these fluid, polar materials remains at an early stage. One can anticipate that, in addition to entropic considerations and the ease at which fluids can reorganize themselves, electrostatic molecular interactions will be of primary importance. From this it necessearily follows that intermolecular distances will also be a dominating factor.

In this work, we report the synthesis, phase behavior, and critical physical properties - polarization value, mass density, and nanostructural organization - of a series of novel thiophene based, three ring ferronematic compounds. Polarized optical microscopy (POM), X-ray scattering, ferroelectric polarization and mass density measurements are used to demonstrate the existence of the $N_F$ phase and characterize its nanostructure. The materials exhibit larger ferroelectric polarization and mass density than previously reported in the $N_F$ phase. In some cases, evidence for a lower temperaturea smectic phase with a unique 1/3 molecular length layering is observed, and a tentative model for this phase is described.

## 2. Results

### *A. Materials*

Previously the $N_F$ phase has been reported in compounds like DIO [3], LC1 and LC2 [43] which have saturated heterocyclic rings. The new compounds we prepared are of heteroaromatic ring-based $N_F$ compounds that incorporate a thiophene ring. The thiophene ring has already been widely utilized in liquid crystal research to obtain conventional nematic[44], smectic, columnar, and blue phases[45–48] and recently ferroelectric nematic phases.[49] Here, the important structure modification implemented is the substitution of a thiophene group as the A-ring in lieu of a benzene A-ring found in the prototypical $N_F$ RM734.[4] The thiophene ring emulates the 2,4-dimethoxybenzene electronically and spatially without any need for additional substituents. Equally important, and in further contrast to most previously reported $N_F$ materials, these compounds have no flexible "tail" groups such as the two methoxy substituents in RM734.

Figure 1 displays the molecular structures of the thiophene compounds studied, together with two reference materials: a prototypical $N_F$ (RM734) and the parent thiophene analog of RM734 (PN03111)[42]. RM734 features an all-benzene core with internal linking groups (two esters) and terminal functional groups (ethers and nitro) arranged such that their bond and group dipoles align, resulting in a large net molecular dipole. Typical RM734-like $N_F$ compounds are characterized by short terminal substituents along the director axis and an unusual ortho substituent on the terminal A-ring. In PN03111, the A-ring of RM734 is replaced by a 2-thiophene ring with no additional substituents.

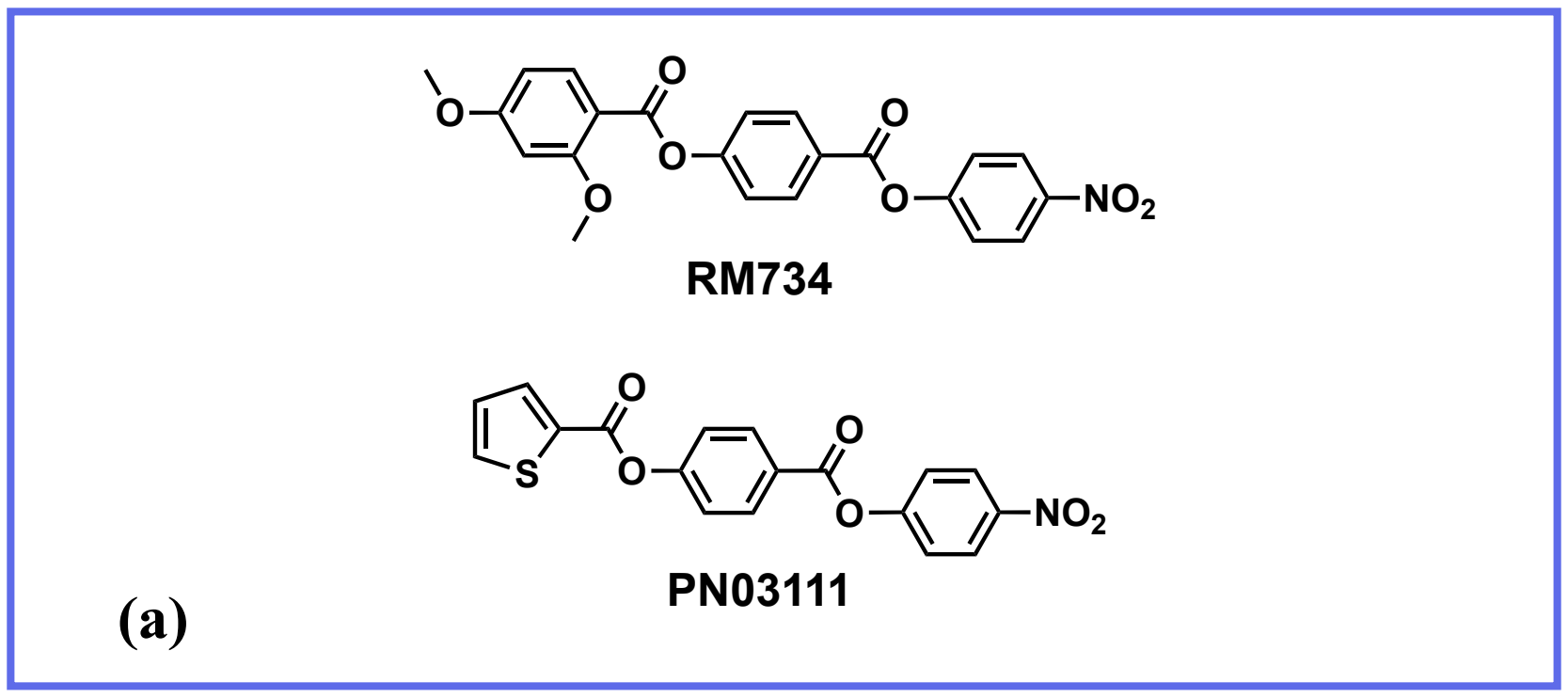

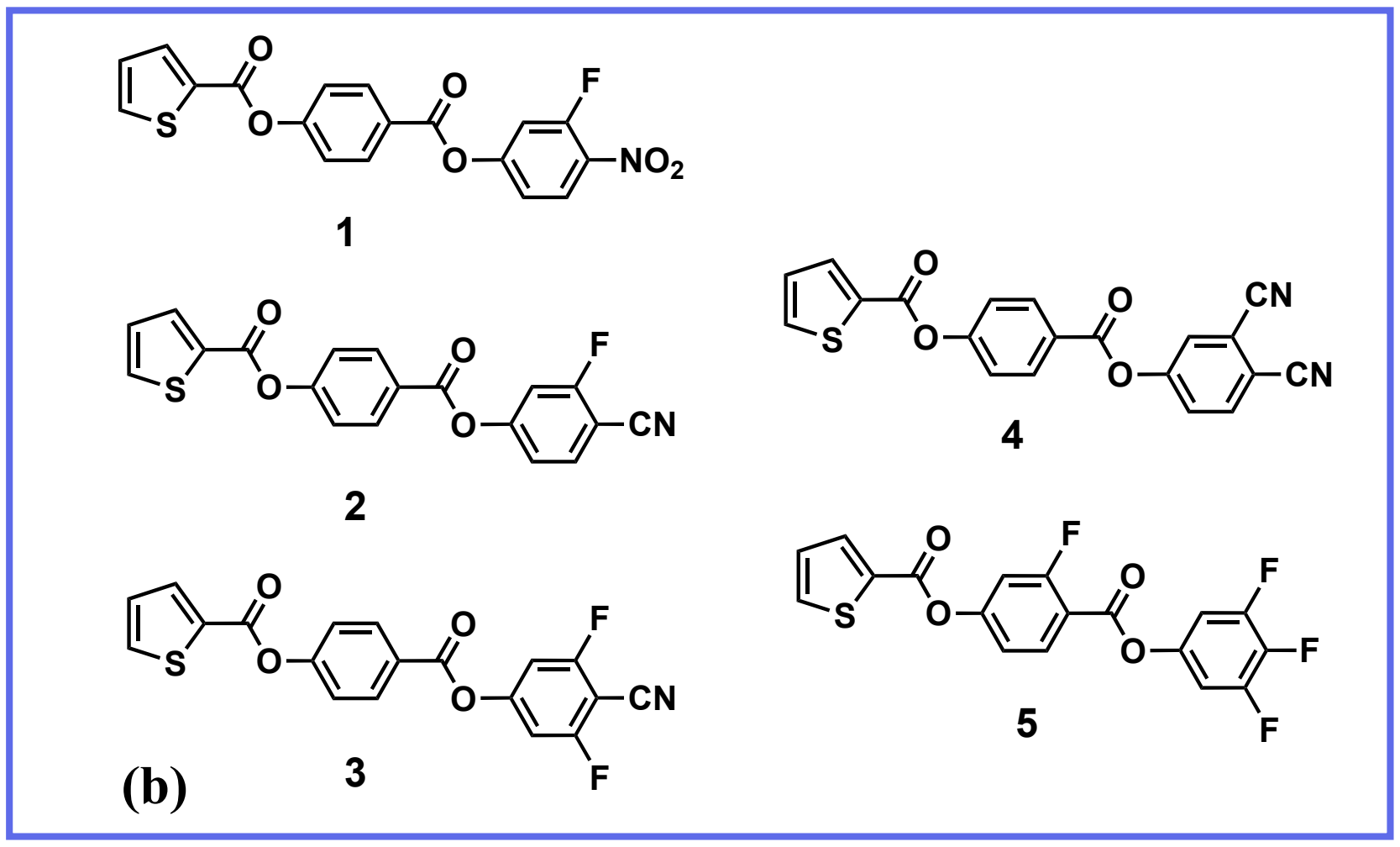

*Figure 1: Molecular structures of the materials relevant to this work. (a) The reference substances: 4-[(4-nitrophenoxy) carbonyl] phenyl 2,4-dimethoxybenzoate (RM734) (top) and 4-[(4-nitrophenoxy) carbonyl] phenyl-2-thiophenecarboxylate PN03111 (bottom). (b) Structures of the five thiophene materials described in detail in this study.*

The five primary materials investigated in this paper (Figure 1(b)) share the same core framework as PN03111, differing only in the substituents positioned on the B and C rings. In contrast to the I-N transition of RM734, these compounds as well as PN03111 have a direct I-$N_F$ transition on cooling. Compared to PN03111, compound **1** features an additional fluorine substituent ortho to the terminal nitro group. This modification lowers the clearing point (transition to isotropic phase) by about 30 °C, avoiding the risk of thermal decomposition over time at elevated temperature, which occurs with PN03111 and which makes its physical property characterization less reliable. Both compound **1** and PN03111 exhibit a positionally ordered (smectic) phase when cooled below the $N_F$ phase.

Compound **2**, structurally resembles compound **1**, but features a less polar cyano group in place of the terminal nitro group. This substitution significantly reduces the molecular dipole moment. Despite this change, the $N_F$ remains present in compound **2**. Interestingly, the the $N_F$ phase range broadens in compound **2**, and the isotropic temperature is further reduced, relative to compound **1**.

Compound **3** adds a second fluorine in the ortho position compared to compound **1**. This modification increases the $N_F$-isotropic transition temperature, while also broadening the $N_F$ phase range. The 130 °C wide $N_F$ temperature range of this compound is noteworthy and highlights the significant impact that an extra polar bond near the terminal groups can have on the $N_F$ behavior of these materials.

Compound **4** contains two adjacent cyano groups at the meta and para sites of the C ring. Compared to the structurally similar compound **2**, it exhibits a slightly lower $N_F$-isotropic tranistion temperature. However, the presence of two cyano groups results in a less stable $N_F$ phase, characterized by a reduced phase range, compared to the difluorocyano system in **3**. Additionally, a smectic phase appears in this compound below the $N_F$ phase.

Compound **5** features a distinct substitution pattern, lacking any large polar group at the terminal position. Instead, it has three highly electronegative fluorine atoms at the para and two meta positions on the C ring and an additional fluorine atom substituent on the B ring, which is unique among the materials studied here. The B-ring substitution is crucial, as the analogous compound (not further discussed here) without this additional fluorine atom exhibits a metastable $N_F$ phase with a range of only 1°C.

The molecular structures, phase sequences and dipole moments of the materials studied are listed in Table 1. The isotropic phase above the $N_F$ phase is denoted by I, and the mesophases observed below the $N_F$ phase for **1, 4** and **PN03111** above the crystal phase are labeled by X.

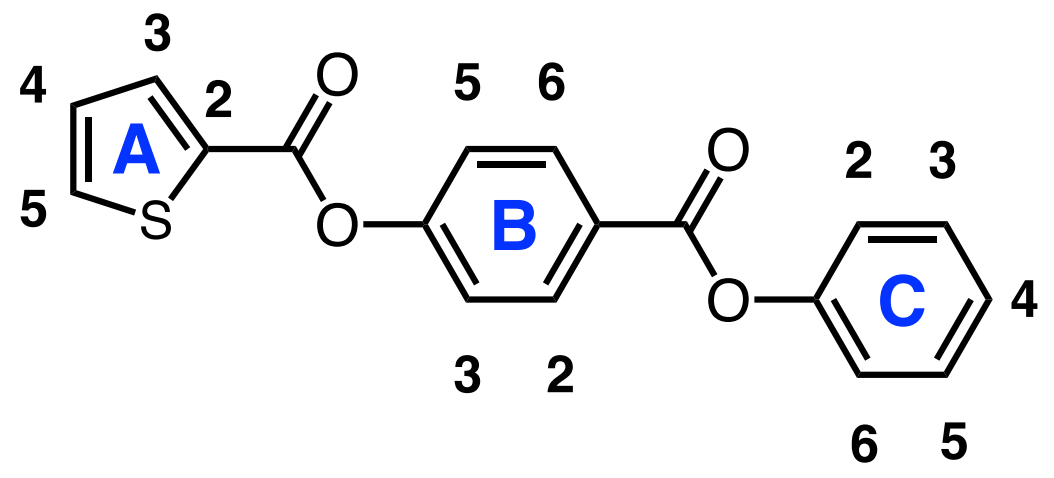

| # | Structure (substituent sites numbered) | | | Physical Properties | | |
|---|---|---|---|---|---|---|
| | Ring A | Ring B | Ring C | Dipole Moment (D) | Molecular Length (nm) | Phase Sequence (°C) |
| **1** | 2-thiophene | - | 3-F, 4-$NO_2$ | 11.19 | 1.83 | Cr 189 N 207 I 204 $\mathbf{N_F}$ 162 X 130 Cr |
| **2** | 2-thiophene | - | 3-F, 4-CN | 8.71 | 1.83 | Cr 167 $\mathbf{N_F}$ 192 I 191 $\mathbf{N_F}$ 101 Cr |
| **3** | 2-thiophene | - | 3,5-F, 4-CN | 9.74 | 1.86 | Cr 158 $\mathbf{N_F}$ 213 I 210 $\mathbf{N_F}$ 83 Cr |
| **4** | 2-thiophene | - | 3,4-CN | 9.97 | 1.86 | Cr 182 $\mathbf{N_F}$ 184 I 182 $\mathbf{N_F}$ 149 X 80 Cr |
| **5** | 2-thiophene | 2-F | 3,4,5-F | 8.81 | 1.75 | Cr 139 $\mathbf{N_F}$ 153 I 152 $\mathbf{N_F}$ 123 Cr |
| | | | | | | |
| **PN03111** | 2-thiophene | - | 4-$NO_2$ | 10.28 | 1.81 | Cr 195 N 230 I 219 $\mathbf{N_F}$ 165 X 140 Cr |
| **RM734** | 2,4-dimethoxy benzene | - | 4-$NO_2$ | 11.40 | 2.05 | Cr 140 N 190 I 189 N 133 $\mathbf{N_F}$ 75 Cr |

*Table 1: List of structural features, the phase sequence, dipole moments calculated using Chem3D Ultra MOPAC, and the molecular lengths calculated by ChemOfffice of the studied thiophene compounds.*

## *B. Polarizing Optical Microscopy*

Representative polarizing optical microscopy (POM) images of 5 µ*m* thick films of compounds **1-5** contained between planar substrates treated with anti-parallel rubbed polyimide alignment layers are shown in Figure 2.

On cooling from the isotropic phase, the $N_F$ phase appears via nucleation of circular shaped domains as recently described for other materials having direct I-$N_F$ transitions.[27,50] In this I/$N_F$ coexistence range, for compounds **1** and **3**, the texture of the $N_F$ domains with circular cross-sections has dark brushes that meet at the domain center: the brushes are along crossed polarizer and analyzer directions. This texture corresponds to tangential polarization direction with a circular line defect at the center.[50] In compounds **4** and **5**, the horizontal brushes are replaced by green brushes in a pink background. Such textures may be indicative of a biaxial director structure, whose origin is beyond the scope of this paper. The texture of the $N_F$ droplets coexisting in the isotropic phase of compound **2** is even more complicated.

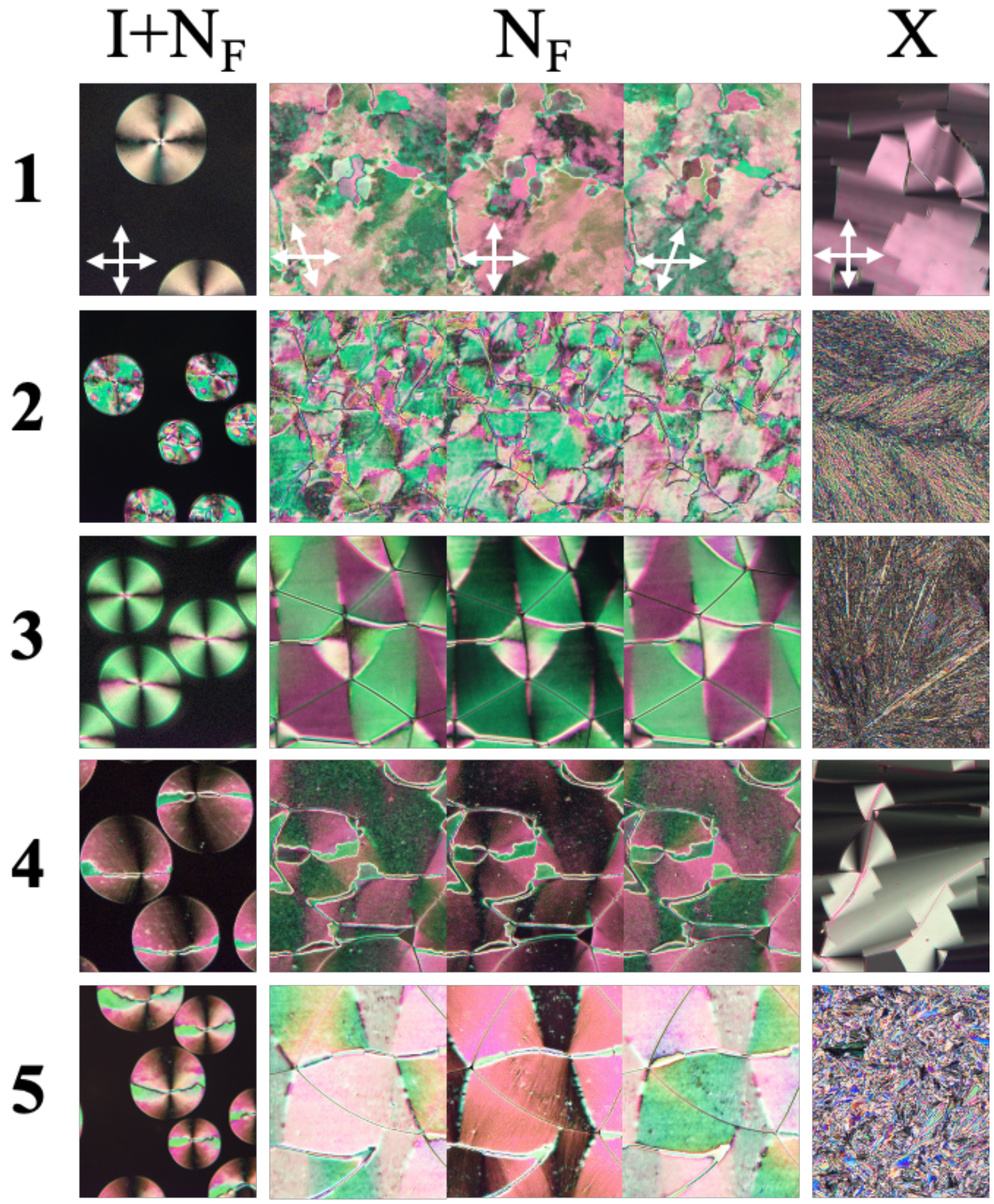

*Figure 2: Representative Polarizing Optical Microscopy (POM) images of compounds* ***1-5*** *in the I/$N_F$ coexisting range, in the $N_F$ phase, and in the respective lower temperature phase forming below the $N_F$ phase. The length of scalebar on image* ***1*** *is 50 µm.*

The textures in the fully formed $N_F$ phase are characterized by various marble or mosaic-type patterns in which the initial $N_F$ domains are separated by straight (compound **3**) or arc shaped (compound **5**) defect lines, while for **1**, **2** and **4** the defect lines are irregular. Looking at the $N_F$ textures between oppositely uncrossed polarizers, we note that the same birefringence colors of certain domains split into different colors that exchange under opposite uncrossing directions. This can be best seen in compound **3**, where the original dark green domains split into magenta and light green areas. These observations indicate oppositely twisted domains along the film thickness, as expected for $N_F$ materials sandwiched between substrates with anti-parallel alignment layers. Below the $N_F$ phase the textures indicate crystal phases for compounds **2**, **3** and **5**, and either columnar or smectic phases for compound **1** and **4**, respectively. Looking at Table 1, we can see that these two compounds (**1** and **4**) have the largest dipole moments, which may be related to the formation of the higher ordered LC phase(s).

## *C. X-ray Diffraction*

Small-angle X-ray scattering (SAXS) results on our thiophene compounds **1-5** and (for comparison) RM734 are summarized in Figures 3-5.**Error! Reference source not found.Error! Reference source not found.** Figure 3 shows two-dimensional SAXS patterns recorded at representative temperatures in the $N_F$ phase of compounds **1-5** and (for comparison) RM734, as well as in the X phase observed in compounds **1** and **4** only. In the $N_F$ phase (top two rows of Figure 3), all compounds produce three diffuse arcs of scattering along the magnetically-aligned director, indicating short-range positional correlations at roughly one, one half, and one third the molecular length (scattering wavenumbers $Q_1, Q_2$ and $Q_3$, respectively). However, particularly in compound **4**, the diffraction at the lowest angle ($Q_1$) appears to be split into two lobes along the azimuthal direction. Along the axis orthogonal to the director and at wider angle are diffuse peaks associated with short-range, liquid-like lateral correlations between the molecules. For similar scaling of the intensity levels relative to the maximum intensity in the images, the thiophene compounds generally produce stronger small angle diffraction along the director axis than RM734 in the $N_F$ phase, particularly in the peak a$t$ $Q_3$ corresponding to the positional correlations at ~1/3 the molecular length.

In the X phase of compounds **1** and **4** (bottom row of Figure 3), the width of this peak narrows considerably, and its intensity saturates in the images at bottom left and center in Figure 3., This signifies the development of long-range positional order, which, particularly in compound **4**, is accompanied by a rotation of the axis of the diffuse wide angle diffraction off its original direction in the N$_F$ phase. After rescaling the intensity levels in the image at the bottom right in Figure 3, we see that the wide angle rotation is associated with an azimuthal splitting of the smaller angle $Q_3$ peak, which indicates the presence of differently oriented domains in the X phase.

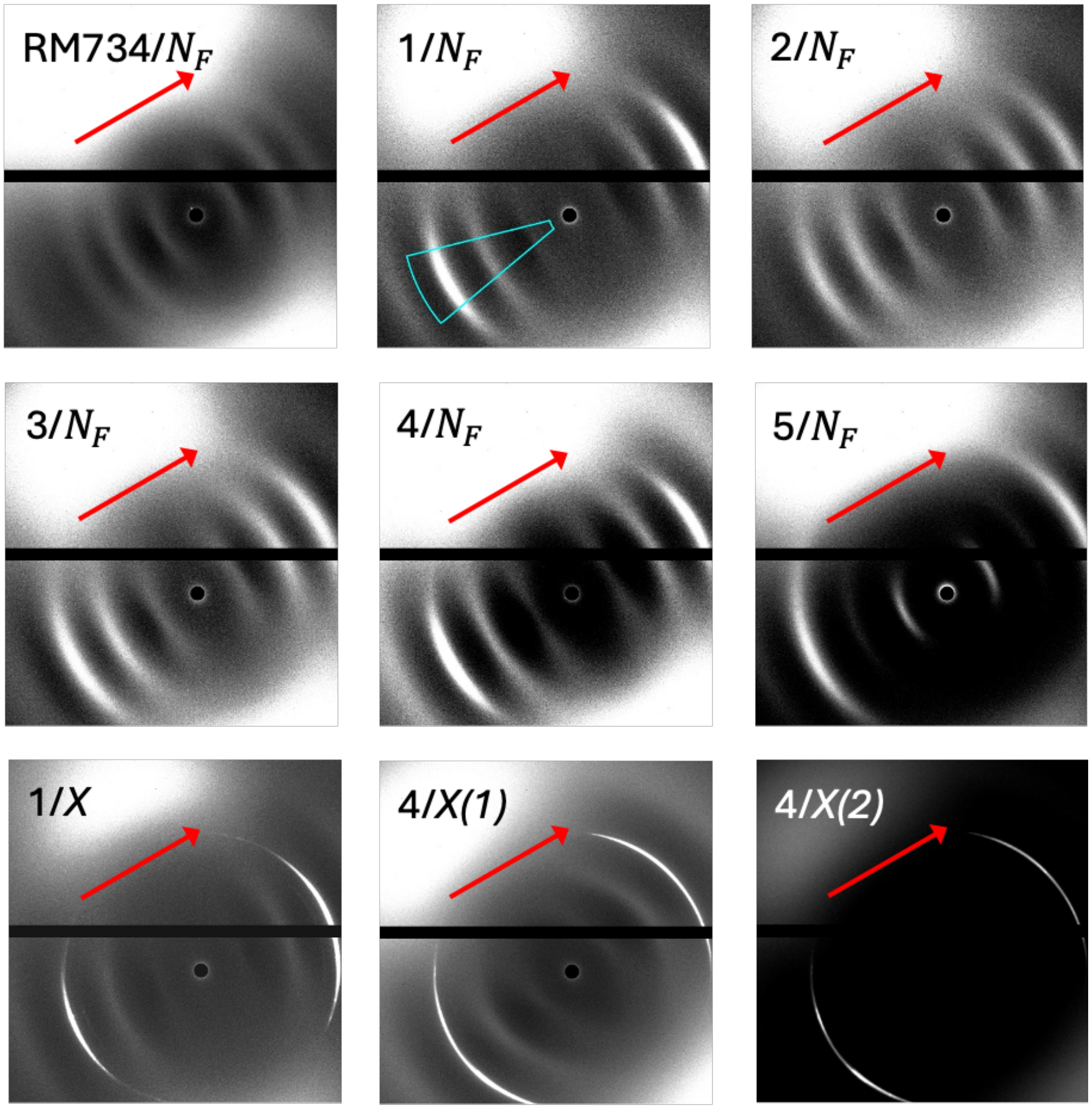

*Figure 3: Two dimensional SAXS patterns recorded on thiophene compounds **1 - 5** and (for comparison) on RM734 at representative temperatures in the middle of the N$_F$ phase (top two rows). In the bottom row also shown are SAXS patterns in the X phase, recorded at temperatures just below the N$_F$ – X transition in compounds **1** and **4**. The red arrows indicate the direction of a magnetic field applied to align the nematic director. The blue outline shows the section of the annulus in which angular averages were taken to produce the radial profiles (Q dependence) of the scattered X-ray intensity in Figure 4.*

Figure 4 plots the wavenumber ($Q$) dependences of the scattering intensity averaged azimuthally over the angular range indicated by the blue annulus outlines in Figure 3. Results are shown for RM734 and for compounds **1-5** at four different relative temperatures ($\Delta T = T_{I-N_F} - T$) below the isotropic to N<sub>F</sub> phase transition. The positions of the three small angle peaks (wavenumbers $Q_1, Q_2$ and $Q_3$) show no temperature dependence in the N<sub>F</sub> phase. For compounds **1-4** they are $Q_1 \approx 3.2\, nm^{-1}, Q_2 \approx 2Q_1 \approx 6.6\, nm^{-1}$ and $Q_3 \approx 3Q_1 \approx 10.0\, nm^{-1}$, with $Q_1$ corresponding to a length $L = 2\pi/Q_1 \approx 1.95\, nm$, while for compound **5** they are slightly larger, $Q_1 \approx 3.5\, nm^{-1}, Q_2 \approx 2Q_1 \approx 7.0\, nm^{-1}$ and $Q_3 \approx 3Q_1 \approx 10.7\, nm^{-1}$, corresponding to $L \approx 1.8\, nm$. These lengths are slightly (~3%) larger than the the stretched molecular lengths $L_S$ calculated by ChemOffice and shown in Table 1.**Error! Reference source not found.**

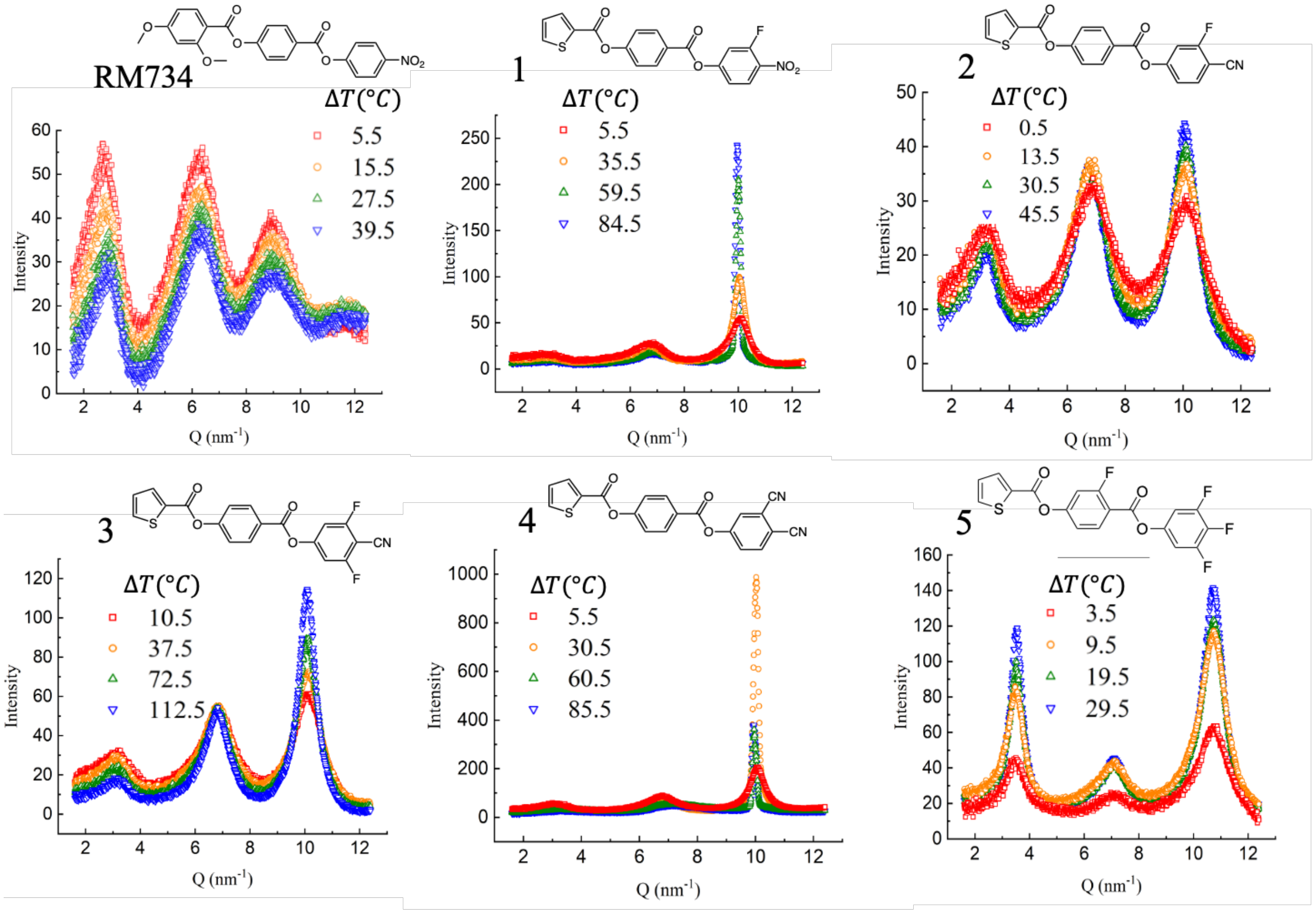

*Figure 4: Molecular structures and wavenumber (Q) dependences of the scattering intensities in* $1.5 - 12\, nm^{-1}$ *range for RM734 and for compounds **1-5** at 4 different relative temperatures* ($\Delta T = T_{I-N_F} - T$) *below the isotropic – N<sub>F</sub> phase transition.*

Strikingly, while for RM734 and other materials with an $N_F$ phase[42,51–54] the scattering intensity ($I$) of the small angle peaks decreases monotonically at larger $Q$ values, for the thiophene compounds **1-5** the peak at $Q_3$ is stronger than at $Q_1$ and $Q_2$. In fact, except for compound **5**, $I_{Q_3} > I_{Q_2} > I_{Q_1}$. In the higher order LC phase (X phase) observed in compounds **1** and **4**, the width of $Q_3$ peak is less than our resolution limit (~0.18 $nm^{-1}$). At the transition to the X phase, $Q_1$ and $Q_2$ increase by $\Delta Q_1 \approx \Delta Q_2 \approx 0.2\ nm^{-1}$, while $Q_3$ is unchanged. As pointed out in connection to the bottom pictures of Figure 3, the X-ray patterns in the X phase in compounds **1** and **4** mainly differ from the $N_F$ phase by the intensity and sharpness of the $Q_3$ peak. In compound **1** the peak remains centered on the axis of the aligning magnetic field, which would indicate an orthogonal smectic A phase. As pointed out above an apparent rotation of the diffraction pattern in the X phase of compound **4** is actually the result of a breakup into domains with different orientations of a smectic A type layer structure.

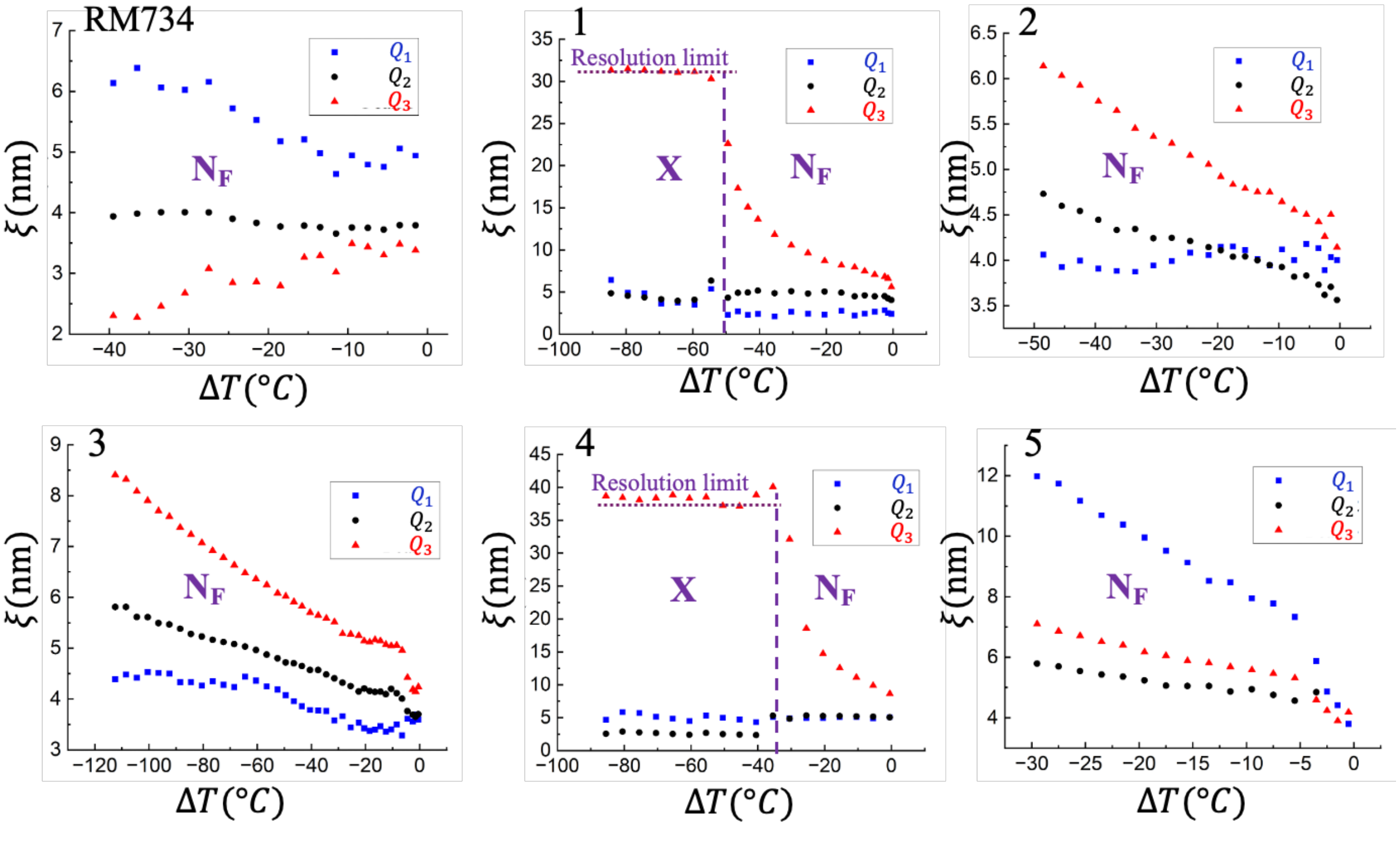

*Figure 5: Relative temperature ($\Delta T = T - T_{I-N_F}$) dependences of the correlation lengths $\xi = 2\pi/FWHM$, where FWHM stands for full width at half maximum of the peak, for RM734 and for the compounds **1-5**.*

We calculate correlation length, ξ, as 2π divided by full width at half maximum of the peak. The temperature dependences (relative to the $N_F$-Iso transition temperature) of all three correlation lengths are shown inFigure 5. In the case of RM734, on cooling, the correlation length associated with the peak at $Q_1$ increases from 2 $nm$ to 6 $nm$, i.e., from about 1 to 3 molecular lengths, while the correlation length associated with the peaks at $Q_2$ and $Q_3$ remain constant at $\sim 4\ nm$ and decrease from $\sim 4\ nm$ to $\sim 2\ nm$, respectively. These values are typical for nematic materials with very short spatial correlations.[55] In contrast to RM734 and similar $N_F$ compounds, the correlation length for short range ordering in the thiophene compounds **1** – **5** is the greatest at the length scale *L*/3 (i.e., the width of the peak at $Q_3$ is the narrowest). Except for compound **5**, the correlations associated with ordering at $L$ have the shortest range (of the three), the opposite of what we observe in RM734, and what has been reported for another prototypical $N_F$, the compound DIO.[24] Compounds **2**, **3** and **5**, i.e., those that have a crystal phase below the $N_F$, have relatively short (between 2nm and 12nm ) correlation lengths; these typically increase as the temperature is decreased. In compounds **1** and **4,** the correlation length associated with the peak at $Q_3$ shows a strong increase in the $N_F$ phase on cooling, reaching $\sim 30\ nm$ before transitioning to the highly ordered LC phase where the peak width reaches our experimental ability to measure, i.e. resolution limit, $\Delta Q \approx 0.018\ nm^{-1}$. On the other hand, the correlation lengths associated with the peaks at $Q_1$ and $Q_2$ are in the $2 - 8\ nm$ range and are almost temperature independent .

### *D. Polarization and density measurements*

Figure 6(a) shows the relative temperature $(T - T_{I-N_F})$ dependences of the ferroelectric polarizations of compounds **1-5** determined from the area of the current peak above the baseline determined by the ohmic current measured under 50 $Hz$ triangular voltages. It is found that $P_S$ is independent of the frequency in the range $40\ Hz \leq f \leq 100\ Hz$. A representative time dependence of the current flowing throught the sample is shown in the lower inset for compound **4** at 176 °C under 50 $Hz$ $30V$ amplitude triangular voltage. It can be seen that the current peak due to polarization switching is dominating over the resistive and capacitive currents. For all compounds $P_S$ increases sharply below the I-$N_F$ transition reaching about 90% of the highest value within 3 °C below $T_{I-N_F}$. Further away from $T_{I-N_F}$, $P_S$ increases only slightly, reaching $P_S \approx 7.6 - 8\ \mu C/cm^2$ by $T - T_{I-N_F} \approx -15\ °C$. The upper inset shows the sketch of the in-plane cell used for the polarization measurements. The red arrow shows the rubbing direction of the PI deposited on the bottom plate and of the direction in which the compound is filled into the cell.

Figure 6(b) shows the voltage dependence of the apparent polarization for compound **4** at 176 °C. The polarization peak becomes symmetric similar to the situation shown in the inset to Figure 6(a) above 2.5 V. Below that the time dependent current is asymmetric: there is an increase of the slope under $V > 0V$ when the voltage increases from $V < 0V$, whereas the current slope is constant under decreasing voltages, but reaching minimum above the lowest voltage value. This asymmetric behavior might indicate some polar arrangement below 2.5 $V$. This is likely related to a small pretilt $\alpha$ of the director yielding a small vertical component of the polarization $P = P_s sin\alpha$, and consequently a depolarization field $E_{dep} = -\frac{P_s sin\alpha}{\varepsilon_o \varepsilon}$.

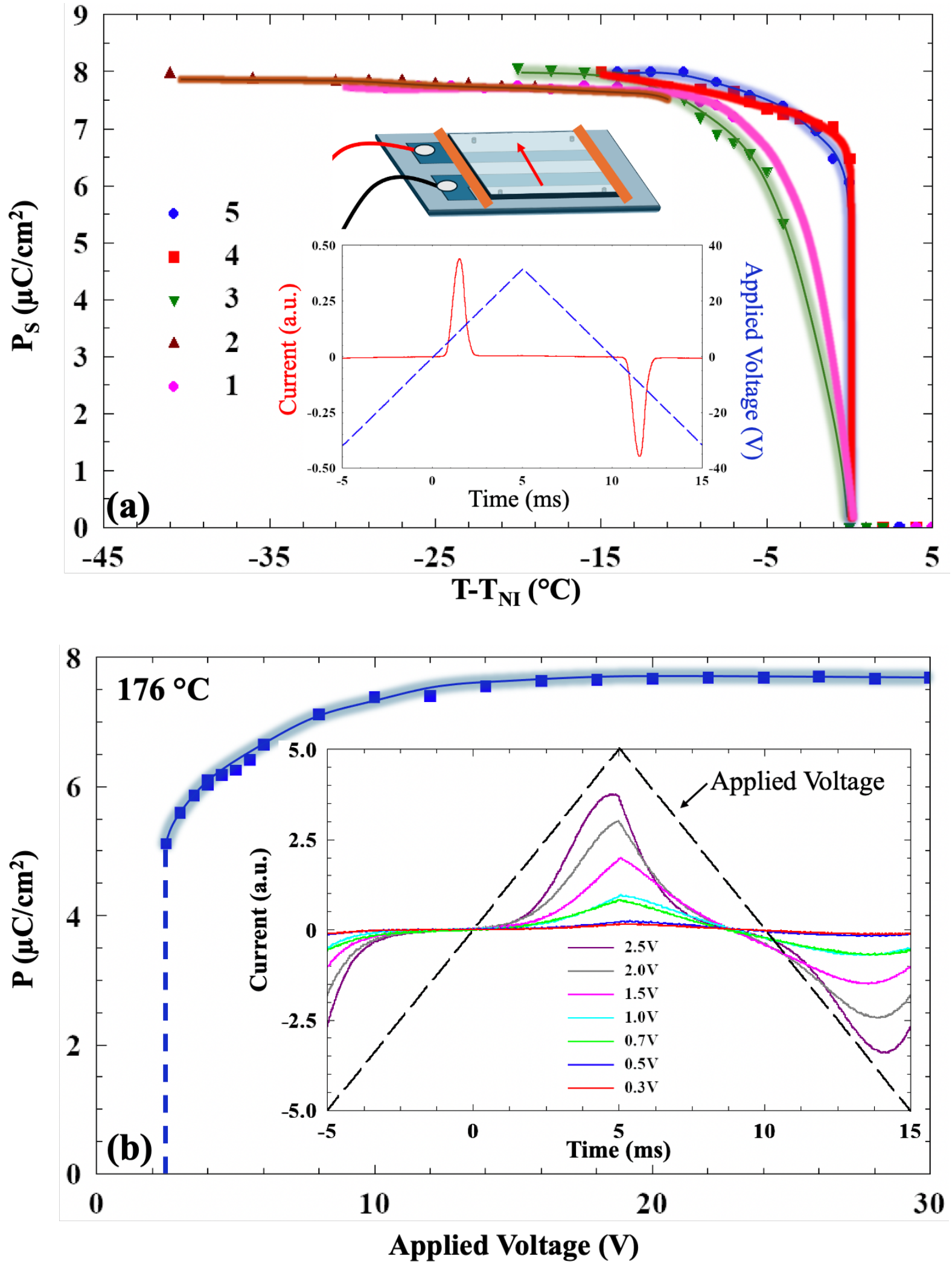

*Figure 6: Summary of the polarization measurements of the studied thiophene compounds. Main pane: Relative temperature dependences of the measured polarizations of compounds **1-5** obtained at $f = 50\ Hz$ calculated from the area of the current peak above the baseline determined by the ohmic current as shown for compound **4** at 176 °C in the lower inset. Upper inset illustrates the geometry of the cells used for the measurements.*

It is significant that the measured saturated polarization values $P_s$ are consistently larger than $P_S \approx 6.4\ \mu C/cm^2$ that has been both reported[1] and confirmed by us for RM734, which actually has a greater dipole moment than any of compounds **1-5**. To resolve this apparent contradiction we look to Eq. 1. With similar polar order parameter and molecular mass, a greater polarization with lesser dipole moment is only possible with correspondingly greater mass density. Numerical estimates for RM734[1], and precision density measurements on another room temperature $N_F$ mixture[5] give their densities $\varrho \approx 1.3\ \mathrm{kg/m^3}$; values that are already 30% larger than for typical nematic liquid crystals.

Without having sufficient quantities of our compounds for commercial density instruments[5], we conducted a preliminary density determination by measuring column length of a known mass of material confined without air bubble in a uniform cylindrical capillary. These measurements give $\varrho_{1-5} \approx 1.54 \pm 0.03\ \mathrm{kg/m^3}$ for compounds **1-5**, and $\varrho_{RM734} \approx 1.3 \pm 0.03\ \mathrm{kg}/m^3$ for RM734. Thus, the compounds reported here, even with lower dipole moment have sufficiently greater density to explain their larger polarization. The implications of this are discussed below.

## 3. Discussion

We synthesized and studied the physical properties of highly polar rod-shaped compounds where a thiophene ring replaced the 2,4-dimethoxybenzene ring of RM734. This change made the following notable changes compared to RM734 and most other $N_F$ materials in this class. (i) On cooling there is no paraelectric nematic phase below the isotropic phase; instead, the studied thiophenes transition directly from the isotropic to $N_F$ phase. (ii) There are strong spatial correlations corresponding to one third of the molecular length instead of the weak spatial correlations with periodicity corresponding to the full molecular length. (iii) The thiophene compounds with a nitro group or two cyano polar groups on the C ring have a smectic phase instead of the usual crystal structure below the $N_F$ phase. (iv) All studied thiophene compounds have ferroelectric polarizations of $P_S \approx 7.8 \pm 0.3\ \mu C/cm^2$, which is about 20% larger than of RM734. (v) These compounds have about 20% greater mass density compared to RM734 and most of the $N_F$ materials with short terminal chains.

We propose that all these features can be attributed to a denser packing of our thiophene compounds **1-5** compared to other $N_F$ materials having short terminal chains. When molecules are more closely packed, the interaction distance decreases which correspondingly increases the electrostatic forces that prefer the permanent dipoles to be oriented either oppositely in the same plane, or in a common direction when not positioned in the same plane. This has been discussed in various phases,[56–58] including recently by Madhusudana for the $N_F$ phase.[59] In Madhusudana's model, the $N_F$ molecules are assumed to a) freely rotate about their axes, and b) have non-uniform electric charge density along their axis due to the dipolar groups attached to the highly polarizable phenyl rings. In this scenario, the electrostatic interaction energy between laterally adjacent molecules is most favorable when lower electron densities (a net positive charge) and higher electron densities (a net negative charge) are nearest.

Our thiophene molecules have three major dipoles: the first one corresponds to a dipole ($NO_2$, CN or F) attached to the C-ring; the second one is an ester group connected to the B ring; and the third one is another ester group connected to the thiophene ring. We used the molecular structure of **1** to sketch the possible molecular packing in the $N_F$ phase and in the underlying smectic phase (see Figure 7(a and b), respectively). The arrangement corresponding to the lowest electrostatic energy would result in a crystal structure and is not nematic. However, in the $N_F$ temperature range, the thermal motion tends to randomize the positive-negative arrangement, yet still provide ferroelectric packing as as we illustrate in Figure 7(a). Whenever the positive-negative arrangement shift in the same direction for larger number of molecules, temporary skewed smectic-like correlations are realized with $L$ periodicity. These are illustrated by dotted blue lines in Figure 7(a). The observed short-range order with dominating $L/3$ periodicity can be explained by cybotactic clusters with alternating dipole directions in adjacent molecules within a layer, as shown inside a brown frame in Figure 7(a). The pink dotted lines with glows indicate these $L/3$ length periodicities. Additionally, as illustrated with glowing light blue dotted lines, these clusters also lead to orthogonal $L$ periodicities. To explain the $L/2$ periodicity, we have to assume the presence of other type of clusters shown inside a purple ellipse, where the laterally neighbor molecules are shifted by half molecular length with alternating dipole directions. The fact that compound **5** with a fluorine atom in the central ring has a smaller peak with $L/2$ periodicity than with $L$ periodicity, might indicate that the fluorine atom of ring B of one molecule is avoiding the sulphur atom in ring A of the neighbor molecules. While in the majority $N_F$ volumes we assume the molecules freely rotate about their long axes (uniaxiality), in the cybotactic clusters we assume the planes of the aromatic rings tend to pack parallel to each other (in contrast to the sketch in Figure 7, where they shown in the same plane for simplicity). Accordingly, the smectic clusters represent biaxial domains.

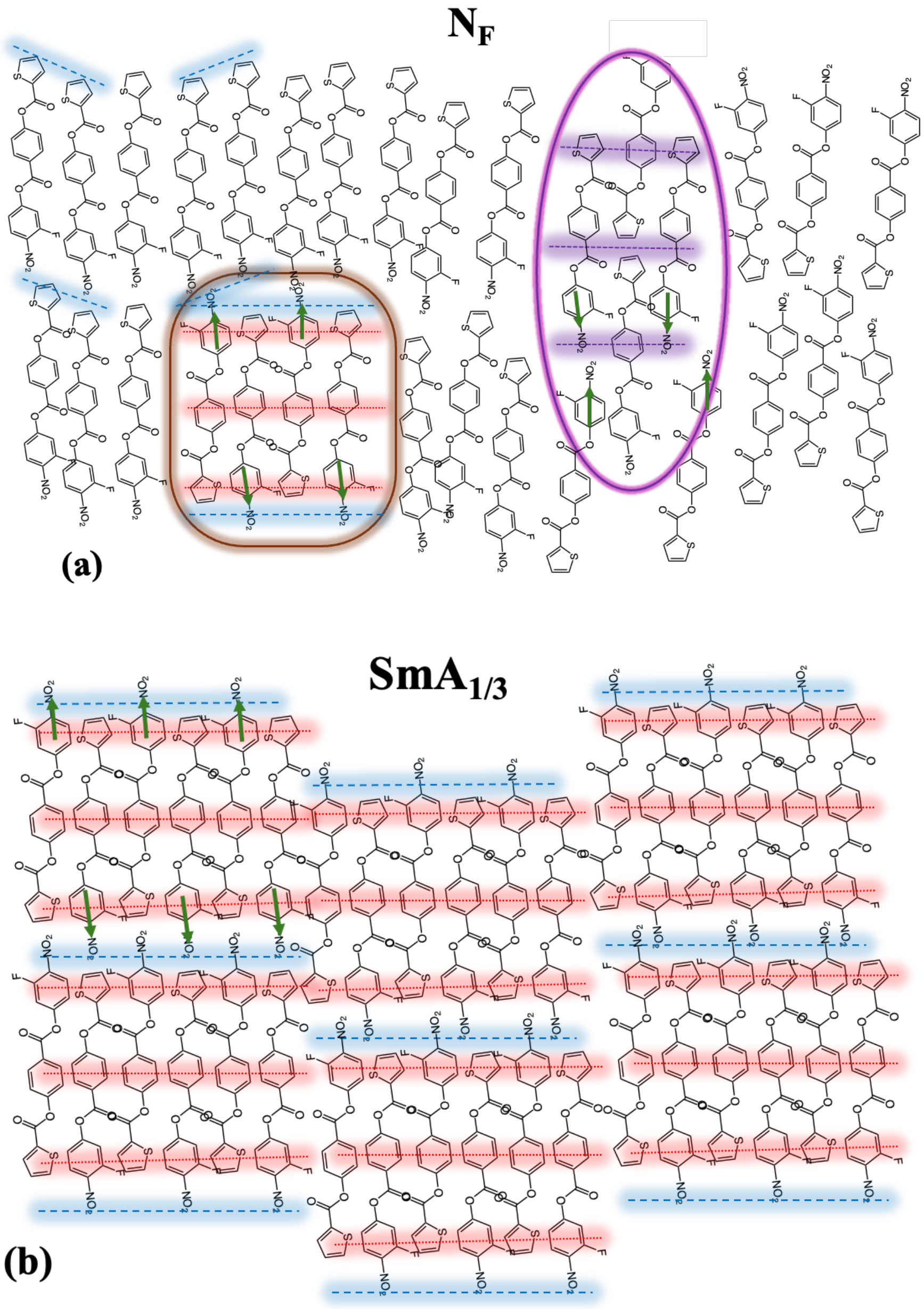

*Figure 7: Schematic illustration of the molecular packing in the $N_F$ phase of compounds* ***1-5*** *(a) and in the smectic phase of compounds* ***1*** *and* ***4*** *(b) using the molecular structure of compound* ***1****. There are three aromatic rings C, B and A separated by ~L/3 distances. The pink, purple and light blue lines with glows indicate the L/3, L/2 and L length periodicities. Green arrows indicate the dipole moment of a molecule. In the majority $N_F$ volumes of (a) we assume the molecules freely rotate about their long axes (uniaxiality), while in the cybotactic clusters we assume the planes of the aromatic rings tend to pack cofacial to each other although, in the sketch they shown in the same plane (non-cofacial) for simplicity.*

A tentative and simplified model of the smectic phase with $L/3$ long-range periodicity in compounds **1** and **4** is shown in Figure 7(b). Although in the neighbor domains the aromatic rings with alternating dipole directions pack together, they do so randomly, i.e., the A ring can stick to the B or C, etc. Such arrangement will provide a long-range $L/3$ periodicity and only short-range order with $L$ and $L/2$ periodictity. Clusters with $L/2$ periodictity are not shown in Figure 7(b) for simplicity. This is a good approximation for compound **4** where the diffraction associated with the $L/2$ periodictity is extremely weak. In addition to neglecting the domains with $L/2$ periodicity, we also have neglected showing domains with slightly different orientations that we observed by X-ray diffraction (see bottom-right image of Figure 3).

To emphasize the 1/3 molecular length periodicity of this paraelectric smectic phase, we denote it as $SmA_{1/3}$ phase. The proposed structure of this phase contrasts sharply with the polar orthogonal smectic ($SmA_F$) phase observed previously for several other $N_F$ materials[52,60] with one molecular length ($L$) periodicity and with macroscopic polarization. To the best of our knowledge such a $SmA_{1/3}$ smectic phase has hitherto never been observed. Note, that for all $N_F$ materials with a flexible terminal chains, the separation between neighboring molecules (and consequently between neighboring dipoles) is larger, and the correlation lengths (even between the full molecules) are shorter than for the thiophene molecules lacking flexible tails.

## 4. Methods

### *A. Chemical Synthesis*

The synthesis of these materials generally follows the methodology outlined in the scheme shown in Figure 8. This scheme illustrates two distinct approaches employed to synthesize the reference compound PN03111.

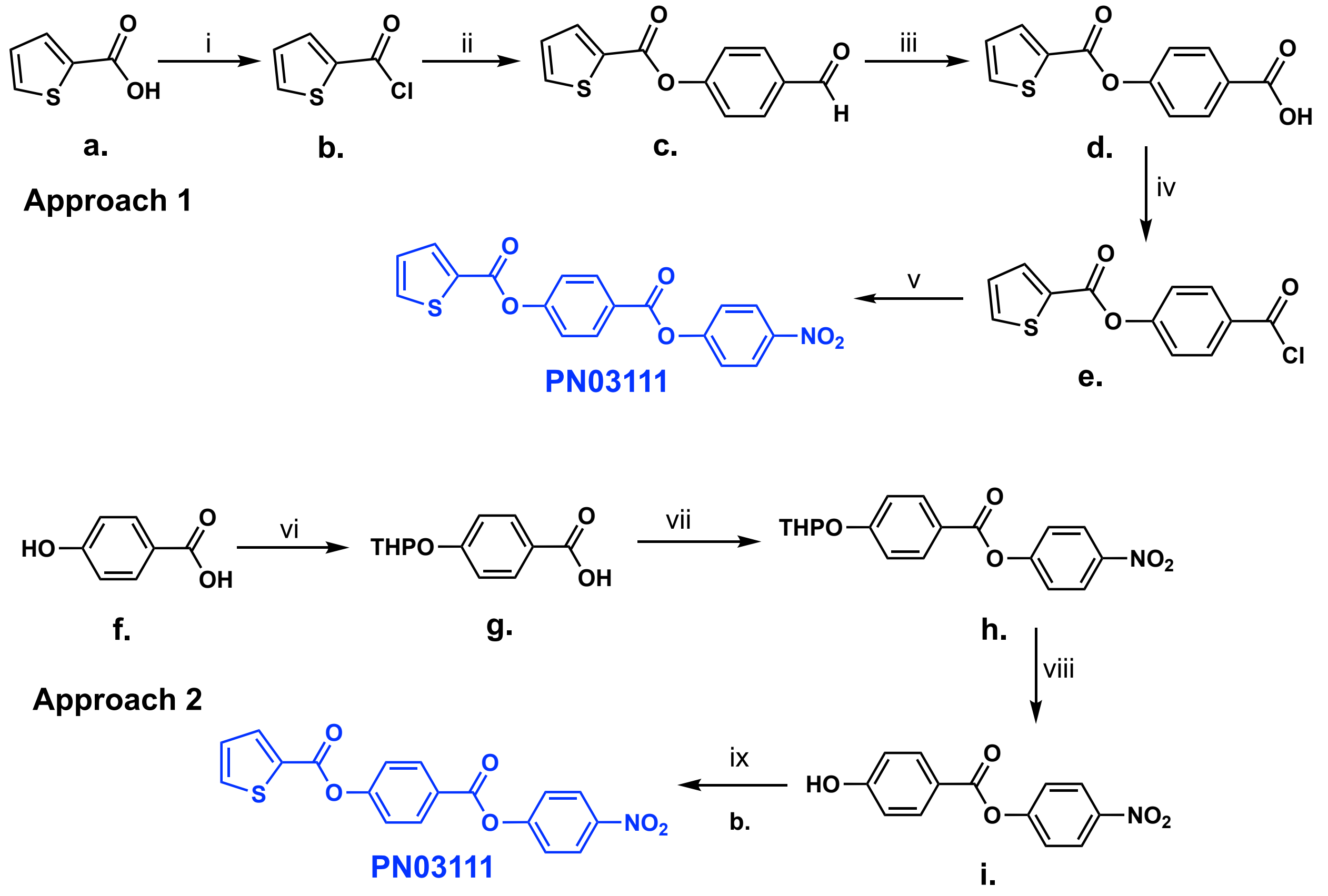

i. Oxalyl chloride, DCM (quantitative)
ii. 4-hydroxybenzaldehyde, pyridine (85%)
iii. Oxone (quantitative)
iv. Oxalyl chloride, DCM (quantitative)
v. 4-nitrophenol, pyridine (80%)
vi. DHP, PPTS, THF (quantitative)
vii. DCC, DMAP, DCM (70%)
viii. PTSA, Ethanol (quantitative)
ix. Pyridine (75%)

*Figure 8: Synthetic routes that are implemented for the synthesis of PN03111 material.*

In **Approach 1**, the synthesis begins with the assembly of the target molecule from the left side (A-ring), sequentially adding the central (B-ring) and right-side (C-ring) components. Conversely, **Approach 2** involves initial assembly of the central (B-ring) and right-side (C-ring), with the A-ring introduced in the final step. The primary synthetic methodology employed in both approaches is esterification. Notably, even when the acid chloride precursor is not commercially available and must be prepared, the acid chloride route is often preferred due to a significant reduction in the need for chromatographic purification during synthesis.

**Approach 1** utilizes the commercially available precursor 2-thiophenecarboxylic acid (A-ring). This was first converted to its corresponding acid chloride using oxalyl chloride (reaction not shown). The acid chloride was then reacted with 4-hydroxybenzaldehyde in an esterification step to yield intermediate **b**. The aldehyde functional group of **b** was subsequently oxidized to a carboxylic acid using Oxone, producing the monoester carboxylic acid **c**. Intermediate **c** was converted to its acid chloride (reaction not shown), which was then subjected to a final esterification reaction with 4-nitrophenol to produce the target compound PN03111. This approach has high efficiency, with reaction yields ranging from 80% to higher, and provided clean products without requiring extensive purification steps.

**Approach 2** starts with the commercially available precursor 4-hydroxybenzoic acid (**d**, B-ring). The phenolic functional group was initially protected with dihydropyran (DHP), yielding the THP-protected intermediate **g**. Esterification of **g** with 4-nitrophenol using the Steglich method followed by subsequent removal of the THP protecting group afforded the phenolic monoester intermediate **h**. Finally, after esterification reaction between **h** and 2-thiophenecarbonyl chloride (commercially available) produced the target compound PN03111. Both approaches provide viable synthetic routes to PN03111, with the key distinctions lying in the order of ring assembly and the handling of intermediates. Synthesis details are provided in the SI.

## *B. Polarizing Optical Microscopy (POM)*

For POM measurements an OLYMPUS BX60 microscope equipped with a Nikon D5600 camera was used. The LC samples were placed in heat stages controlled by METTLER FP90 temperature controller. 5 μm LC films were studied in cooling with 1°C/min cooling rates.

## *C. X-ray measurements*

Small angle X-ray scattering (SAXS) measurements were carried out in the Advanced Materials and Liquid Crystal Institute of Kent State University on a Xenocs Xeuss 3.0 SAXS/WAXS imaging beamline equipped with a Cu $K_\alpha$ source producing $8.05\ keV$ X-rays (1.542 Å wavelength). The beam size at the sample was $0.9\ mm$, the sample-detector distance was $140\ mm$, and the detector's pixel size was $75\mu m\ x\ 75\mu m$. The detector's active area (including detector gaps) was $77.1\ mm\ x\ 79.65\ mm$. The resolution limit in scattering wavenumber $Q$ of the recorded SAXS patterns is estimated to be $\Delta Q{\sim}0.18\ nm^{-1}$, which corresponds to $2\pi/0.18 \sim 35\ nm$ maximum detectable correlation length. The SAXS patterns were processed using Datasqueeze software to obtain the azimutally averaged scattered intensity, *I(Q)* The latter was fit to the sum of three

Pseudo-Voigt functions and a fixed background, which yielded values for the FWHM of the individual peaks at $Q_1$, $Q_2$, and $Q_3$ presented in Figure 4.

### *D. Ferroelectric Polarization measurements*

The ferroelectric polarizations were measured using in-plane cells with $d = 5\ mm$ electrode gap as shown in the inset to Figure 6(a). The electrode width and length are $w = 2\ mm$ and $l \approx 1\ cm$ respectively . The thickness of the cell is achieved by using $h = 10 \pm 0.2\ \mu m$ spacers. Triangular pulses of frequency $f$ and peak-to-peak voltage of $V$ were applied at various constant temperatures. The time dependence of the polarization current $I_P(t)$ is calculated by measuring the voltage drop ($V_m$) across a $R = 20\ k\Omega$ resistor. After the polarization current drops back to the baseline with slope determined by the ohmic current $I_\Omega$, the area above the ohmic baseline is calculated to give the spontaneous polarization value $P_s$ as $P_s = \frac{\int V_m dt}{2R(l\cdot h)}$.

## 5. Acknowledgement

This work was financially supported by US National Science Foundation grant DMR-2210083.

# Supporting Information

## Ferroelectric Nematic and Smectic Liquid Crystals with Sub-Molecular Spatial Correlations

Parikshit Guragain[a,+], Arjun Ghimire[b,+] Manisha Badu[b], Netra Prasad Dhakal[c], Pawan Nepal[a], James T. Gleeson[b], Samuel Sprunt[b,c,*], Robert J. Twieg[a,*], Antal Jákli[b,c,*]

[a] Department of Chemistry and Biochemistry, Kent State University, Kent, OH 44242, USA
[b] Department of Physics, Kent State University, Kent, OH 44242, USA
[c] Materials Science Graduate Program and Advanced Materials and Liquid Crystal Institute, Kent State University, Kent, OH 44242, USA
*: Corresponding authors: rtwieg@kent.edu, ssprunt@kent.edu and ajakli@kent.edu
+: These authors contributed equally to this work

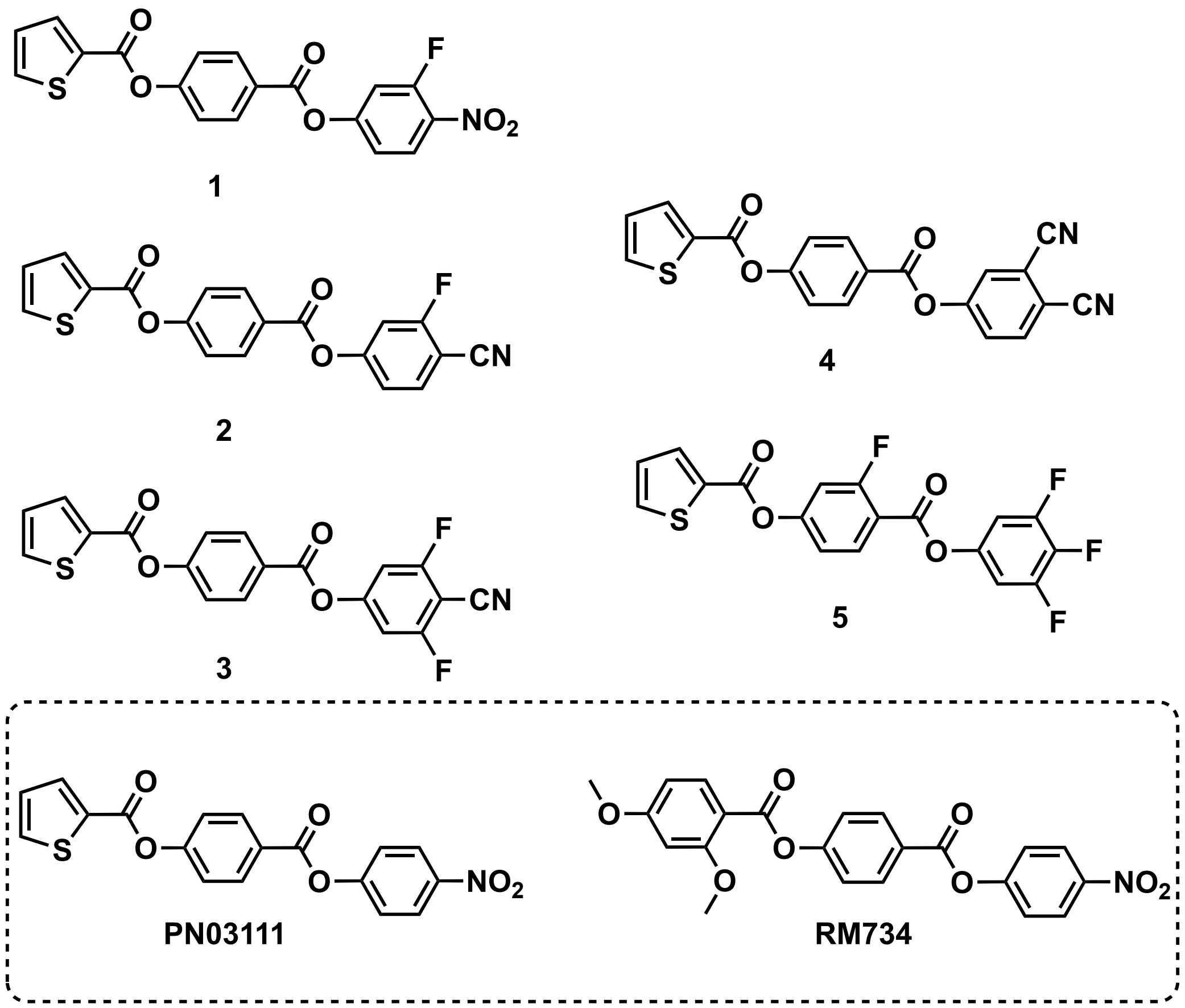

**Figure S1.** *Structures of target compounds **1-5** and reference compounds PN03111 and RM734*

## A. Preparation and characterization of intermediates

*2-thiophenecarbonyl chloride*

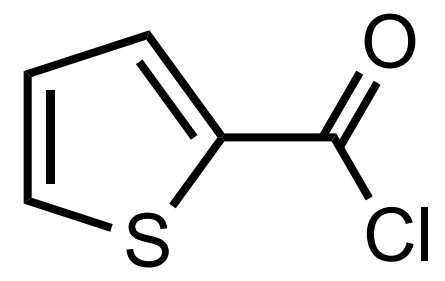

In a 500 ml recovery flask with stir bar, addition funnel and bubbler were placed thiophene-2-carboxylic acid (12.8 gm, 100.0 mmol), dry dichloromethane (100 ml) and dimethylformamide (two drops). The resulting mixture was chilled in an ice water bath and oxalyl chloride (19.05 gm, 150.0 mmol) in dry dichloromethane (25 ml) was added dropwise with stirring over about thirty minutes. The mixture was allowed to warm to room temperature and was stirred overnight at room temperature. The next day completion of reaction was tested by warming the solution with a heat gun. As the mixture cooled there was no gas evolution seen and so the reaction is complete. The mixture was concentrated by rotary evaporation, dry dichloromethane (25 ml) was added and then removed by rotary evaporation. The acid chloride prepared in this fashion (less than or equal to 100 mmol) is suitable for direct use in the subsequent reaction.

All other acid chlorides were prepared by this method and used without further purification.

*4-(thiophene-2-carbonyloxy) benzaldehyde*

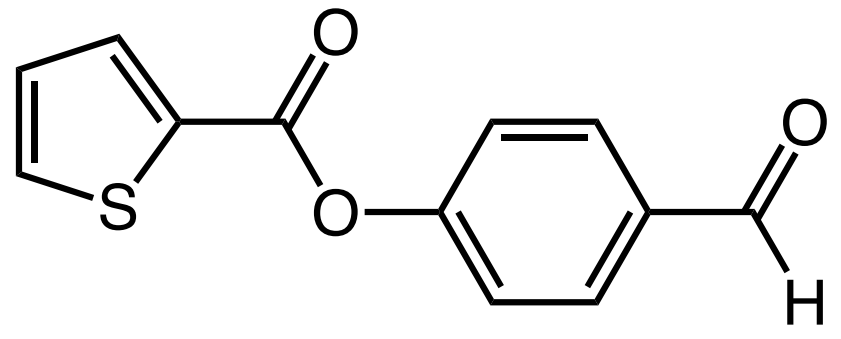

In a 500 ml recovery flask with stirbar and nitrogen inlet was placed 2-thiophenecarboxylic acid (3.84 gm, 30.0 mmol) and dry dichloromethane (100 ml). The resulting mixture was stirred in an ice bath a few minutes and then DCC (7.43 gm, 36.0 mmol) was added all at once with the assistance of some dichloromethane (20 ml). The mixture was stirred a few minutes and then DMAP (185 mg, 1.5 mmol) and 4-hydroxybenzaldehyde (3.66 gm, 30.0 mmol) were added. The mixture was allowed to warm to room temperature and was stirred overnight. TLC indicated the reaction was complete. Silica gel (60 cc) was added, and the mixture was concentrated to dryness by rotary evaporation. The impregnated silica gel was added to a column of silica gel made up with dichloromethane and eluted with dichloromethane. Fractions containing pure product were combined and concentrated to give the product (5.58 gm, 80%) of sufficient purity for subsequent oxidation.

**$^{1}$H NMR (500 MHz, $CDCl_3$)** δ 10.03 (s, 1H), 8.02 (dd, *J* = 3.8, 1.3 Hz, 1H), 7.97 (d, *J* = 8.5 Hz, 2H), 7.72 (dd, *J* = 5.0, 1.3 Hz, 1H), 7.43 (d, *J* = 8.5 Hz, 2H), 7.21 (dd, *J* = 5.0, 3.8 Hz, 1H).
**$^{13}$C NMR (126 MHz, $CDCl_3$)** δ 190.92, 159.84, 155.25, 135.24, 134.19, 134.11, 132.14, 131.26, 128.24, 122.45.

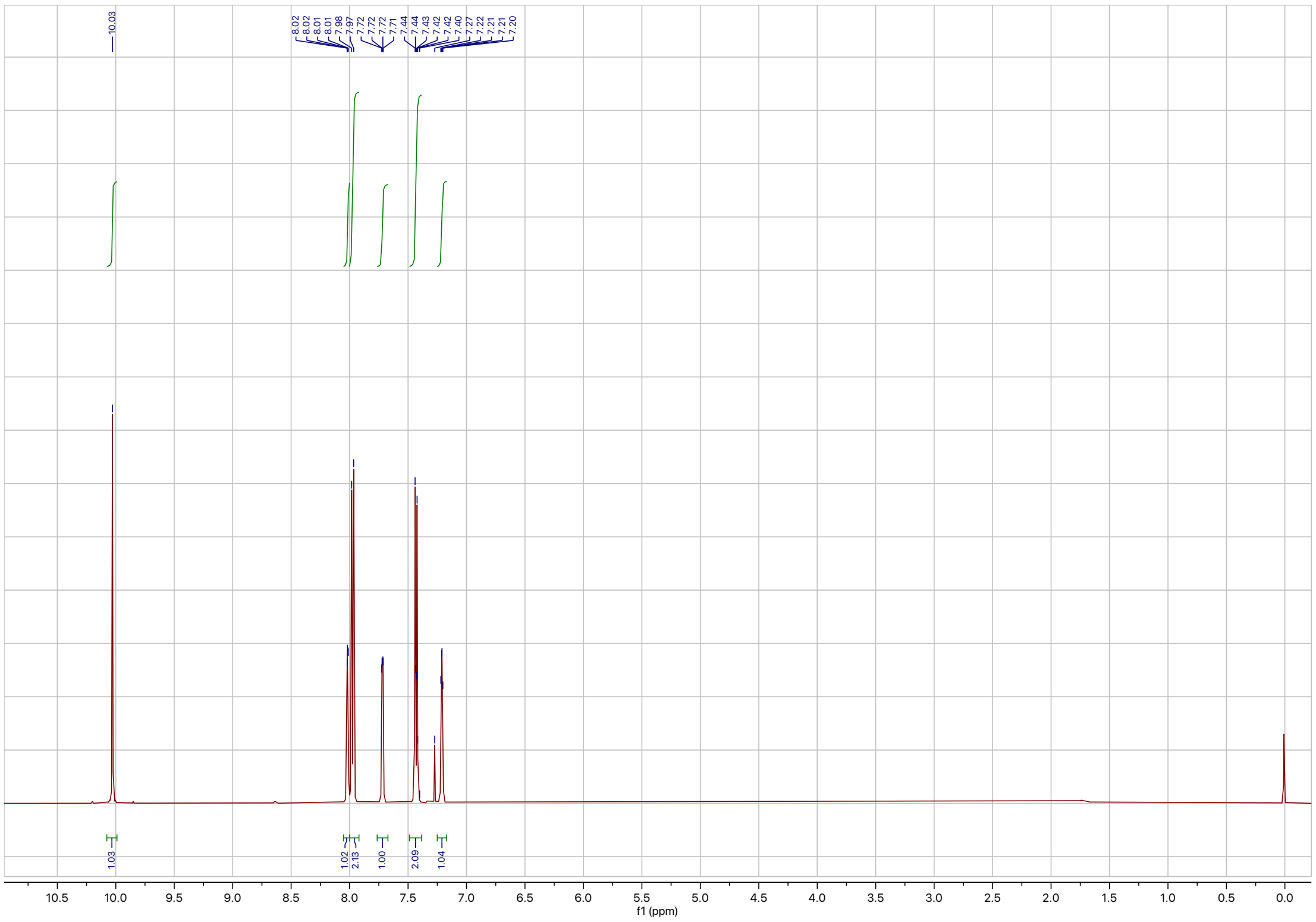

**Figure S2**. *$^{1}$H-NMR of 4-(thiophene-2-carbonyloxy) benzaldehyde*

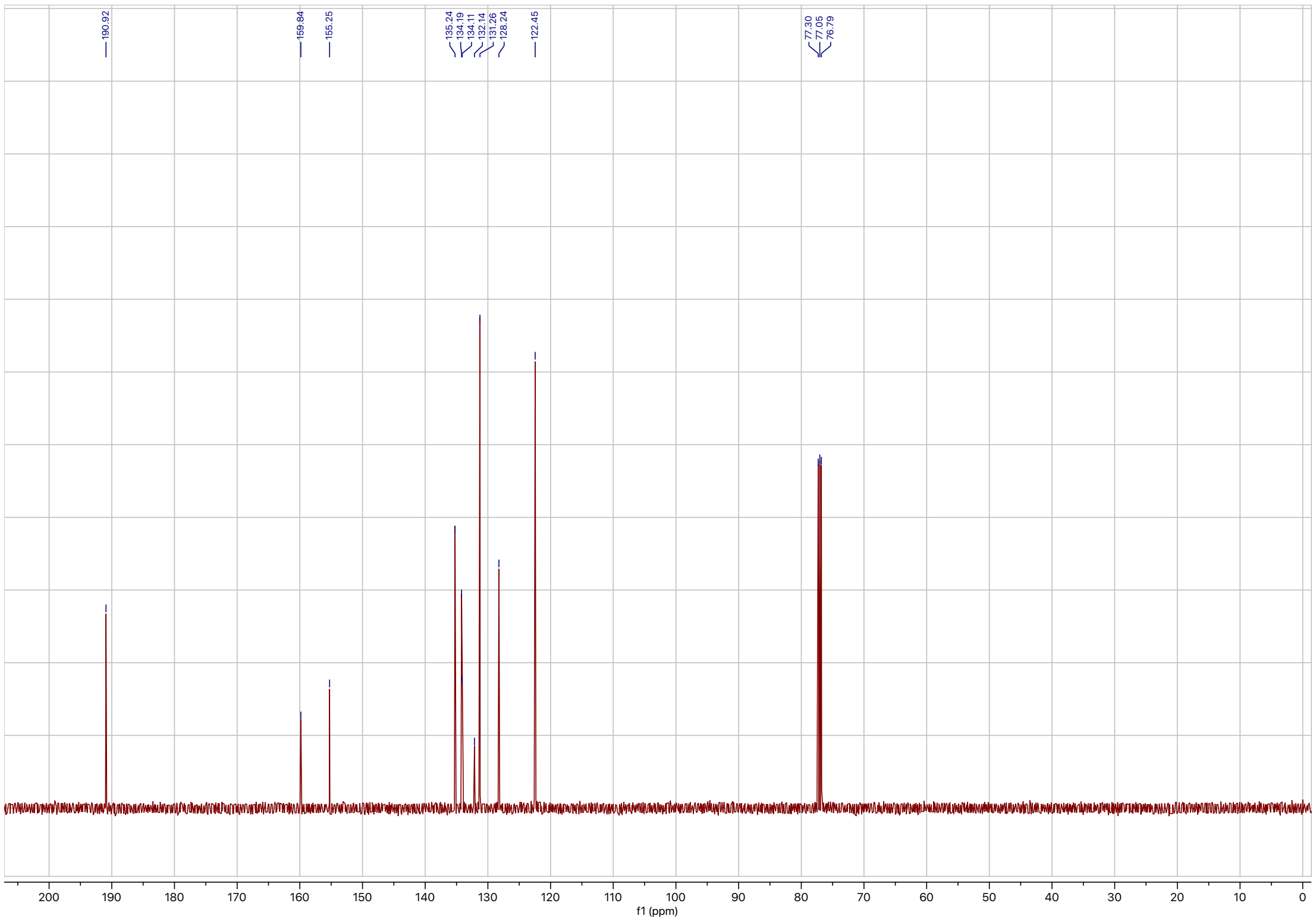

**Figure S3**. *$^{13}$C-NMR of 4-(thiophene-2-carbonyloxy) benzaldehyde*

*2-fluoro-4-(thiophene-2-carbonyloxy) benzaldehyde*

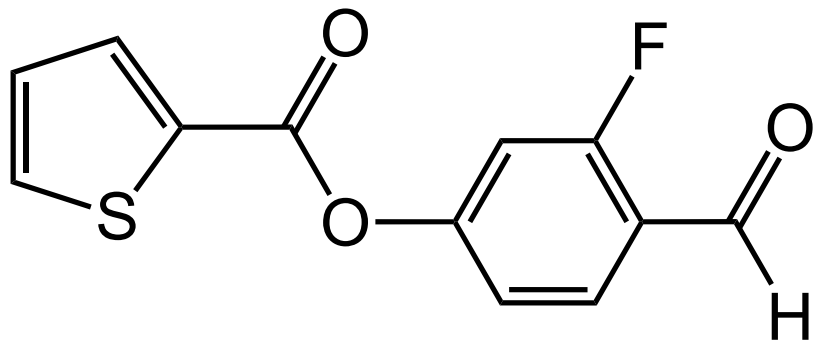

In a 200 ml round bottom flask equipped with a stir bar was placed 2-fluoro-4-hydroxybenzaldehyde (3.15 g, 22.50 mmol) and pyridine (30 ml) under nitrogen. The mixture was stirred until all the solid dissolved and then the flask was placed in an ice bath. The 2-thiophenecarbonyl chloride (3.30 g, 22.50 mmol) in DCM (10.0 ml) was added dropwise. The resulting mixture after thirty minutes was allowed to warm to room temperature and left stirring overnight. After this time, TLC indicated formation of a single less polar product. Solvent was evaporated by rotary evaporation then ice was added with vigorous stirring. More ice water was added to fill the flask. The precipitate obtained was vacuum filtered washed well with water, air dried and used in the subsequent reaction without further purification. Yield- 4.29 g (76 %)

**$^{1}$H NMR (500 MHz, $CDCl_3$)** δ 10.35 (d, $J$ = 0.8 Hz, 1H), 8.02 (dd, $J$ = 3.8, 1.3 Hz, 1H), 7.97 (dd, $J$ = 8.8, 7.8 Hz, 1H), 7.74 (dd, $J$ = 5.0, 1.3 Hz, 1H), 7.24 – 7.15 (m, 3H).
**$^{19}$F NMR (470 MHz, $CDCl_3$)** δ -118.96 (dd, $J$ = 11.2, 7.8 Hz).
**$^{13}$C NMR (126 MHz, $CDCl_3$)** δ 186.04, 186.00, 166.01, 163.94, 159.32, 156.30, 156.21, 135.53, 134.55, 131.65, 129.72, 129.70, 128.33, 122.02, 121.96, 118.37, 118.34, 110.61, 110.42.

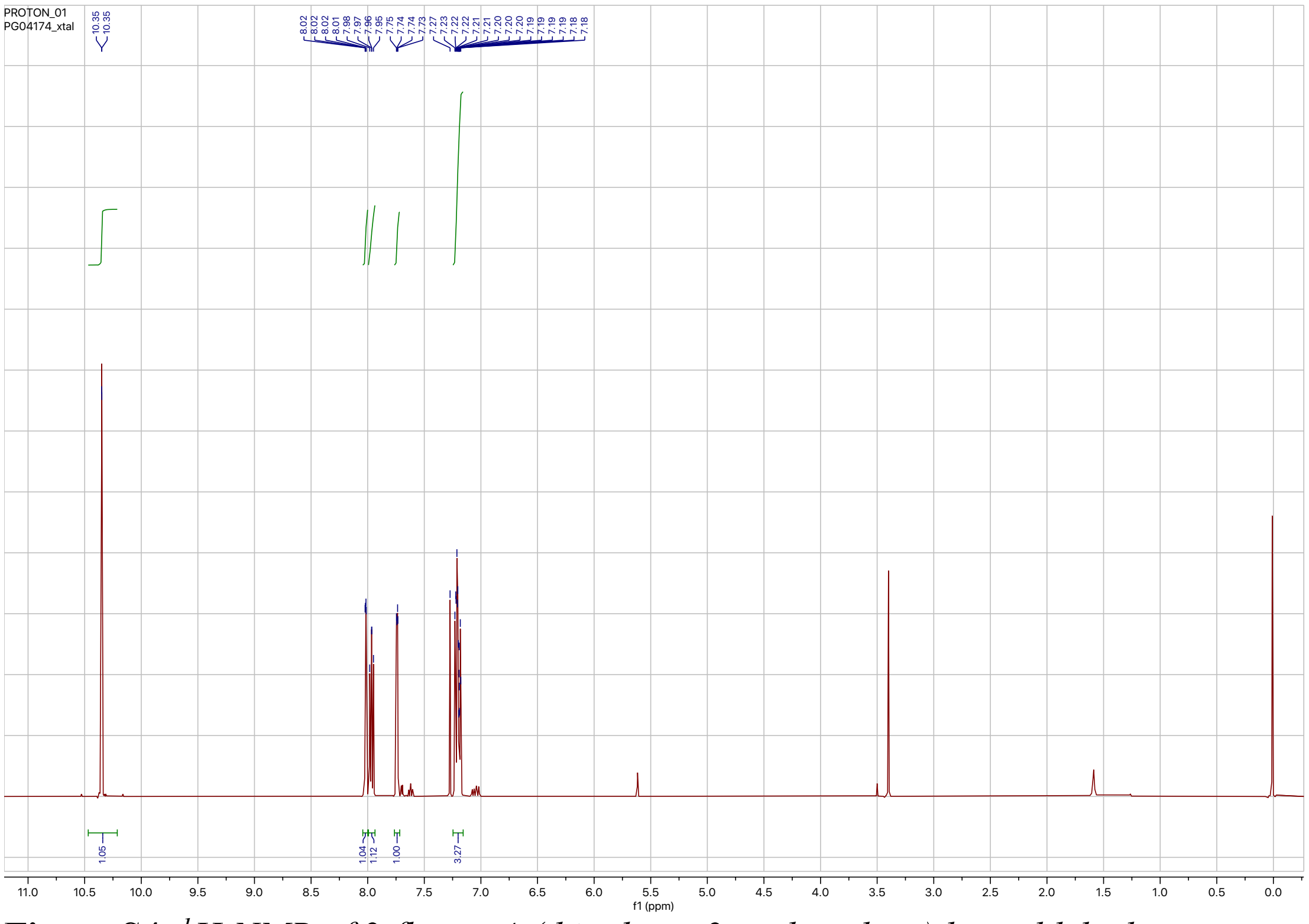

**Figure S4**. *$^{1}$H-NMR of 2-fluoro-4-(thiophene-2-carbonyloxy) benzaldehyde*

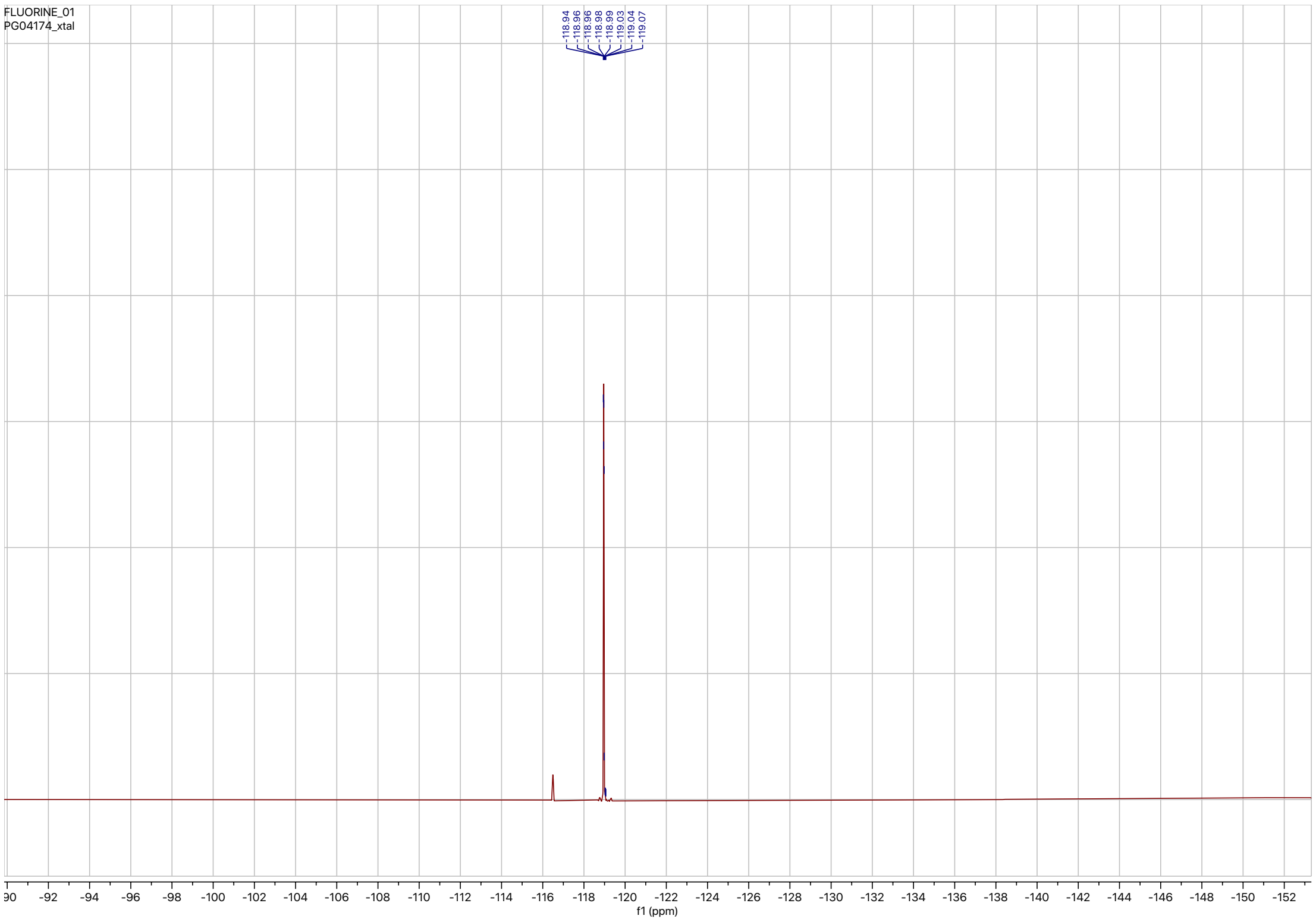

**Figure S5**. *[19]F-NMR of 2-fluoro-4-(thiophene-2-carbonyloxy) benzaldehyde*

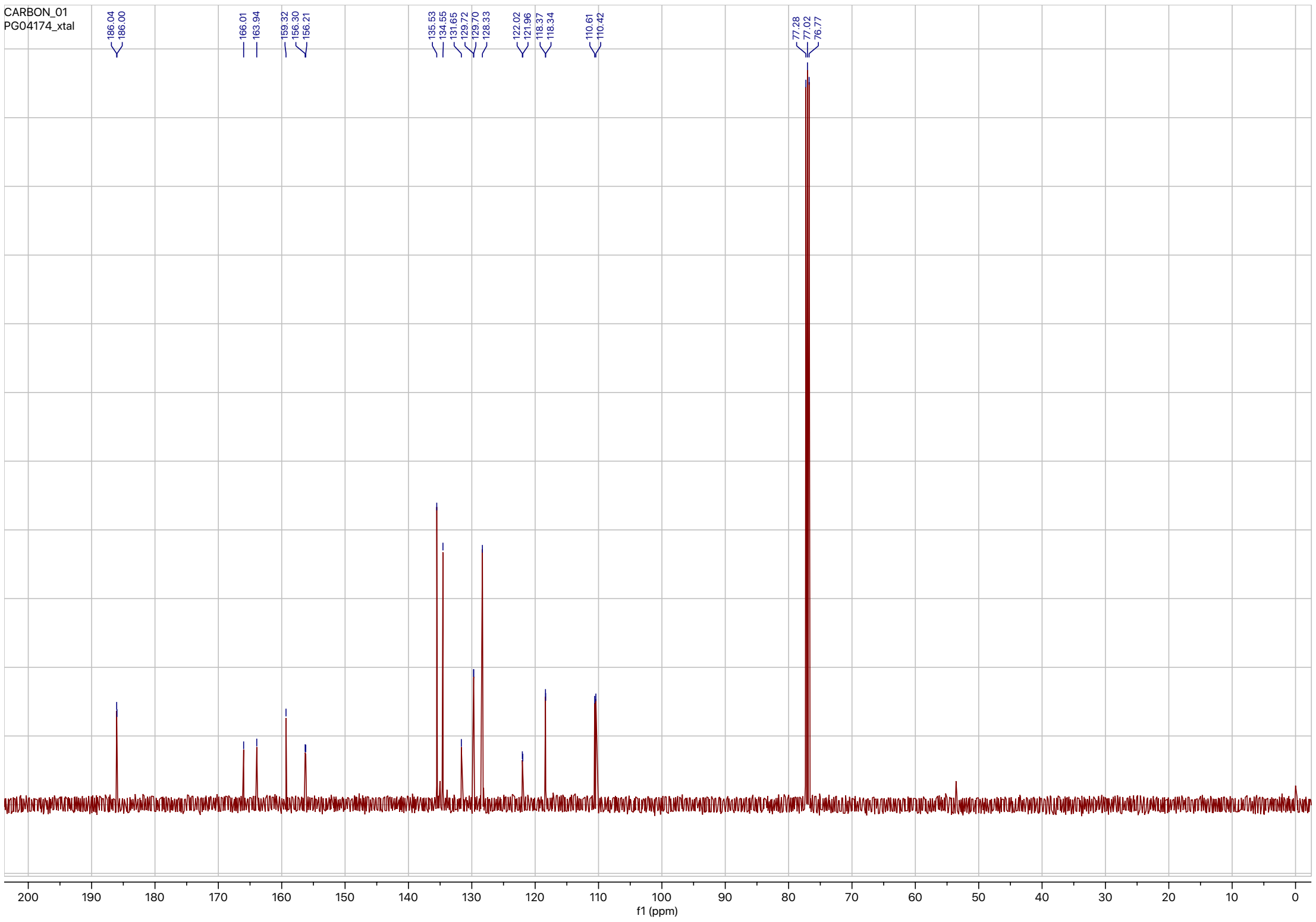

**Figure S6**. *[13]C-NMR of 2-fluoro-4-(thiophene-2-carbonyloxy) benzaldehyde*

*4-(thiophene-2-carbonyloxy) benzoic acid*

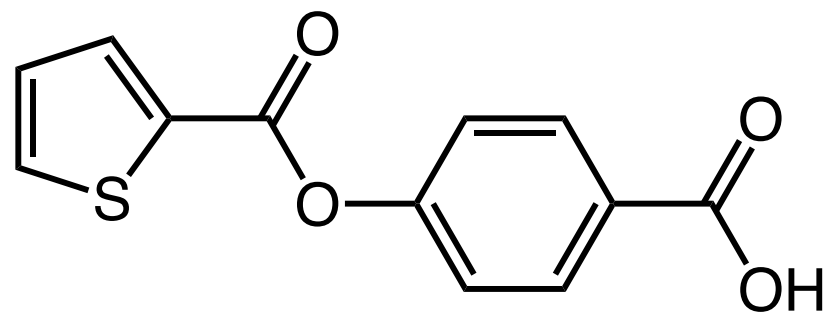

In a 500 ml recovery flask with stirbar and nitrogen inlet was placed the aldehyde (4.64 gm, 20.0 mmol) and dry DMF (50 ml). The mixture was stirred in a room temperature water bath until all dissolved and then Oxone (12.3 gm, 20.0 mmol) was added all at once. The mixture was stirred briefly in the water bath and then at room temperature with the bath removed. After six hours TLC indicated that the reaction was complete, and it was quenched by the dropwise addition of ice cold 2% aqueous hydrochloric acid. After completion of addition the mixture was stirred ten minutes and then the solid was isolated by suction filtration, washed with water and air dried overnight. The product (4.88 gm, 98%) is of sufficient purity for further use.

**$^1$H NMR (500 MHz, DMSO)** δ 8.11 (dd, *J* = 5.0, 1.3 Hz, 1H), 8.07 – 7.99 (m, 3H), 7.45 – 7.37 (m, 2H), 7.32 (dd, *J* = 5.0, 3.8 Hz, 1H).
**$^{13}$C NMR (126 MHz, DMSO)** δ 167.04, 160.14, 154.06, 136.16, 135.99, 131.92, 131.43, 129.29, 129.12, 122.62.

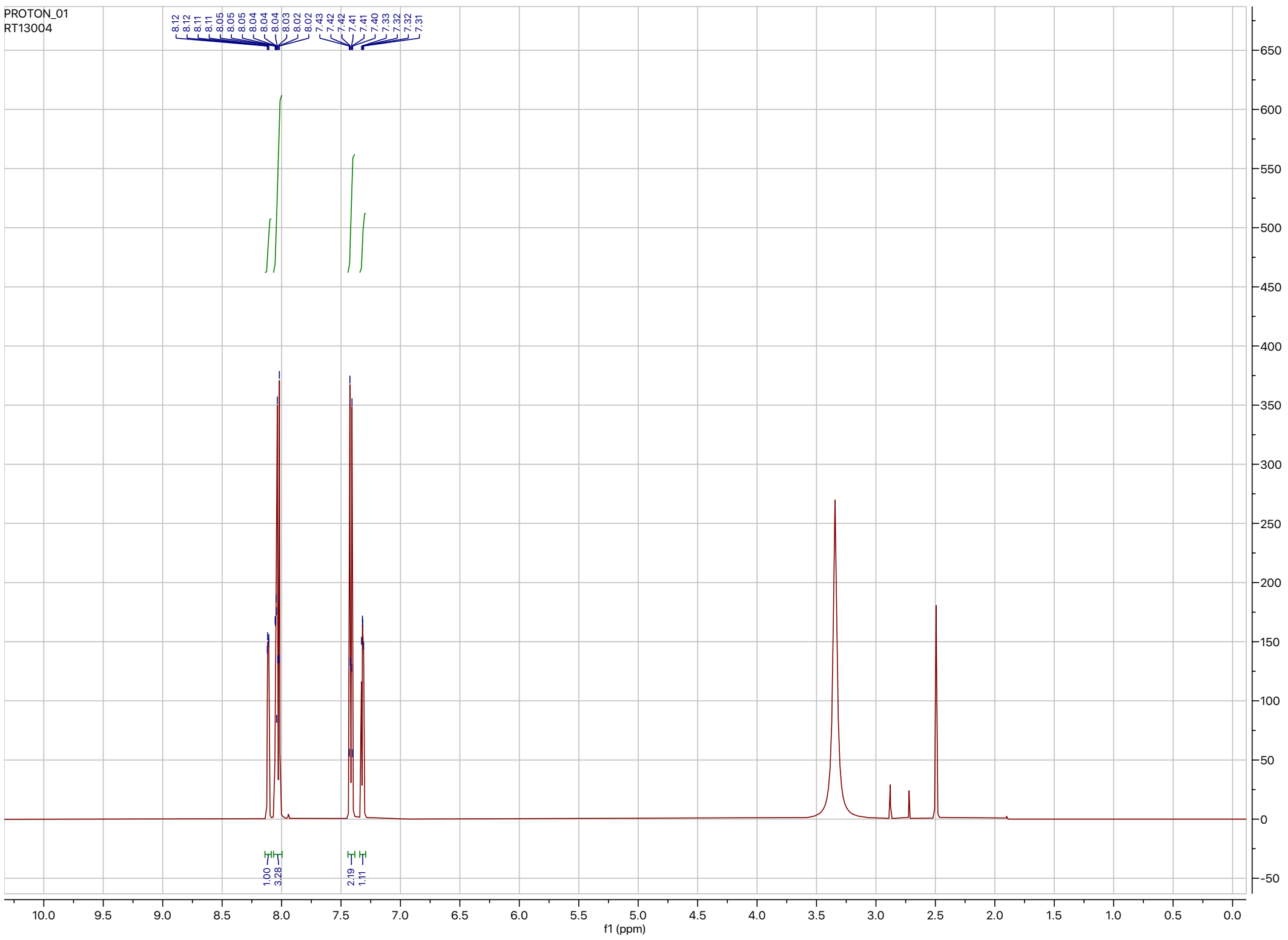

**Figure S7**. *$^1$H-NMR of 4-(thiophene-2-carbonyloxy) benzoic acid*

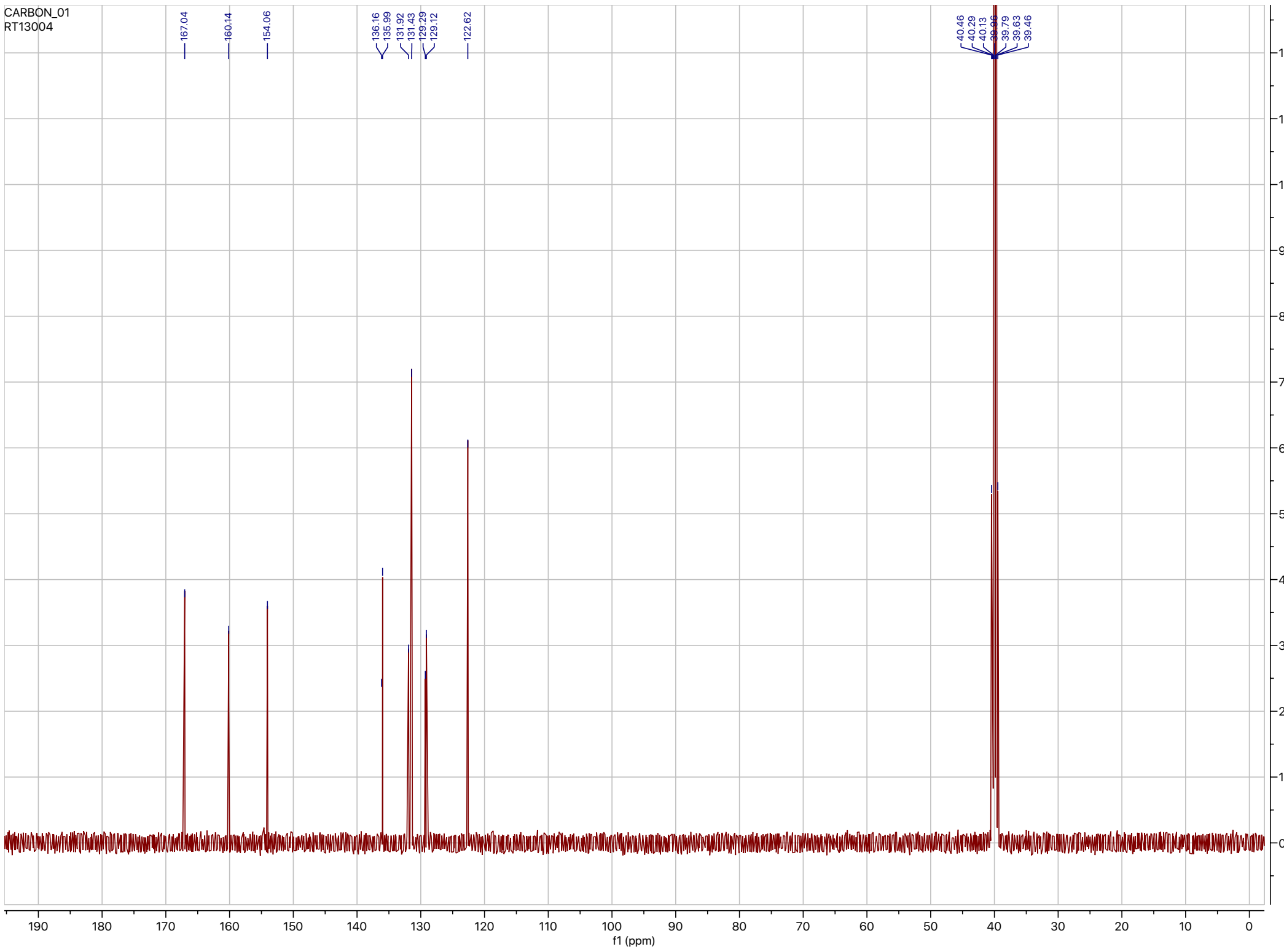

**Figure S8**. *$^{13}$C-NMR of 4-(thiophene-2-carbonyloxy) benzoic acid*

<u>*2-fluoro-4-(thiophene-2-carbonyloxy) benzoic acid*</u>

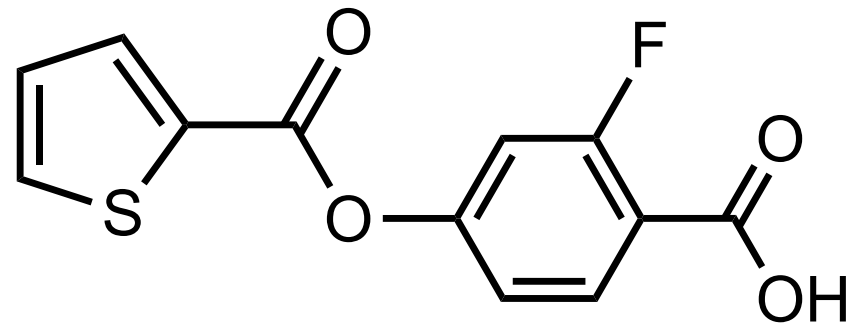

In a 500 ml round bottom flask equipped with a magnetic stir bar was placed 4-carbonyl-3-fluorophenyl-2-thiophenecarboxylate (4.25 g, 17.0 mmol) and DMF (60 ml). The mixture was stirred until all the solid dissolved then solid Oxone (11.0 g, 175.0 mmol) was added. The mixture was left stirring overnight at room temperature when all the starting material was found to be consumed showing a single polar product in TLC. The reaction was quenched by dropwise addition of 5% HCl (150.0 ml) then ice water was added to fill the flask and left stirring. The precipitate obtained was vacuum filtered and washed with water then air dried. The precipitate was recrystallized with a mixture of methanol and 1-propanol to obtain a white solid product. Yield- 4.26 g (94%)

**$^{19}$F NMR (470 MHz, DMSO)** δ -107.32 – -107.42 (m).

**$^{1}$H NMR (500 MHz, DMSO)** δ 8.13 (dd, *J* = 5.0, 1.3 Hz, 1H), 8.05 (dd, *J* = 3.8, 1.3 Hz, 1H), 7.96 (t, *J* = 8.5 Hz, 1H), 7.42 (dd, *J* = 11.4, 2.2 Hz, 1H), 7.32 (dd, *J* = 5.0, 3.8 Hz, 1H), 7.27 (dd, *J* = 8.6, 2.3 Hz, 1H).

**$^{13}$C NMR (126 MHz, DMSO)** δ 164.85, 164.82, 162.88, 160.83, 159.79, 154.58, 154.49, 136.42, 136.24, 133.40, 133.38, 131.58, 129.33, 118.79, 118.76, 117.78, 117.70, 111.95, 111.74.

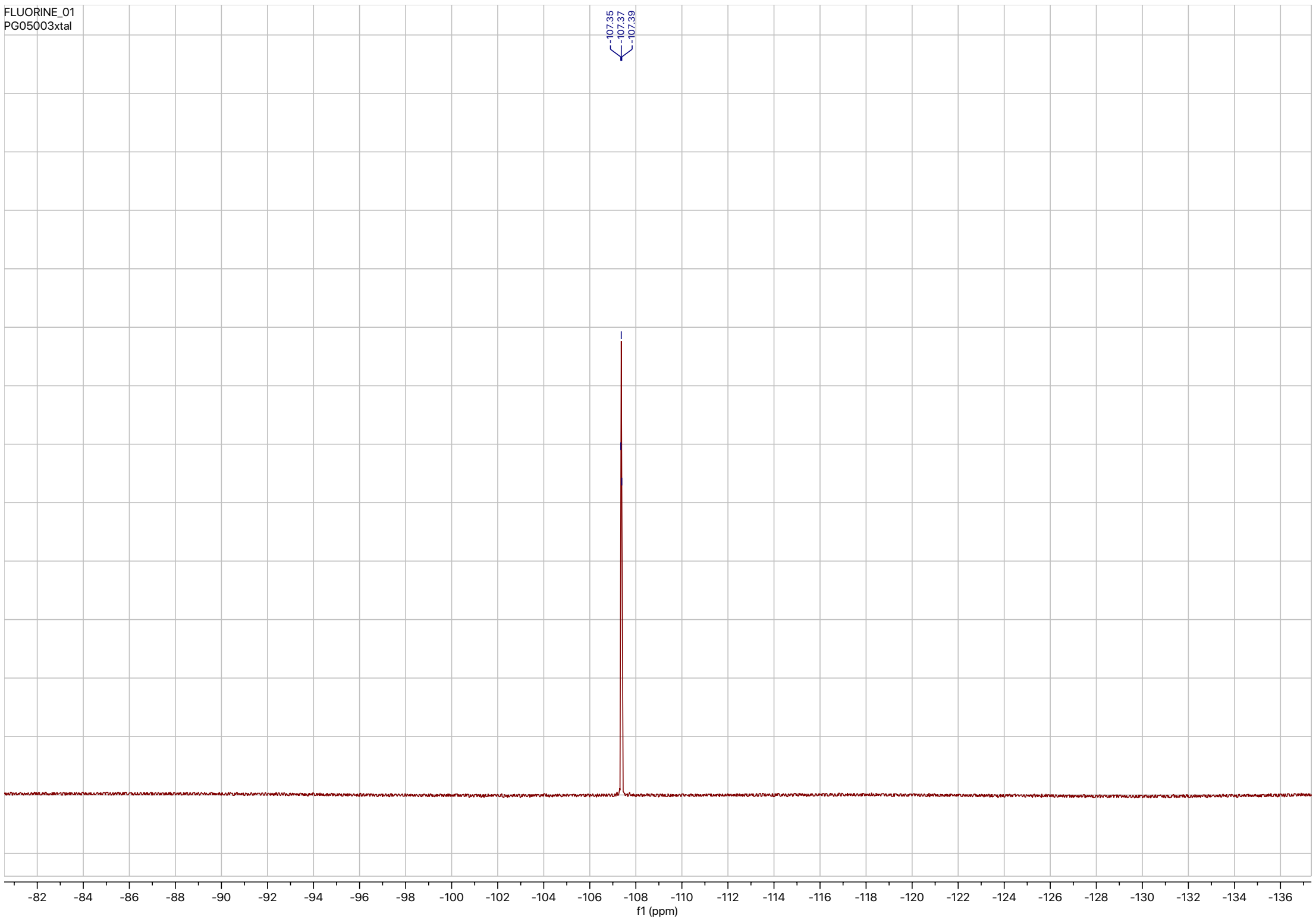

**Figure S9**. *[19]F-NMR of 2-fluoro-4-(thiophene-2-carbonyloxy) benzoic acid*

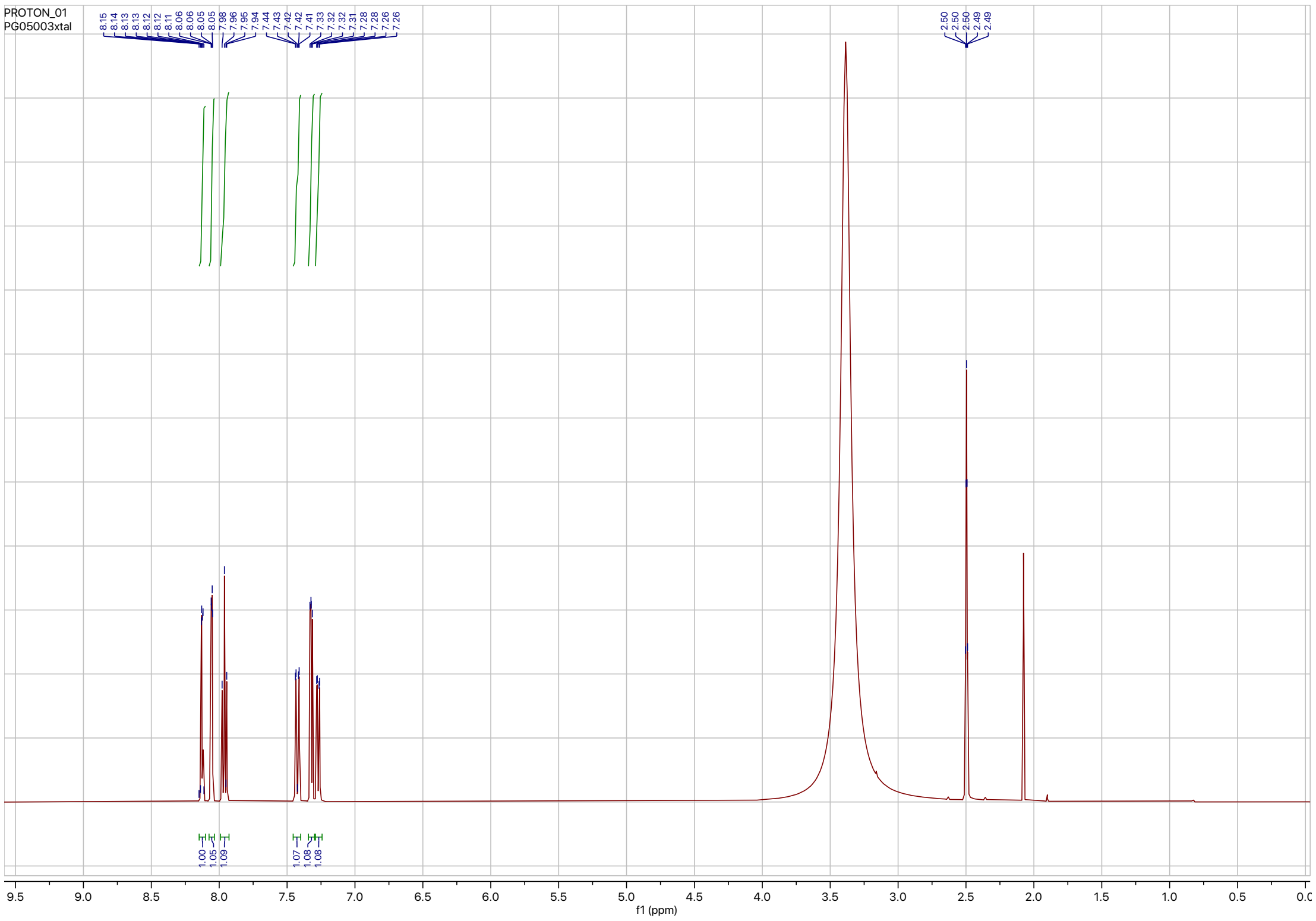

**Figure S10**. *[1]H-NMR of 2-fluoro-4-(thiophene-2-carbonyloxy) benzoic acid*

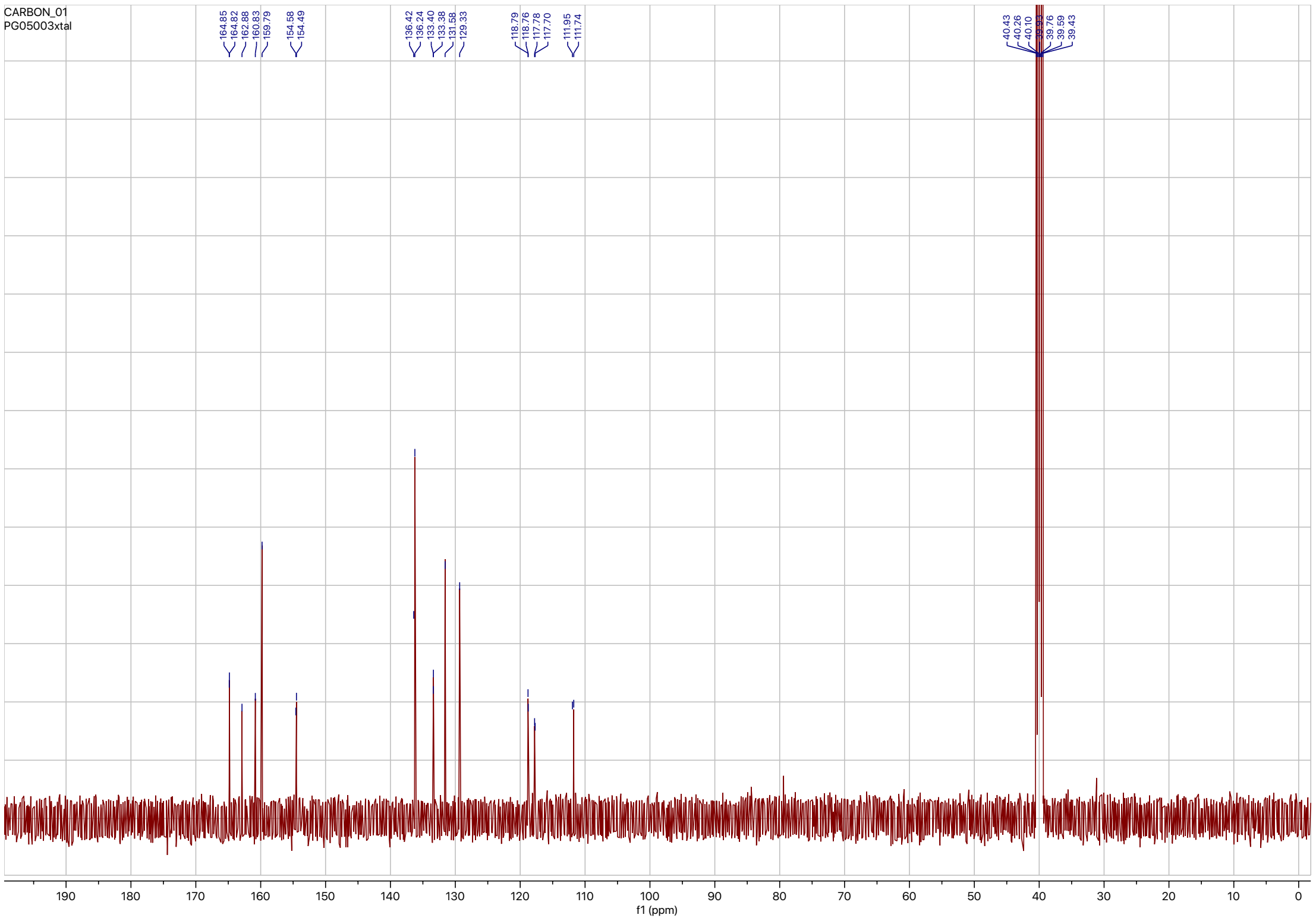

**Figure S11**. *$^{13}$C-NMR of 2-fluoro-4-(thiophene-2-carbonyloxy) benzoic acid*

### B. Preparation and characterization of the Target Compounds

**1**. *4-[(3-fluoro-4-nitrophenoxy) carbonyl] phenyl-2-thiophenecarboxylate*

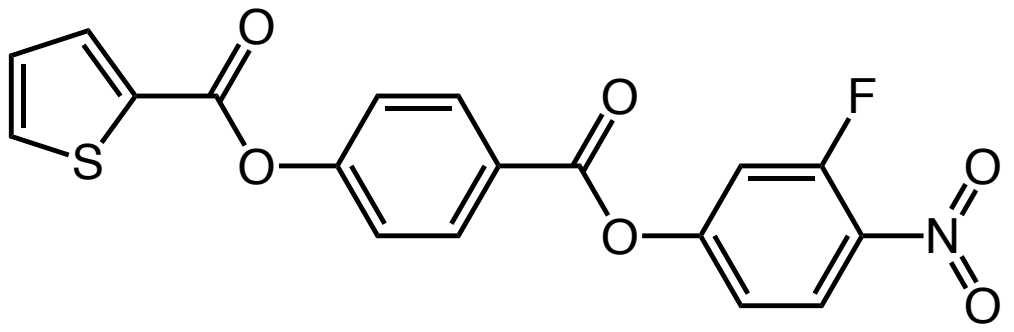

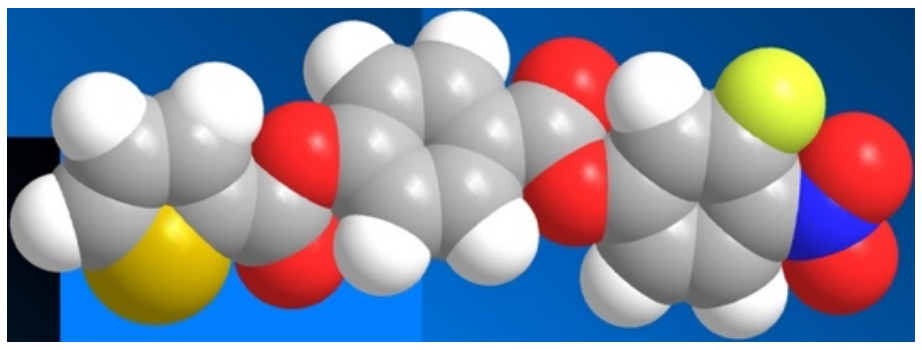

In a 250 ml round bottom flask with stirbar and nitrogen inlet was placed 4-carboxyphenyl 2-thiophenecarboxylate (0.992 gm, 4.0 mmol) and dry dichloromethane (50 ml). The mixture was stirred a few minutes in an ice bath and DCC (1.03 gm, 5.0 mmol) was added. After a few minutes DMAP (25 mg, 0.2 mmol) and 3-fluoro-4-nitrophenol (0.628 gm, 4.0 mmol) were added. The mixture was gradually allowed to warm to room temperature and after four hours TLC indicated the reaction was complete. Silica gel (40 cc) was added, and the slurry was concentrated to dryness by rotary evaporation. The impregnated material was placed at the top of a column made up with dichloromethane and eluted with dichloromethane. Fractions containing the product were combined and concentrated and then recrystallized from 1-propanol to give the pure product (1.169 gm, 75%). **$^{1}$H NMR (400 MHz, tetrachloroethane-*$d_2$*)** δ 8.35 – 8.29 (m, 2H), 8.26 (dd, *J* = 9.1, 8.4 Hz, 1H), 8.09 (dd, *J* = 3.8, 1.3 Hz, 1H), 7.82 (dd, *J* = 5.0, 1.3 Hz, 1H), 7.52 – 7.46 (m, 2H), 7.39 (dd, *J* = 11.3, 2.4 Hz, 1H), 7.33 – 7.27 (m, 2H). **$^{13}$C NMR (101 MHz, tetrachloroethane-*d2*)** δ 163.19, 160.08, 155.42, 135.63, 134.79, 132.27, 131.99, 128.62, 127.47, 127.45, 125.72, 122.52, 118.45, 118.41, 112.76, 112.52, 74.32, 74.05, 73.77.

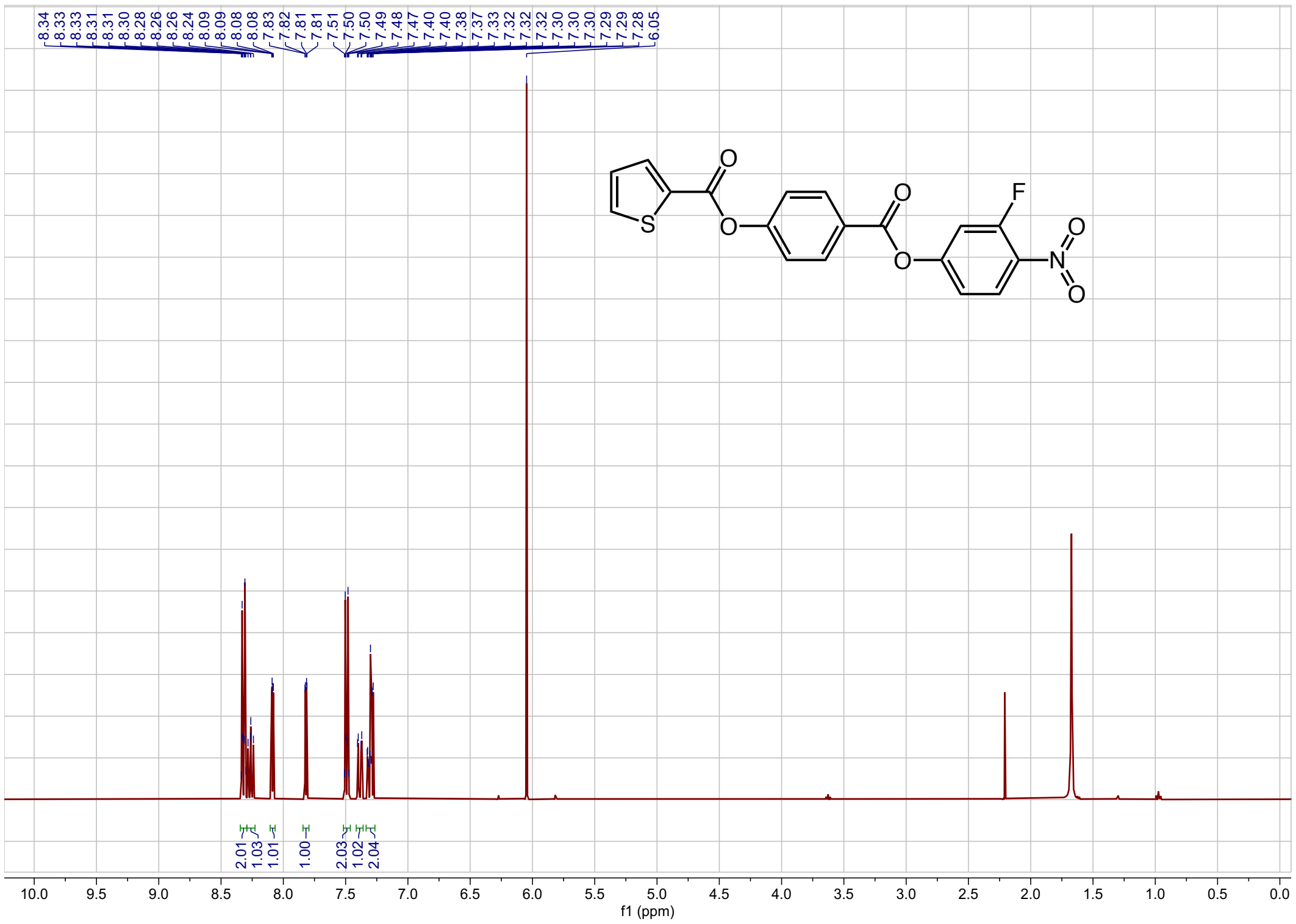

**Figure S12**. *$^{1}$H-NMR of 4-[(3-fluoro-4-nitrophenoxy) carbonyl] phenyl-2-thiophenecarboxylate*

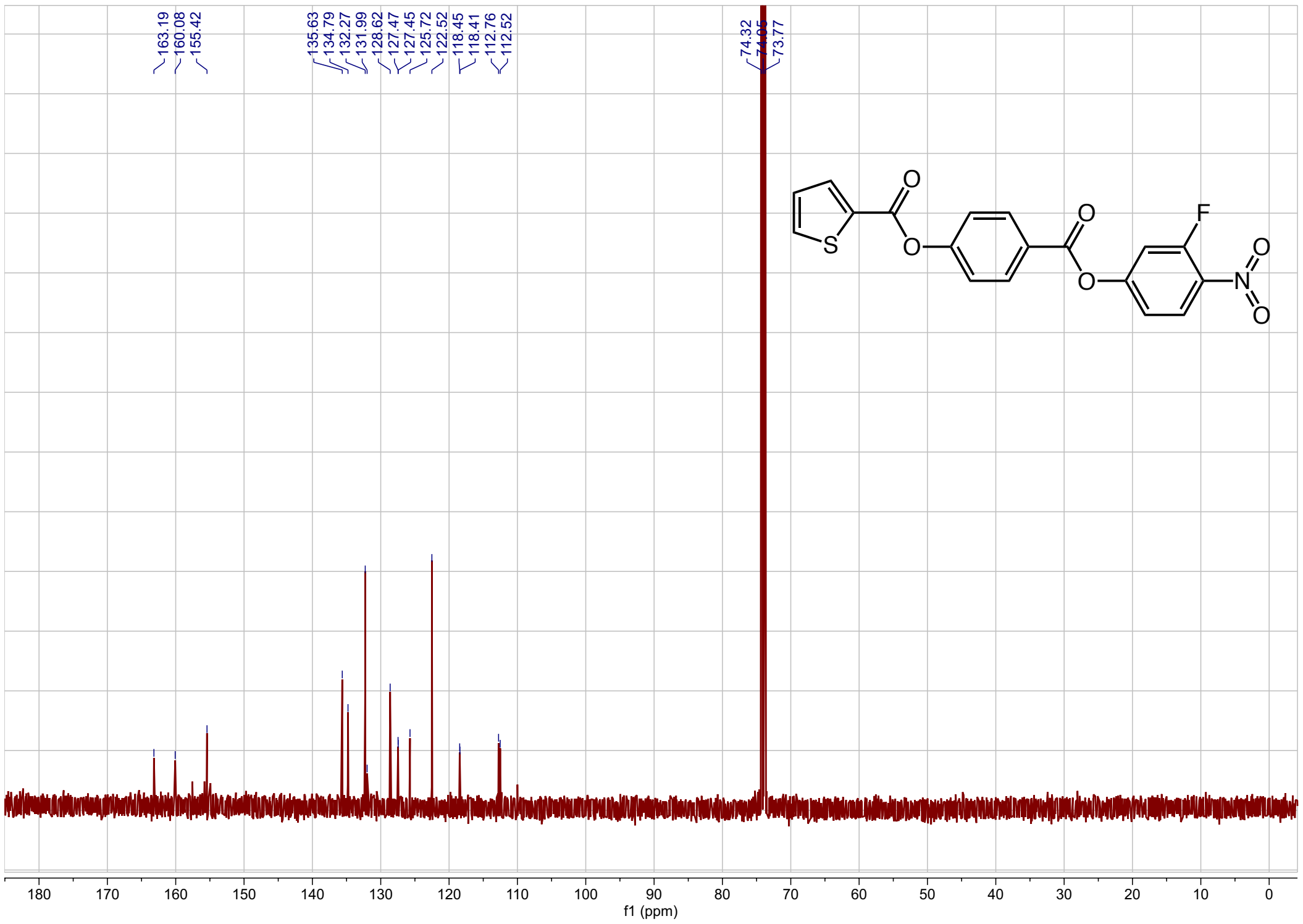

**Figure S13**. *$^{13}$C-NMR of 4-[(3-fluoro-4-nitrophenoxy) carbonyl] phenyl-2-thiophenecarboxylate*

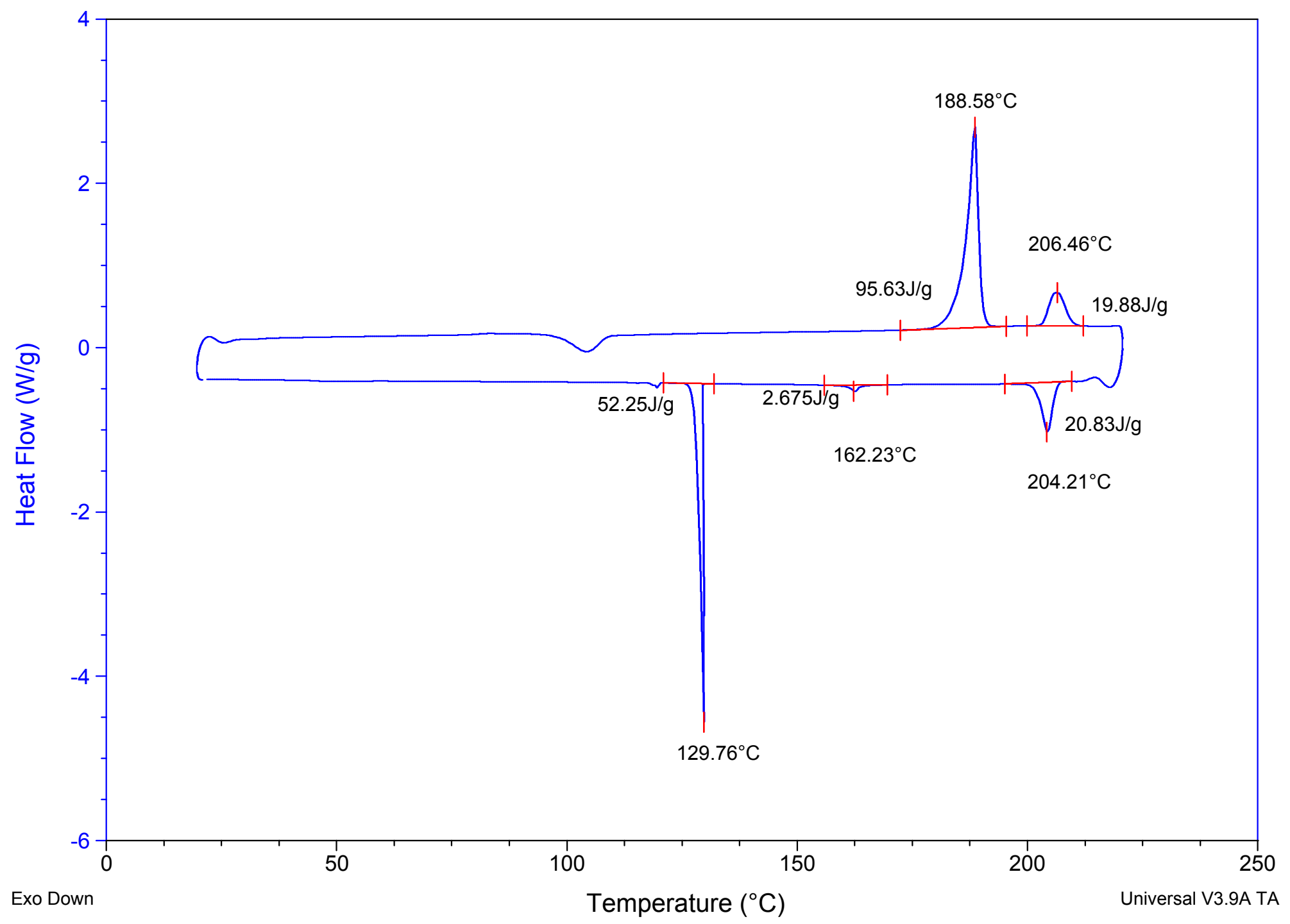

**Figure S14**. *DSC scan at 5°C/min rate of 4-[(3-fluoro-4-nitrophenoxy) carbonyl] phenyl-2-thiophenecarboxylate*

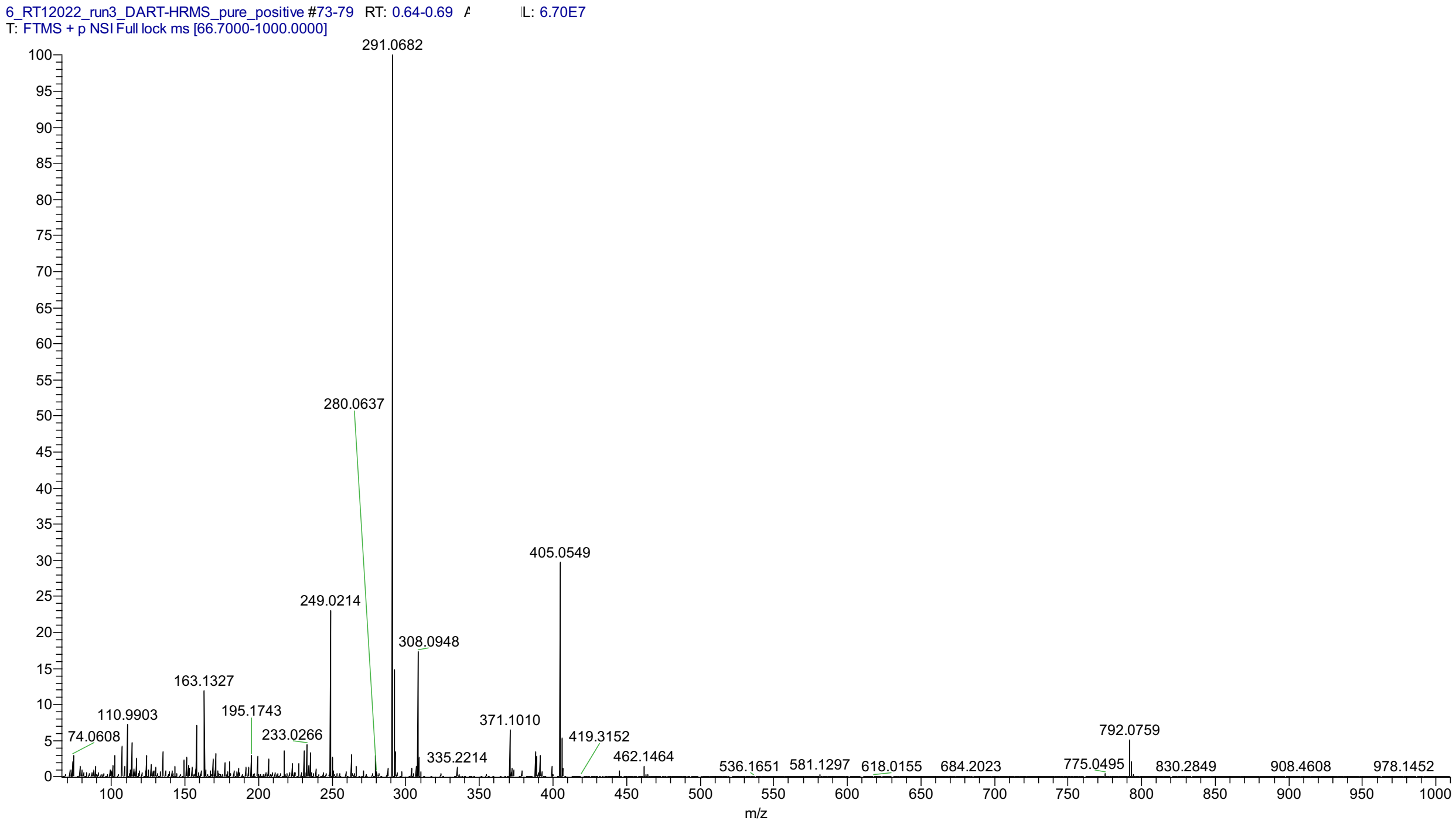

**Figure S15**. *HRMS data for target compound **1**. Base peak at 291.0682 could fit to multiple compositions (contamination or by-product). The second strongest signal at 405.0549 is for* $[C_{18}H_{10}FNO_6S+NH_4]^+$*, with only a small signal at 388.0283 for* $[C_{18}H_{10}FNO_6S+H]^+$.

**2**. *4-[(3-fluoro-4-cyanophenoxy) carbonyl] phenyl-2-thiophenecarboxylatee*

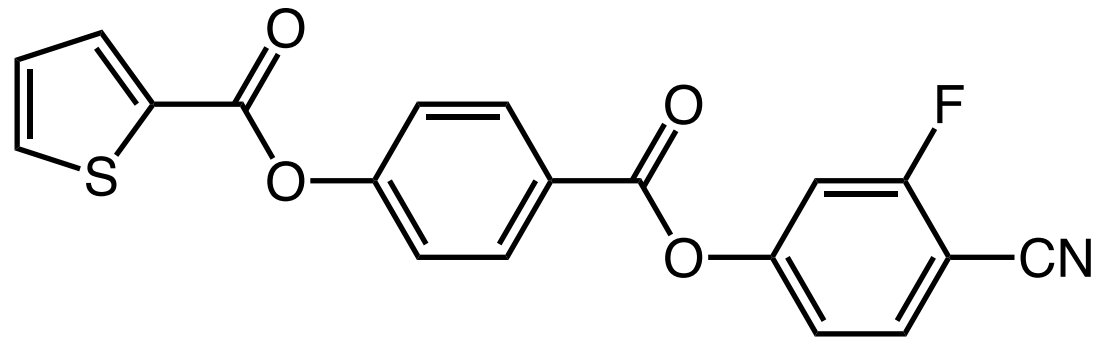

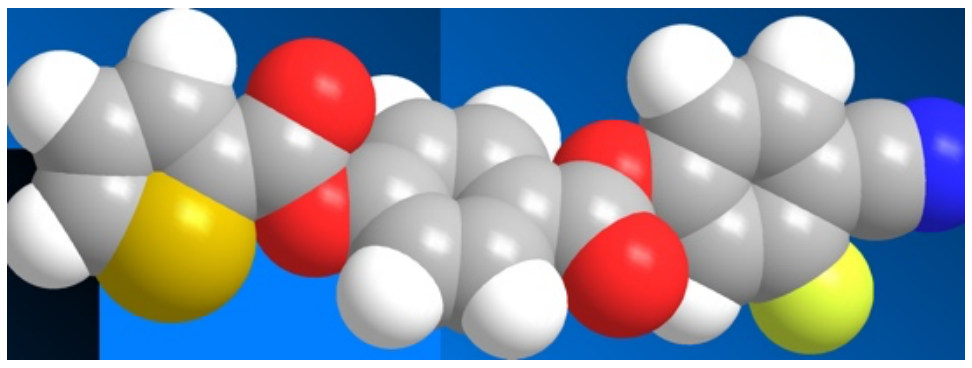

In a 250 ml round bottom flask equipped with stirbar and nitrogen inlet was placed 2-fluoro-4-hydroxybenzonitrile (690 mg, 5.0 mmol) and dry pyridine (20 ml). The mixture was stirred until everything dissolved and was then chilled in an ice water bath and the 4-(thiophene-2-carbonyloxy) benzoyl chloride (5.0 mmol) dissolved in dichloromethane (10 ml) was added in portions by pipette. After completion of addition the mixture was stirred in the ice bath for fifteen minutes and then out of the ice bath for thirty minutes. TLC after this time indicated the reaction was complete. The dichloromethane was removed by rotary evaporation, replaced with tetrahydrofuran (20 ml) which was also removed by rotary evaporation. Ice and water were added in portions to the vigorously stirred residue to nearly fill the flask. After stirring for thirty minutes the crude solid product was isolated by suction filtration, washed well with water and air dried. The solid product so obtained was taken up in boiling 1-propanol, hot filtered through fluted filter paper, and concentrated. Crystals appeared on cooling which were collected by suction filtration, washed with 1-propanol and air dried on the filter. Yield 1.701 gm, 93 (%)

**$^{19}$F NMR (470 MHz, $CDCl_3$)** δ -103.22 (t, $J$ = 8.5 Hz).
**$^{1}$H NMR (500 MHz, $CDCl_3$)** δ 8.30 – 8.23 (m, 1H), 8.03 (dd, $J$ = 3.8, 1.3 Hz, 1H), 7.76 – 7.68 (m, 1H), 7.46 – 7.40 (m, 1H), 7.26 – 7.18 (m, 2H).
**$^{13}$C NMR (126 MHz, $CDCl_3$)** δ 164.69, 163.08, 162.61, 159.82, 155.52, 155.44, 155.36, 135.31, 134.28, 134.15, 134.14, 132.08, 132.05, 128.28, 125.79, 122.26, 118.77, 118.74, 113.45, 111.17, 110.99, 99.13, 99.01.

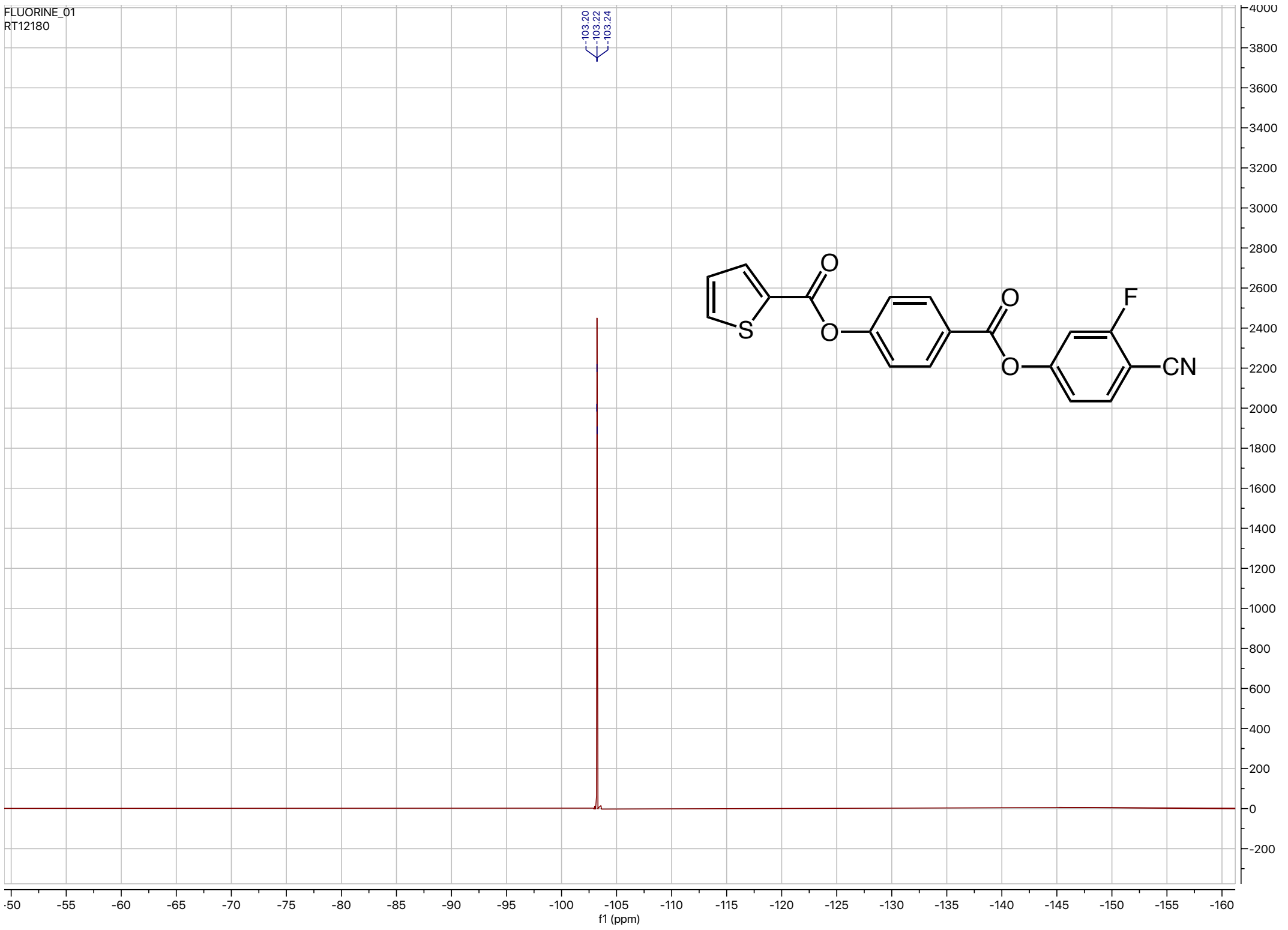

**Figure S16**. *[19]F-NMR of 4-[(3-fluoro-4-cyanophenoxy) carbonyl] phenyl-2-thiophenecarboxylate*

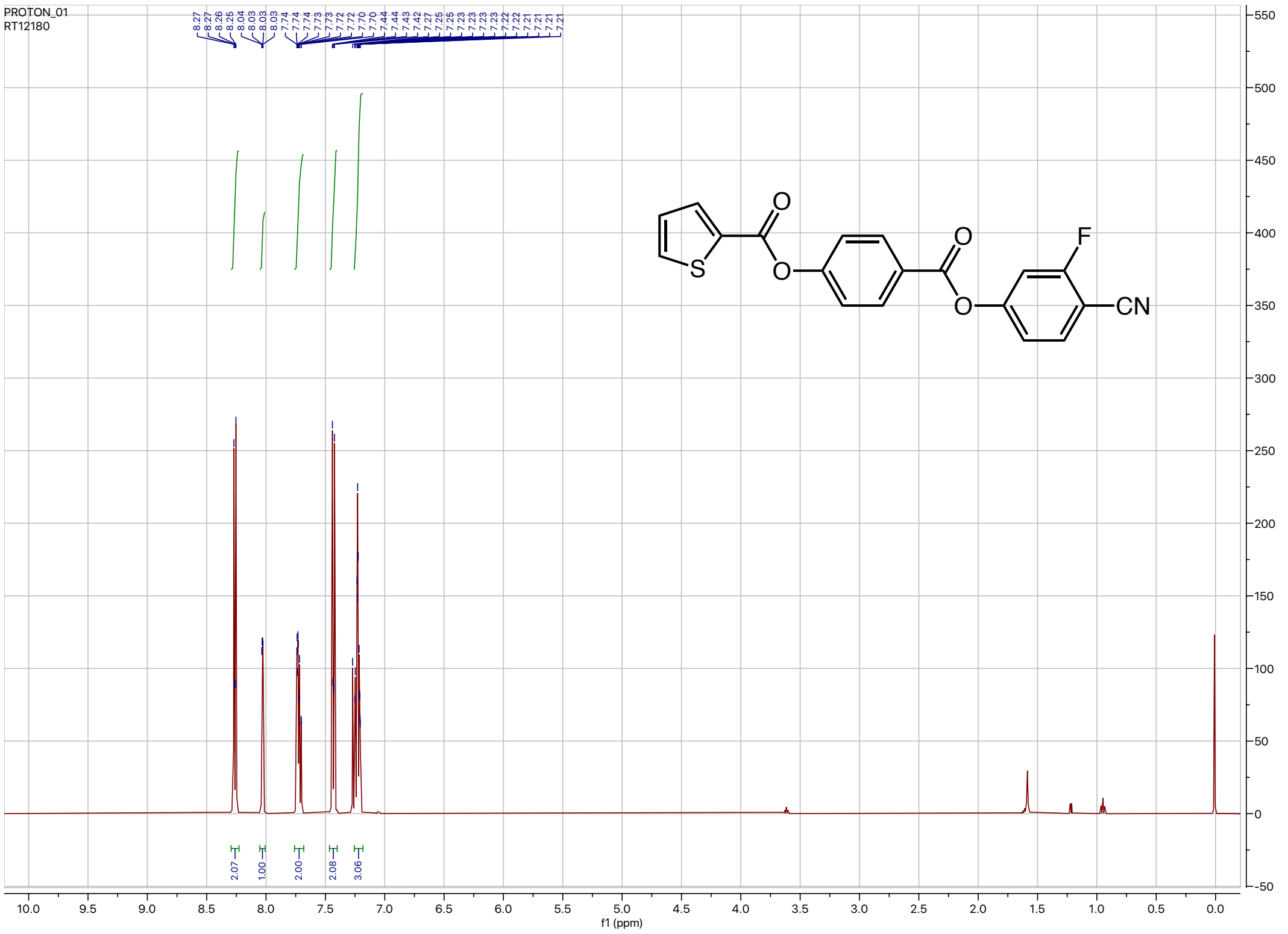

**Figure S17**. *[1]H-NMR of 4-[(3-fluoro-4-cyanophenoxy) carbonyl] phenyl-2-thiophenecarboxylate*

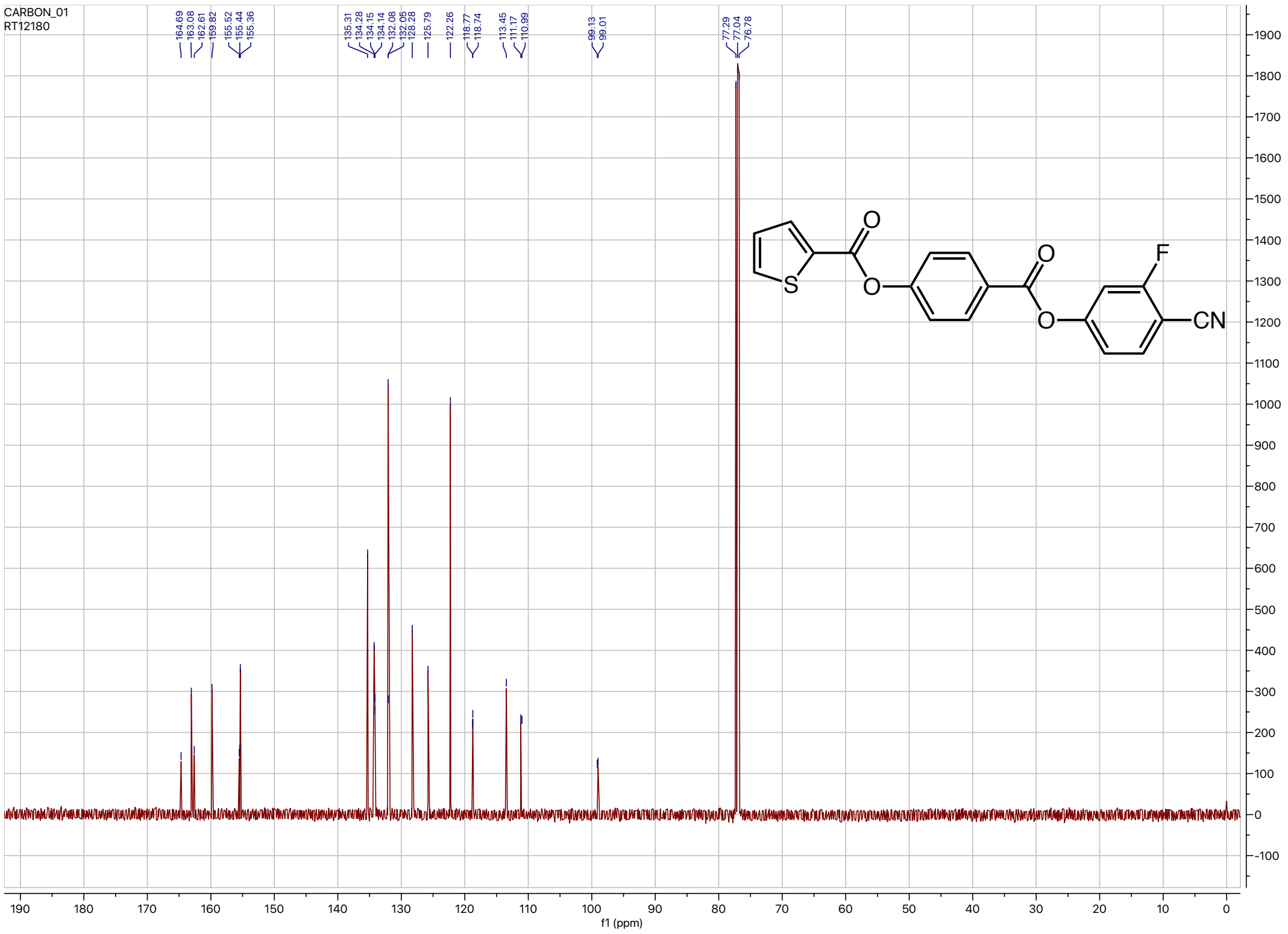

**Figure S18**. *[13]C-NMR of of 4-[(3-fluoro-4-cyanophenoxy) carbonyl] phenyl-2-thiophenecarboxylate*

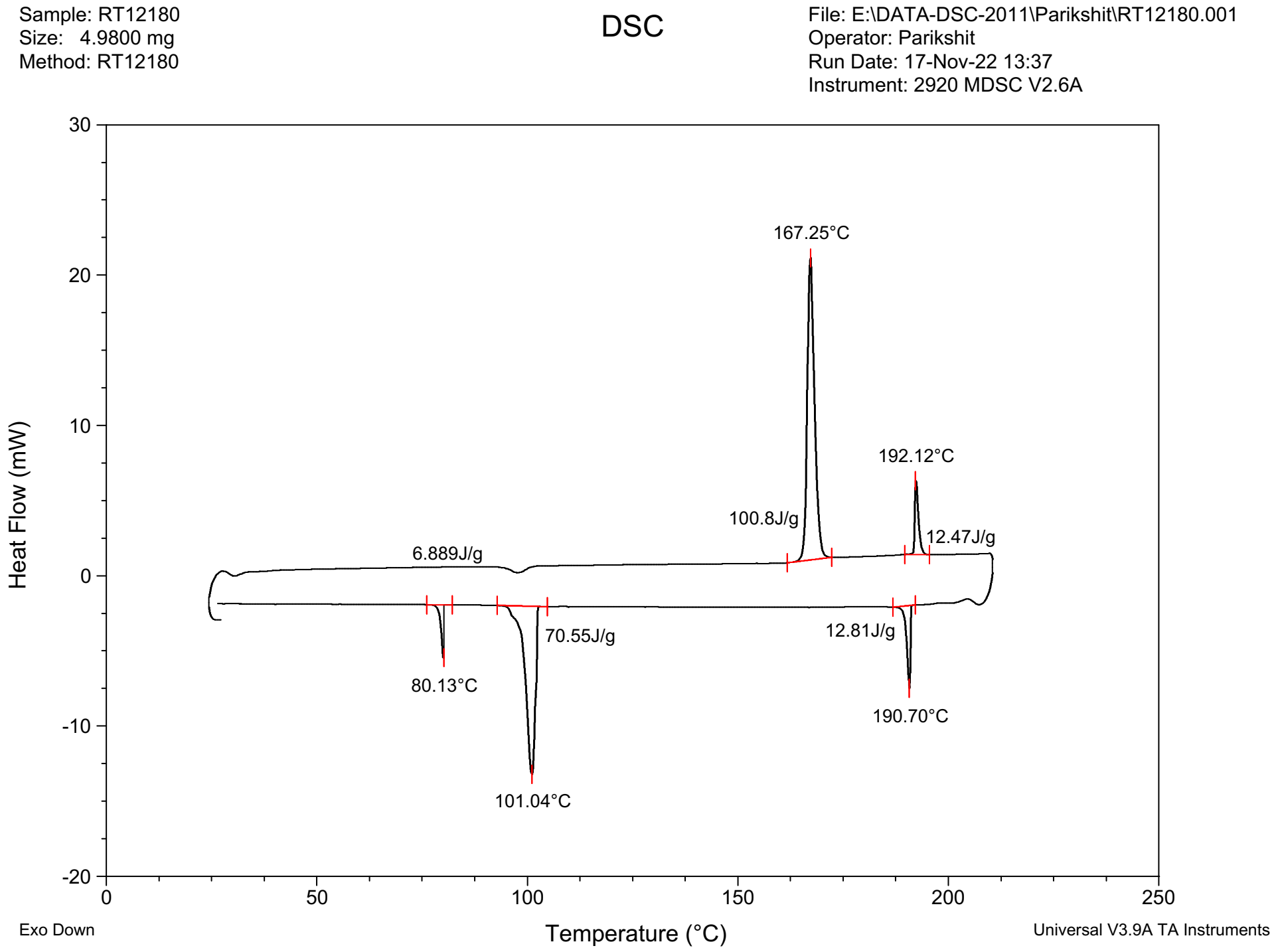

**Figure S19**. *A DSC scan at 5°C/min of 4-[(3-fluoro-4-cyanophenoxy) carbonyl] phenyl-2-thiophenecarboxylate*

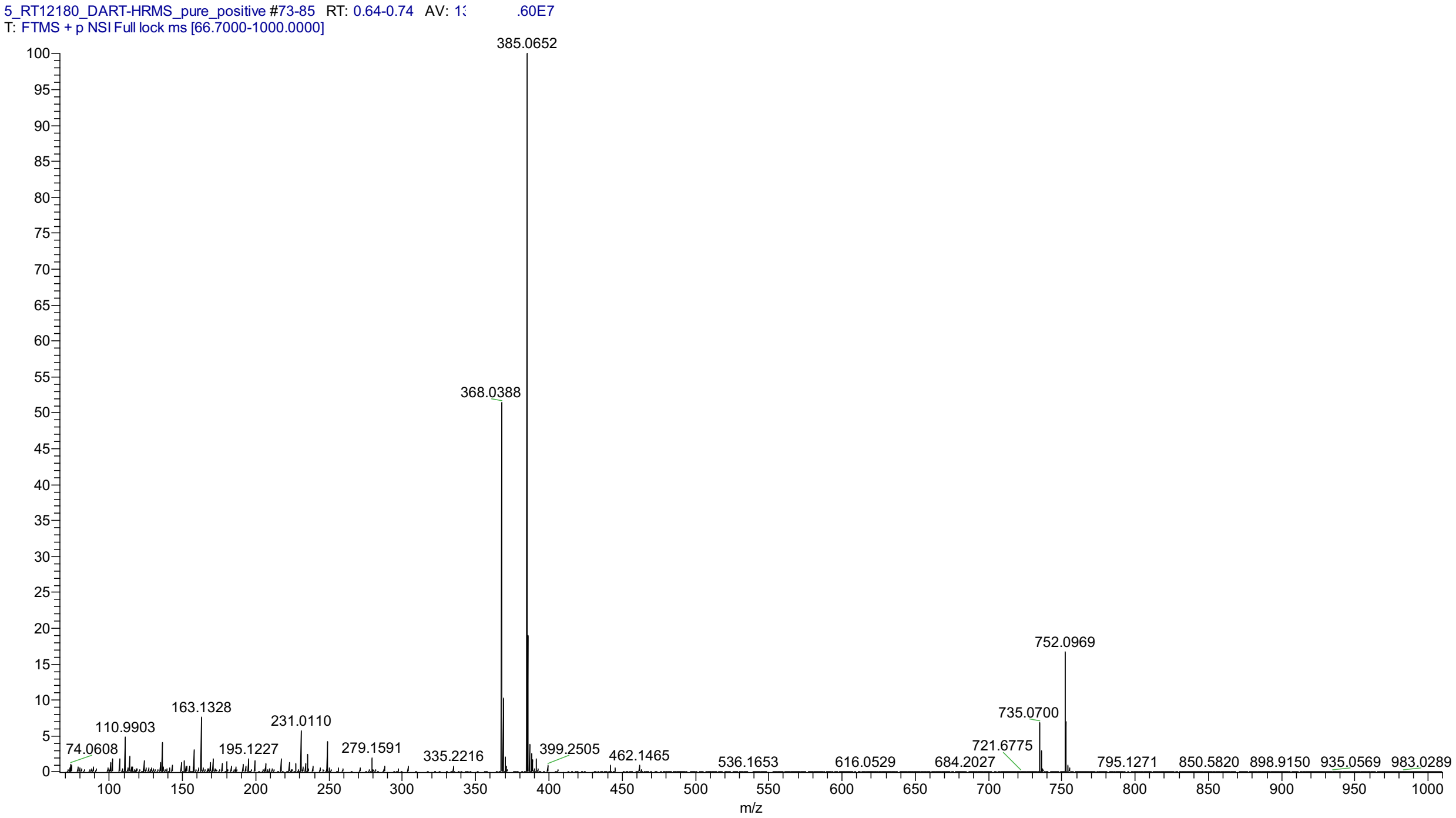

**Figure S20**. *HRMS data for target compound* ***2****. The second strongest signal at 368.0388 is for* $[C_{19}H_{10}FNO_4S+H]^+$ *with base peak at 385.0652 for* $[C_{19}H_{10}FNO_4S+NH_4]^+$.

**3**. <u>*4-[(3,5-difluoro-4-cyanophenoxy) carbonyl] phenyl 2-thiophenecarboxylate*</u>

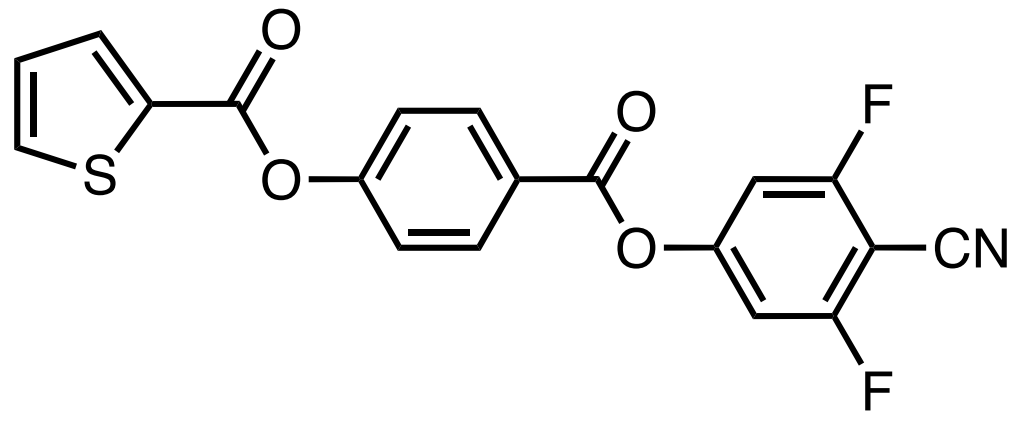

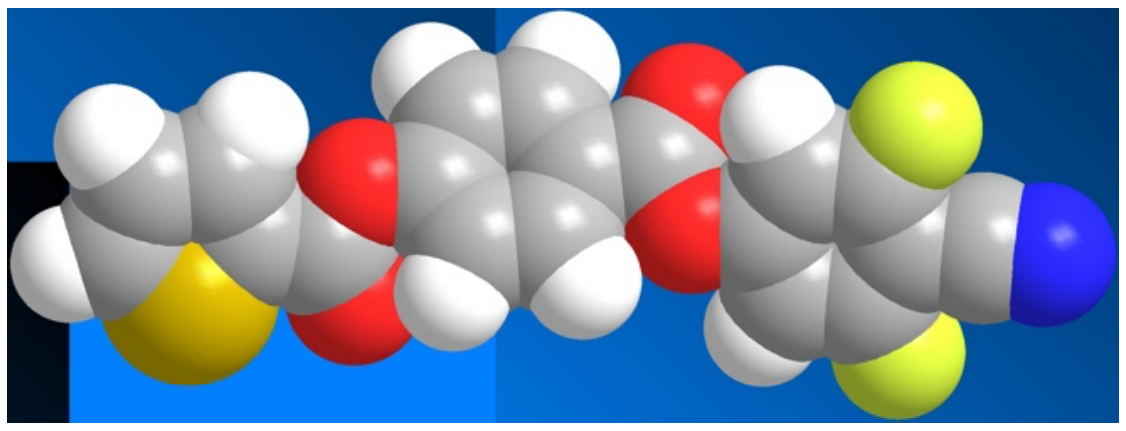

In a 200 ml round bottom flask equipped with a stir bar was placed 2,6-difluoro-4-hydroxybenzonitrile (0.78 g, 5.0 mmol) and pyridine (20 ml) under nitrogen. The mixture was stirred until all the solid dissolved then the flask was placed in an ice bath. The 4-(thiophene-2-carbonyloxy) benzoyl chloride (1.34 g, 5.0 mmol) in DCM (5.0 ml) was added dropwise. After thirty minutes the resulting mixture was allowed to warm to room temperature and stirred overnight. After this time, TLC indicated complete consumption of the starting materials to give a single less polar product. DCM was removed under reduced pressure and the reaction was quenched by adding ice with vigorous stirring and more ice water was added to fill the flask. The precipitate thus obtained was vacuum filtered, air dried and then recrystallized from 1-propanol. Yield- 1.30 g (66%)

**$^{19}$F NMR (470 MHz, $CDCl_3$)** δ -101.84 (d, *J* = 7.7 Hz).
**$^{1}$H NMR (500 MHz, $CDCl_3$)** δ 8.32 – 8.16 (m, 2H), 8.04 (dd, *J* = 3.8, 1.3 Hz, 1H), 7.74 (dd, *J* = 5.0, 1.3 Hz, 1H), 7.51 – 7.37 (m, 2H), 7.23 (dd, *J* = 5.0, 3.8 Hz, 1H), 7.15 – 7.03 (m, 2H). **$^{13}$C NMR (126 MHz, $CDCl_3$)** δ 164.63, 164.58, 162.63, 162.55, 162.50, 159.77, 155.89, 155.55, 135.35, 134.33, 132.15, 131.99, 128.29, 125.35, 122.35, 108.73, 107.08, 107.05, 106.90, 106.86, 90.21.

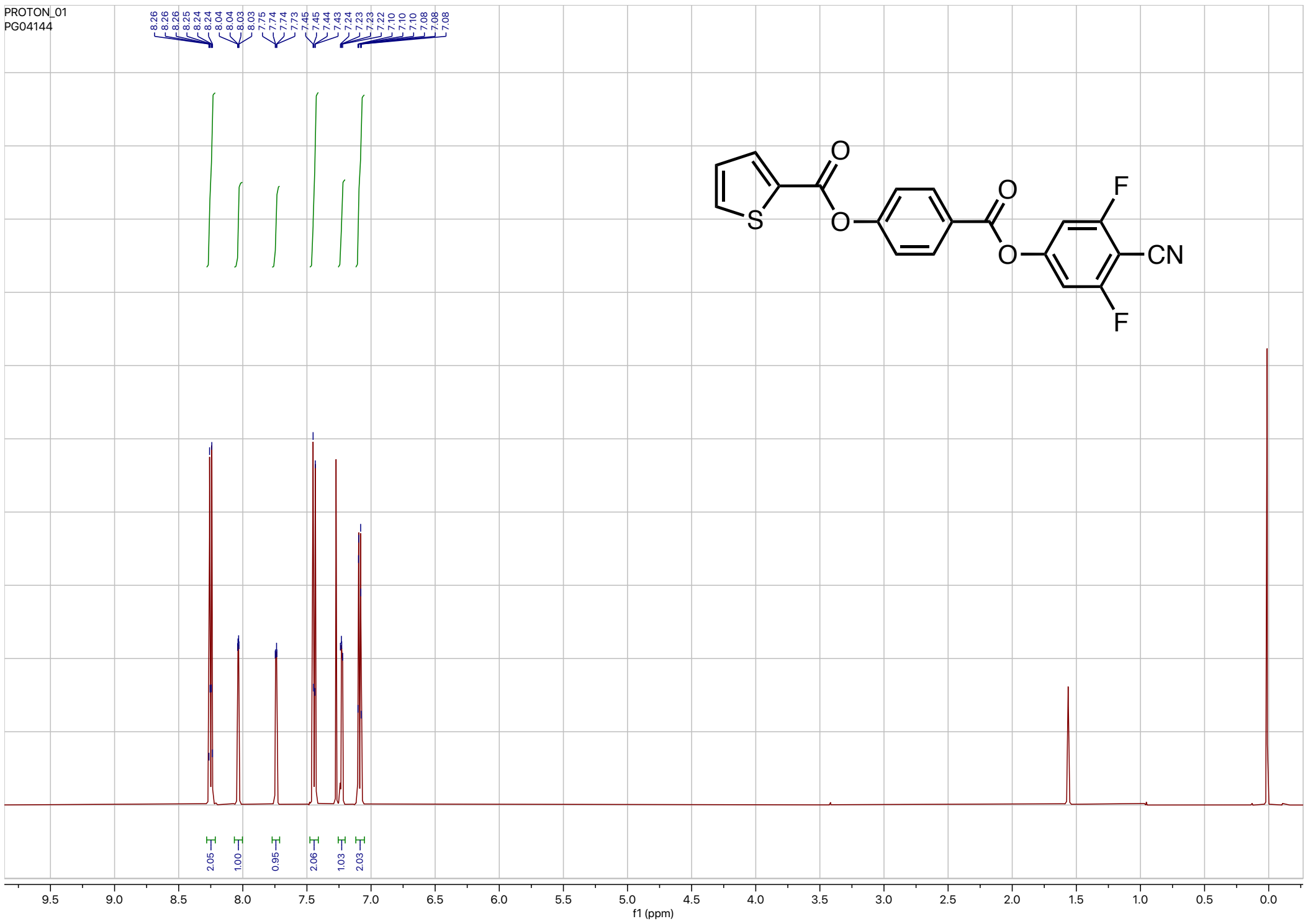

**Figure S21**. *[1]H-NMR of 4-[(3,5-difluoro-4-cyanophenoxy) carbonyl] phenyl 2-thiophenecarboxylate*

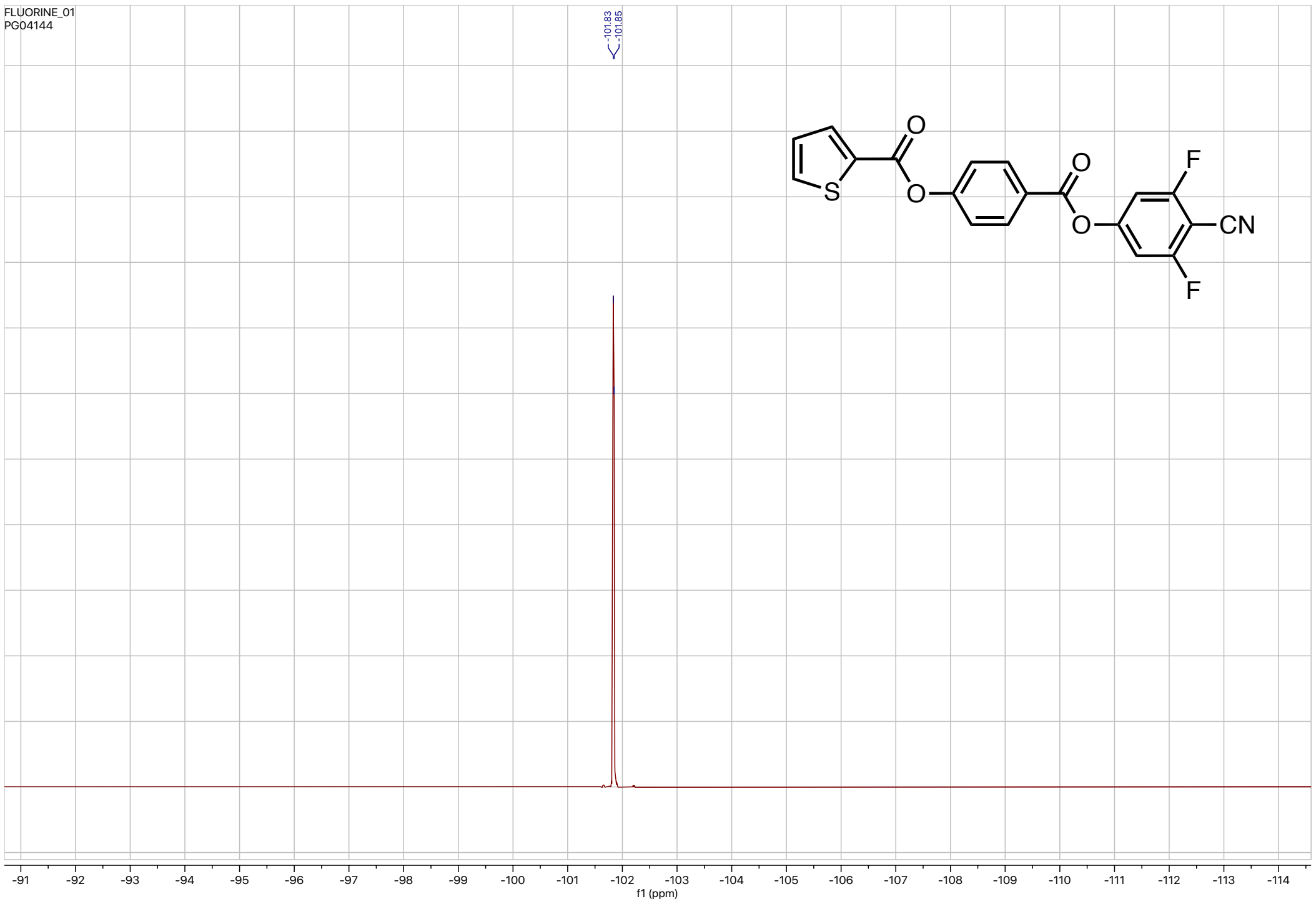

**Figure S22**. *[19]F-NMR of 4-[(3,5-difluoro-4-cyanophenoxy) carbonyl] phenyl 2-thiophenecarboxylate*

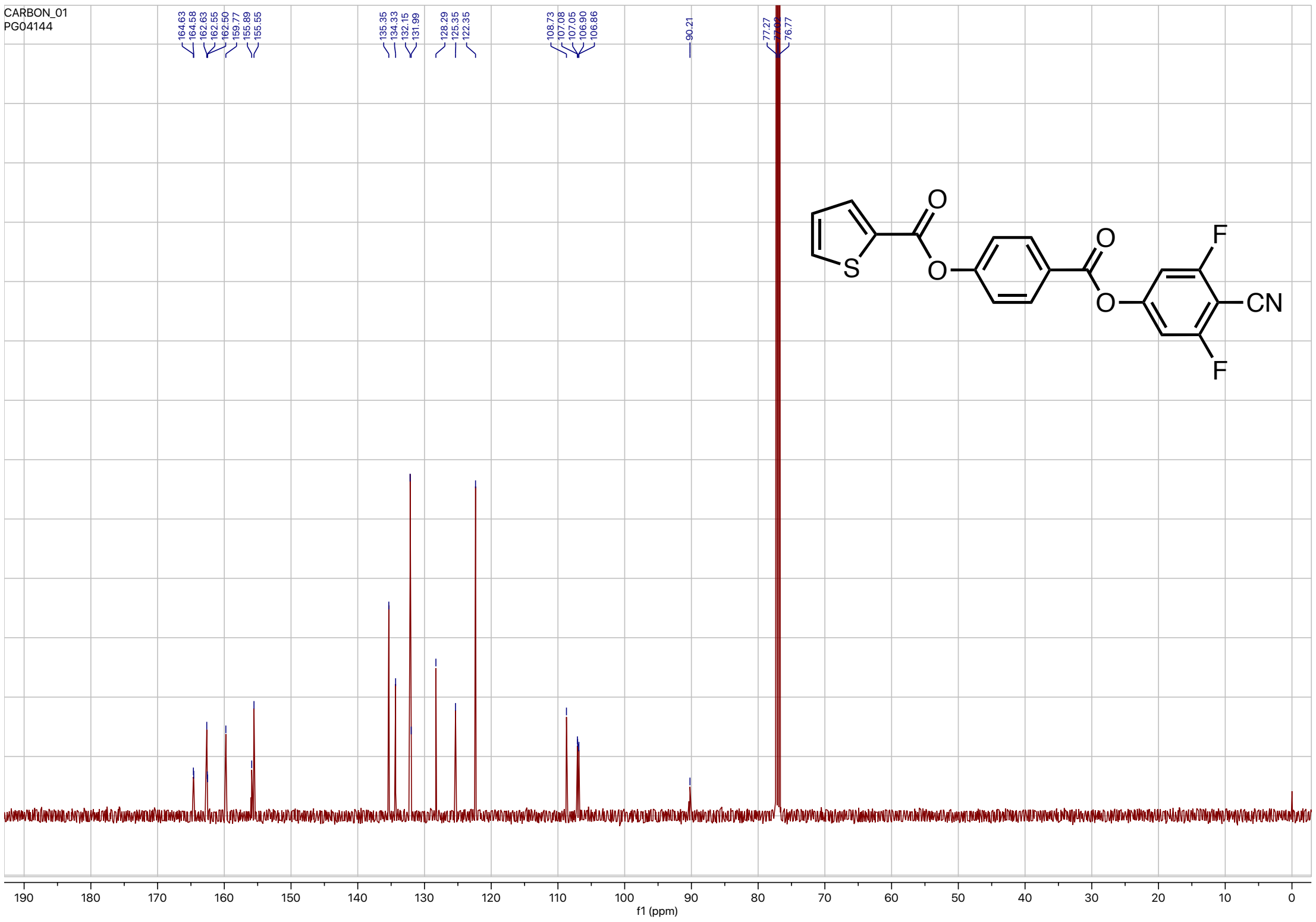

**Figure S23**. *[13]C-NMR of 4-[(3,5-difluoro-4-cyanophenoxy) carbonyl] phenyl 2-thiophenecarboxylate*

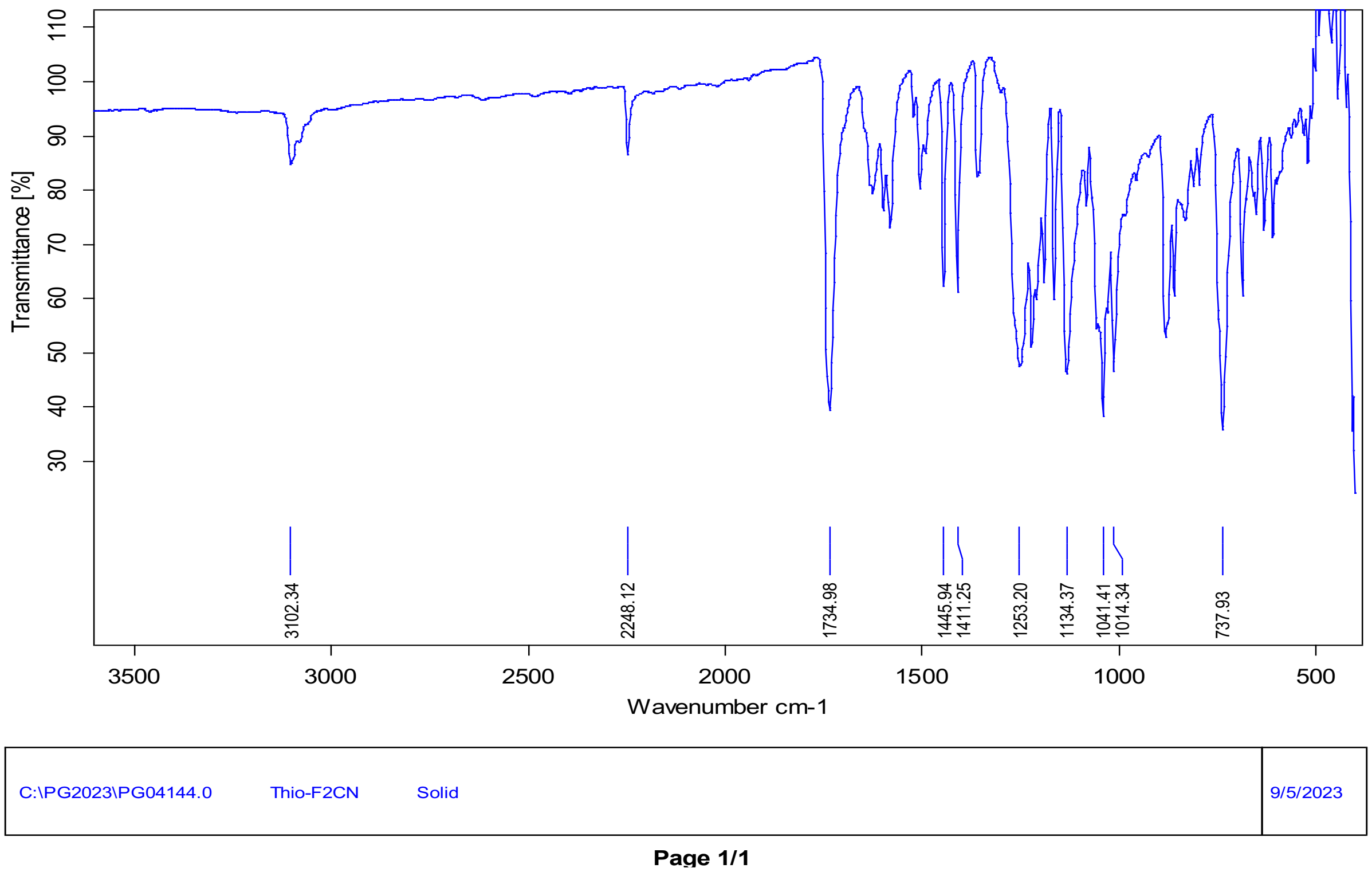

**Figure S24**. *IR spectrum of 4-[(3,5-difluoro-4-cyanophenoxy) carbonyl] phenyl 2-thiophenecarboxylate*

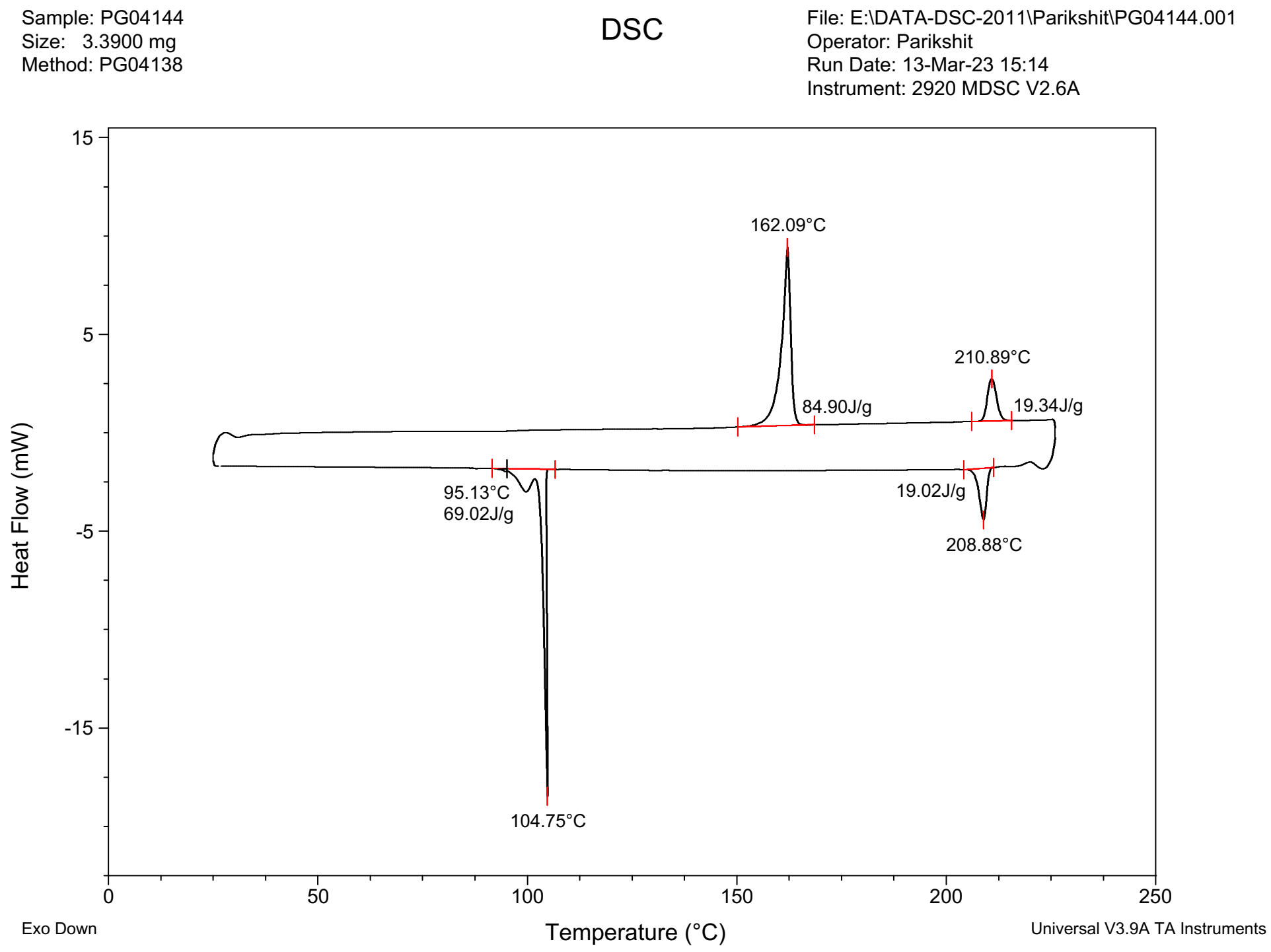

**Figure S25**. *DSC scan at 5$^o$C/min rate of 4-[(3,5-difluoro-4-cyanophenoxy) carbonyl] phenyl 2-thiophenecarboxylate*

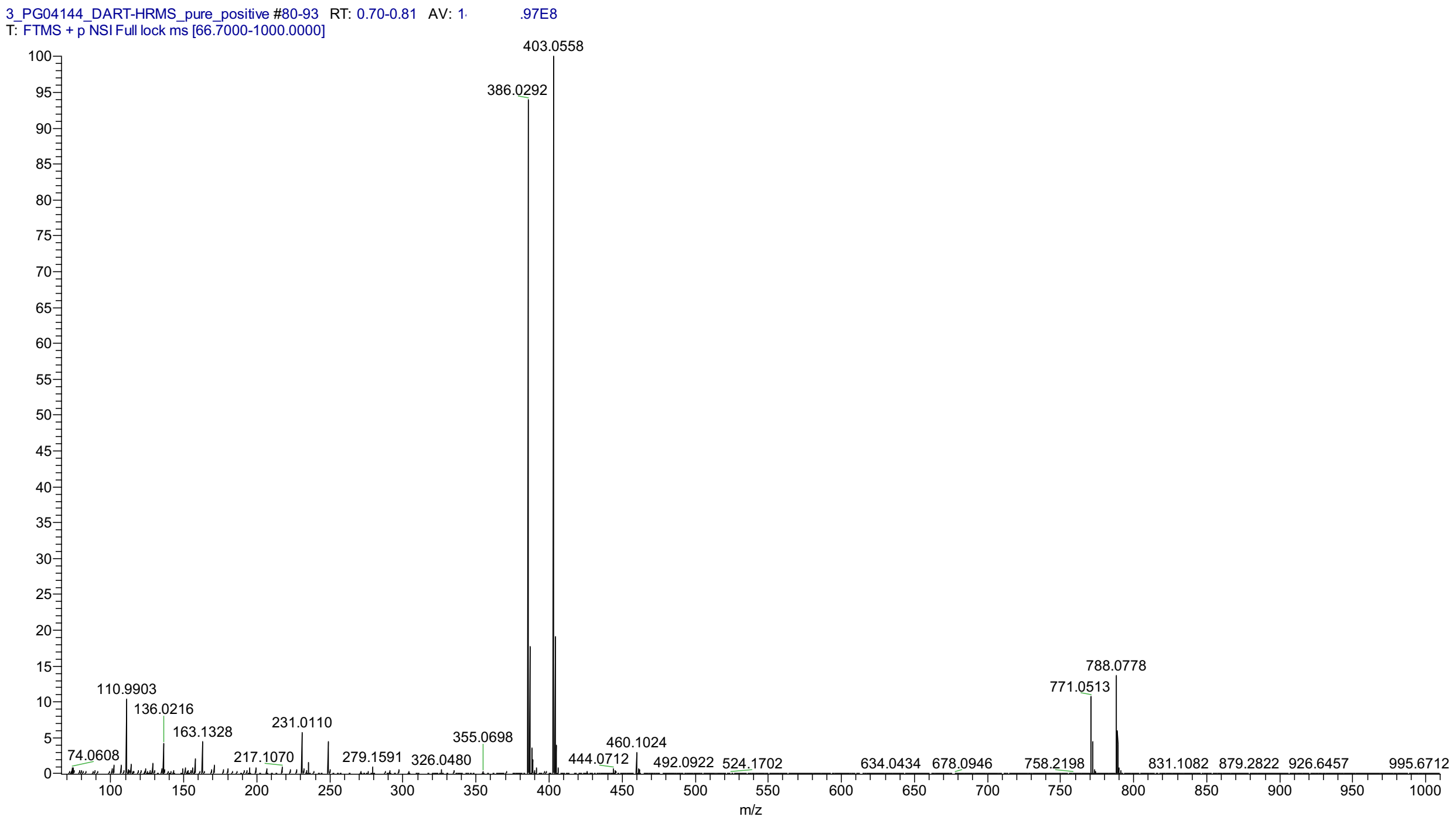

**Figure S26**. *HRMS data for target compound **3**. The second strongest signal at 386.0292 is for [$C_{19}H_9F_2NO_4S$+H]$^+$ with base peak at 403.0558 for [$C_{19}H_9F_2NO_4S$+$NH_4$]$^+$.*

**4**. <u>*4-[(3,4-dicyanophenoxy) carbonyl] phenyl-2-thiophenecarboxylate*</u>

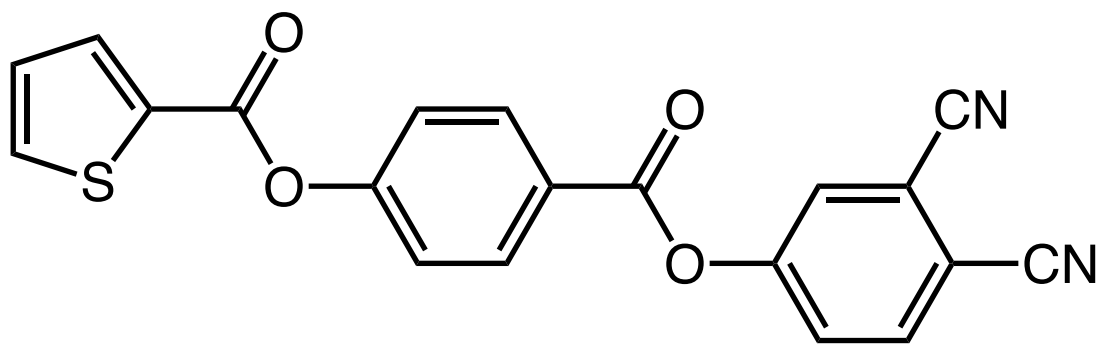

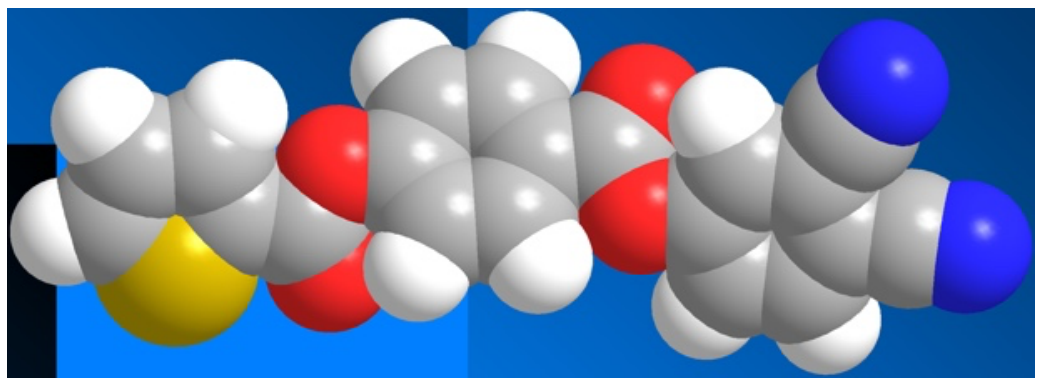

In a 250 ml round bottom flask equipped with stirbar and nitrogen inlet was placed 3,4-dicyanophenol (576 mg, 4.0 mmol) and dry pyridine (20 ml). The mixture was stirred until everything dissolved and then was chilled in an ice water bath and the 4-(thiophene-2-carbonyloxy) benzoyl chloride (less than or equal to 4.0 mmol) dissolved in dichloromethane (10 ml) was added in portions by pipette. After completion of addition the mixture was stirred in the ice bath for fifteen minutes and then out of the ice bath for thirty minutes. TLC after this time indicated the reaction was complete. The dichloromethane was removed by rotary evaporation, replaced with tetrahydrofuran (20 ml) which was also removed by rotary evaporation. Ice and water were added in portions to the vigorously stirred residue to nearly fill the flask. After stirring for thirty minutes the crude solid product was isolated by suction filtration, washed well with water and air dried. The solid product so obtained was taken up in a mixture of boiling 1-propanol and 2-methoxyethanol, hot filtered through fluted filter paper, and concentrated down. Crystals appeared on cooling which were collected by suction filtration, washed with a mixture of 1-propanol and 2-methoxyethanol and air dried on the filter. Yield: 832 mg (56%). A second crop of product was also obtained 280 mg (19%).

**¹H NMR (500 MHz, CDCl₃)** δ 8.27 (d, *J* = 8.8 Hz, 2H), 8.04 (dd, *J* = 3.8, 1.3 Hz, 1H), 7.91 (d, *J* = 8.6 Hz, 1H), 7.80 – 7.72 (m, 2H), 7.68 (dd, *J* = 8.6, 2.4 Hz, 1H), 7.48 – 7.41 (m, 2H), 7.23 (dd, *J* = 5.0, 3.8 Hz, 1H).
**¹³C NMR (126 MHz, CDCl₃)** δ 162.92, 159.77, 155.59, 153.98, 135.36, 135.05, 134.34, 132.18, 131.98, 128.29, 127.24, 127.06, 125.29, 122.38, 117.51, 114.88, 114.53, 113.19.

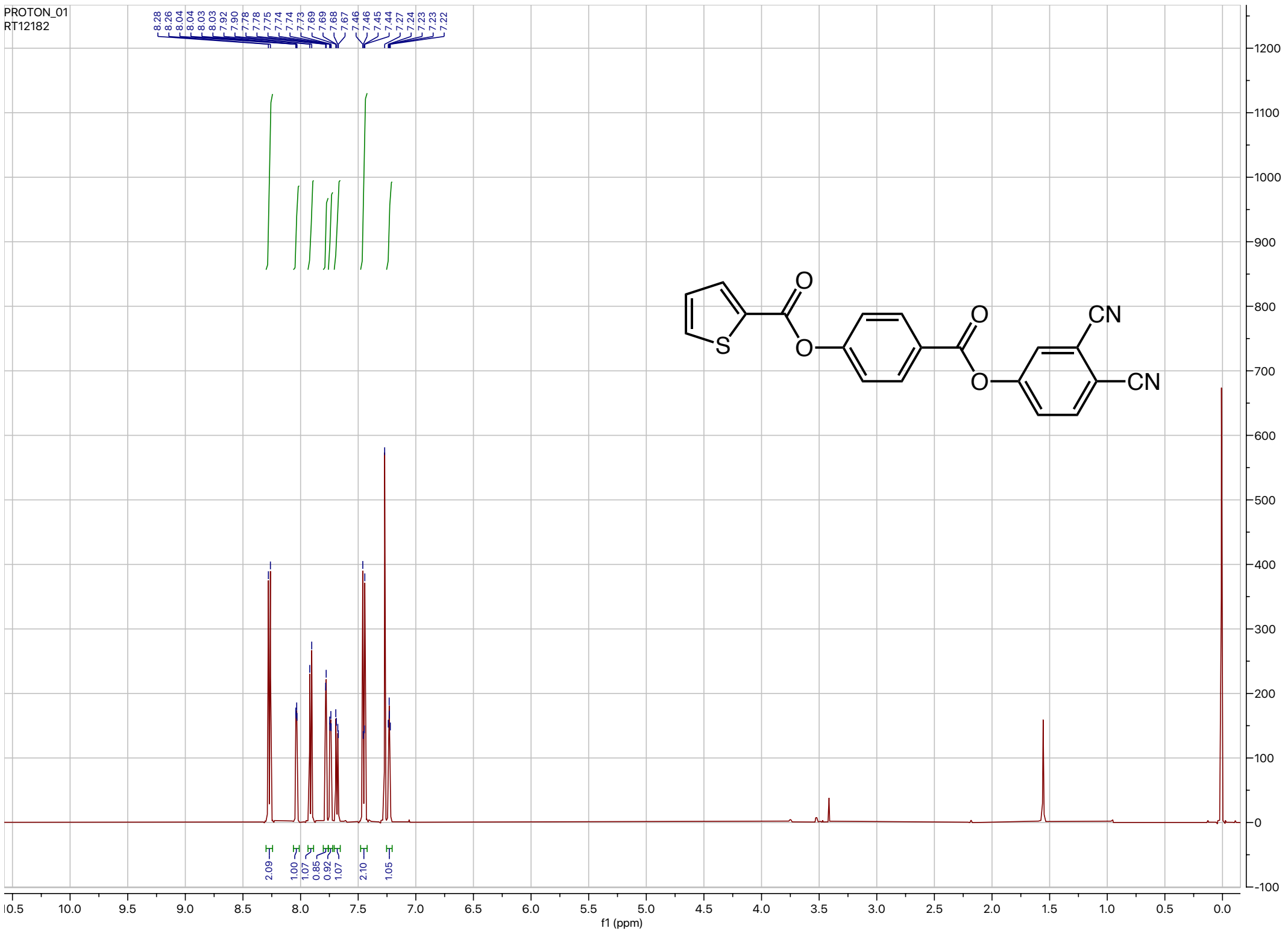

**Figure S27**. *¹H-NMR of 4-[(3,4-dicyanophenoxy) carbonyl] phenyl-2-thiophenecarboxylate*

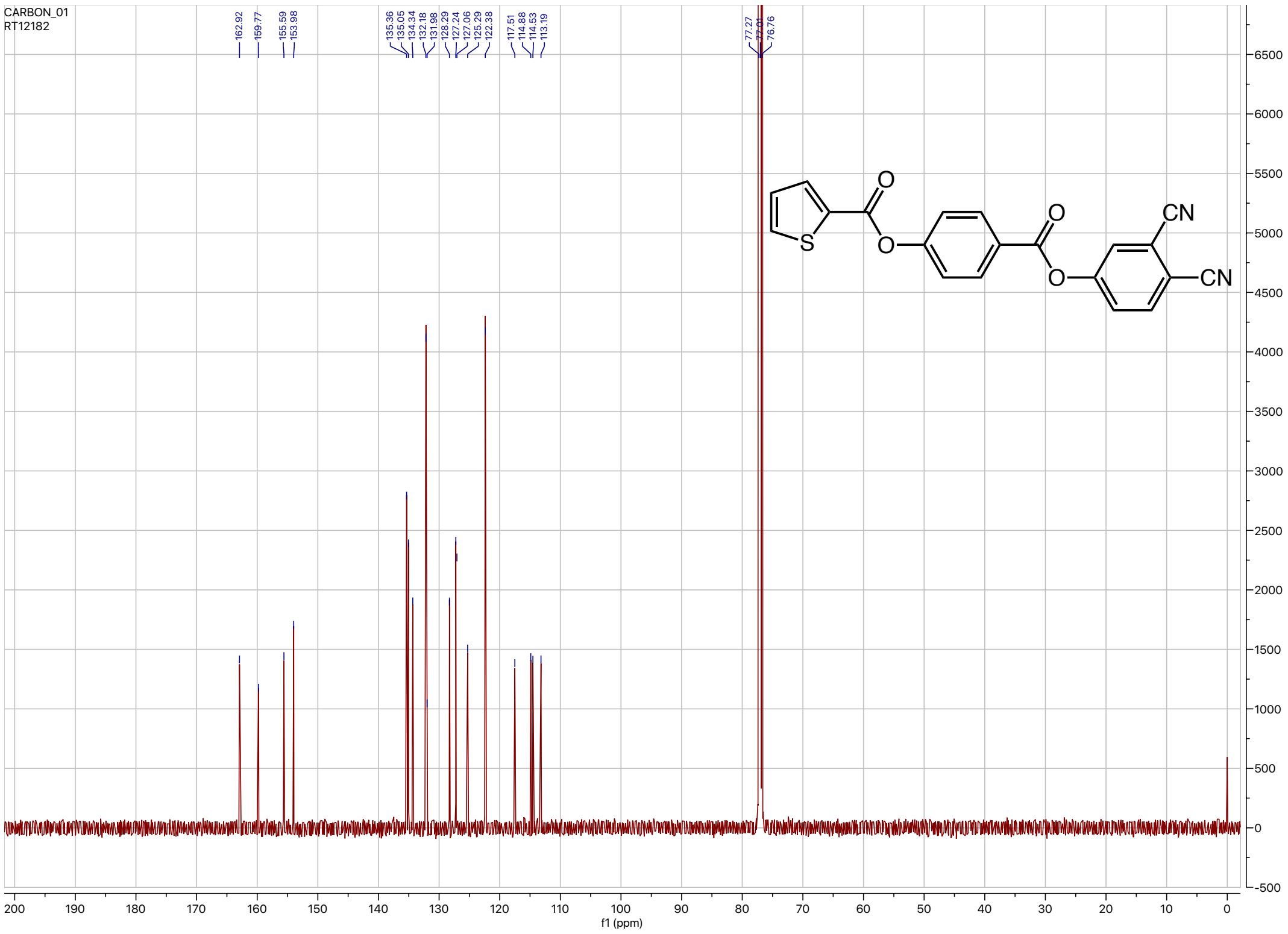

**Figure S28**. *[13]C-NMR of 4-[(3,4-dicyanophenoxy) carbonyl] phenyl-2-thiophenecarboxylate*

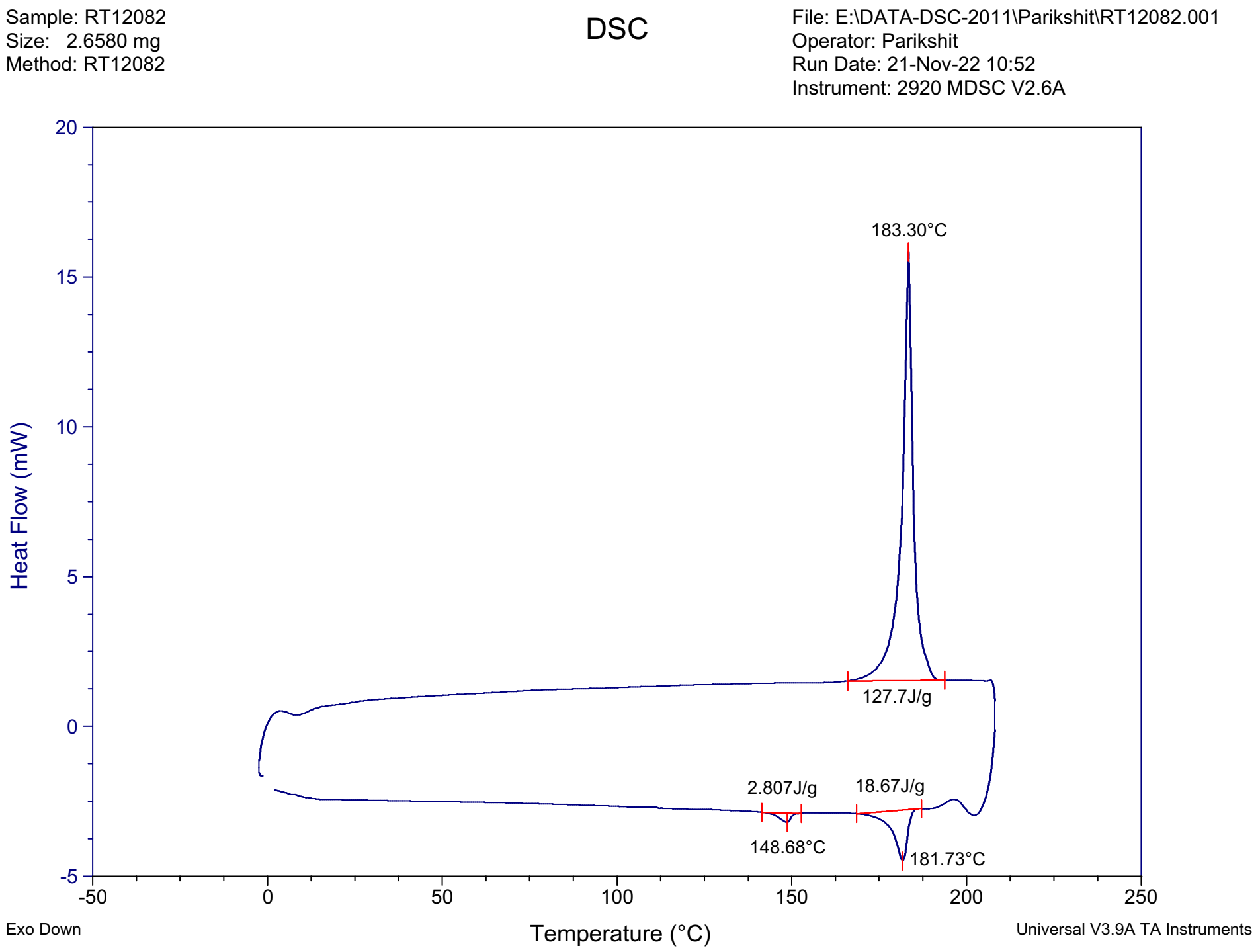

**Figure S29**. *DSC (first scan at 10$^o$C/min rate) of 4-[(3,4-dicyanophenoxy) carbonyl] phenyl-2-thiophenecarboxylate*

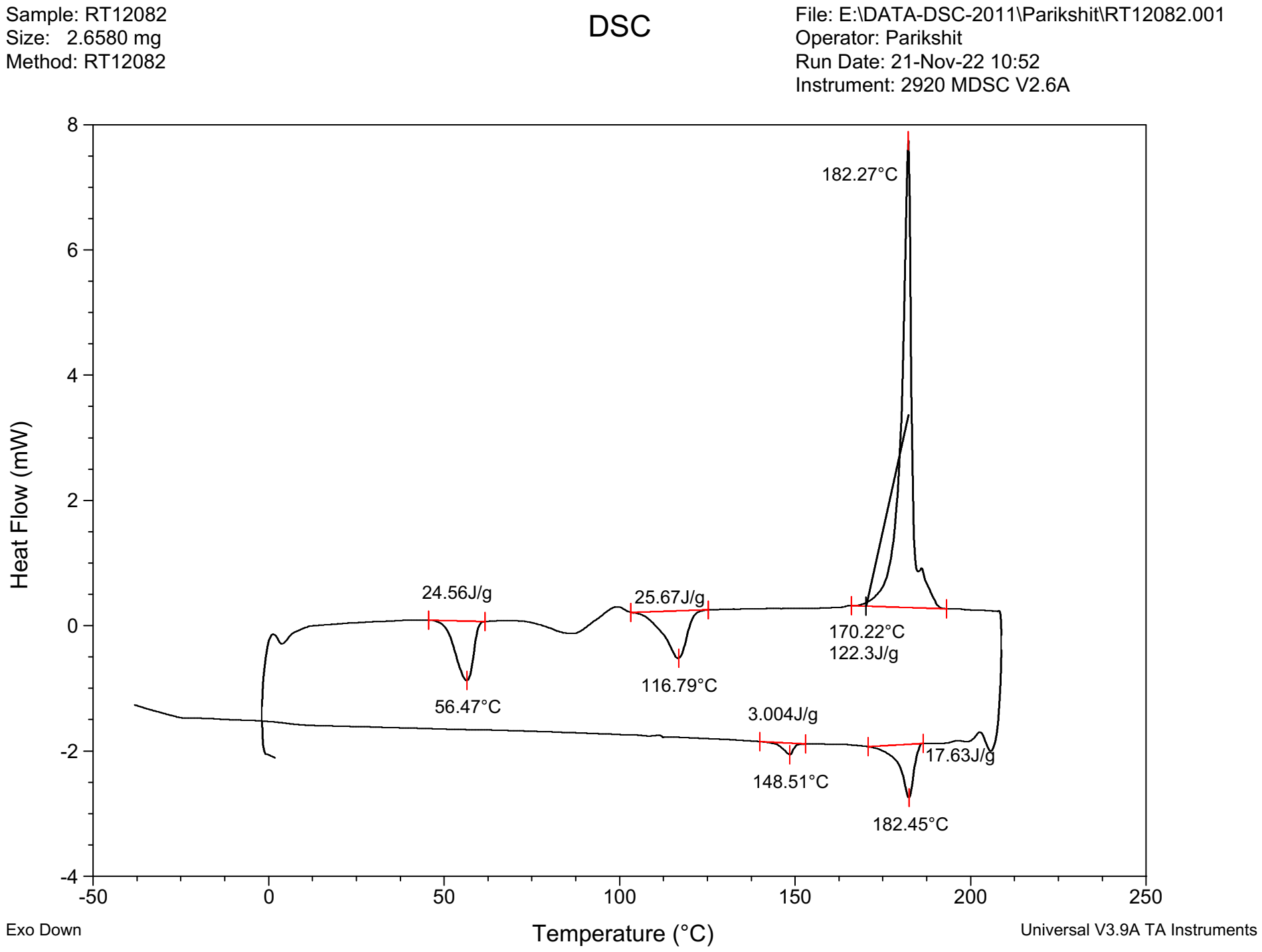

**Figure S30**. *DSC (second scan at 5°C/min rate)) of 4-[(3,4-dicyanophenoxy) carbonyl] phenyl-2-thiophenecarboxylate*

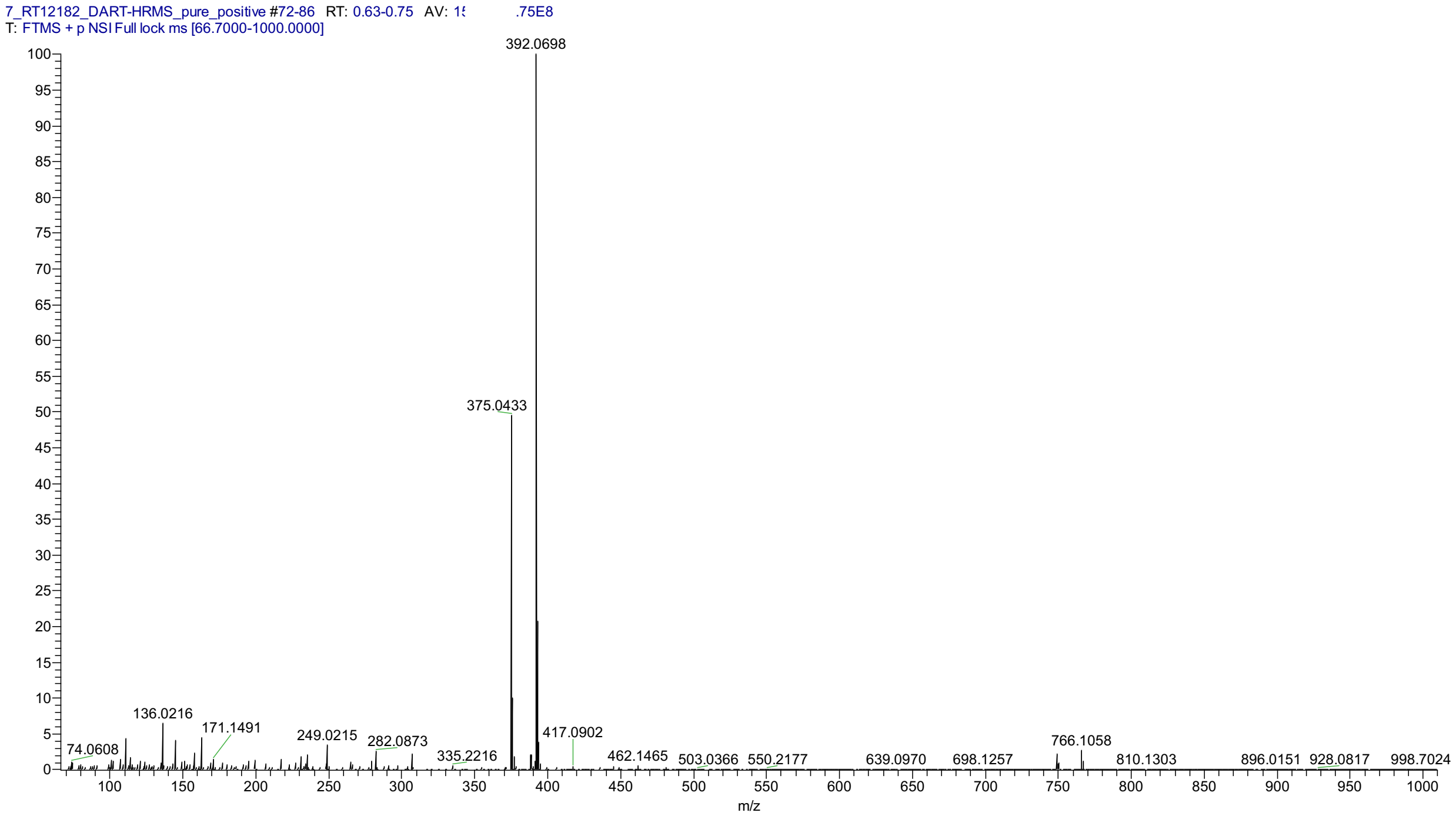

**Figure S31**. *HRMS data for target compound **4**. The second strongest signal at 375.0433 is for $[C_{20}H_{10}N_2O_4S+H]^+$ with base peak at 392.0698 for $[C_{20}H_{10}N_2O_4S+NH_4]^+$.*

**5**. *3-fluoro-4-[(3,4,5-trifluorophenoxy) carbonyl] phenyl-2-thiophenecarboxylate*

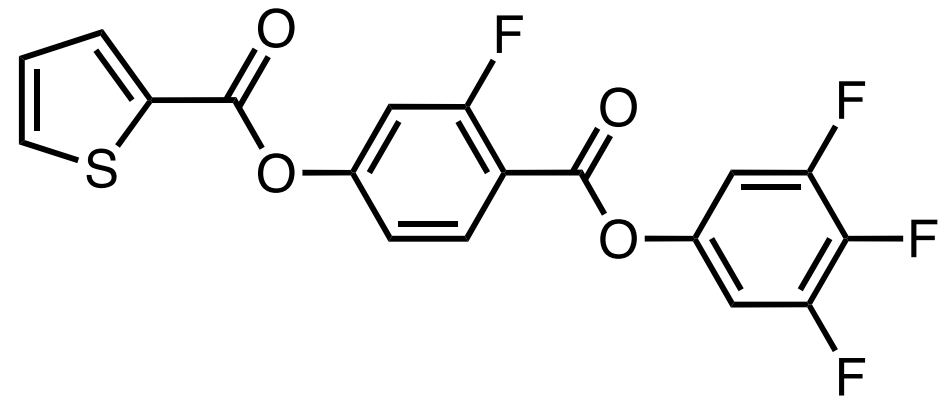

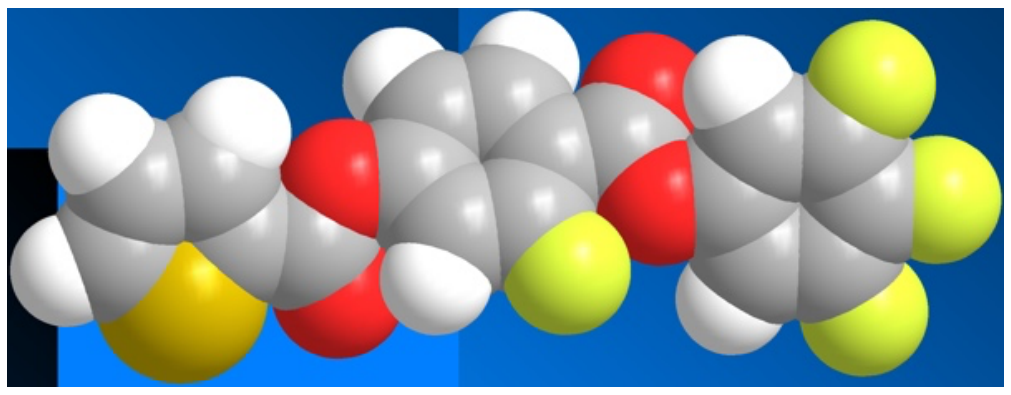

In a 200 ml round bottom flask equipped with a stir bar was placed 3,4,5-trifluorophenol (0.37 g, 2.50 mmol) and pyridine (20 ml) under nitrogen. The mixture was stirred and then the flask was placed in an ice bath. Next, 2-fluoro-4-(thiophene-2-carbonyloxy) benzoyl chloride (0.71 g, 2.50 mmol) in DCM (5.0 ml) was added dropwise. After thirty minutes the resulting mixture was allowed to warm to room temperature and was stirred overnight. After this time, TLC indicated formation of a less polar product with some remaining starting material. The reaction was terminated, and DCM was removed by rotary evaporation. Ice was added to the mixture with vigorous stirring then more ice water was added to fill the flask. The precipitate thus obtained was vacuum filtered and air dried then recrystallized from 1-propanol. Yield- 0.34 g (34%)

**$^{19}$F NMR (470 MHz, $CDCl_3$)** δ -103.69 (dd, *J* = 11.4, 7.9 Hz), -132.49 – -132.65 (m), -163.05 (tt, *J* = 21.0, 5.9 Hz).
**$^{1}$H NMR (500 MHz, $CDCl_3$)** δ 8.15 (dd, *J* = 9.0, 7.9 Hz, 1H), 8.03 (dd, *J* = 3.8, 1.3 Hz, 1H), 7.75 (dd, *J* = 5.0, 1.3 Hz, 1H), 7.25 – 7.20 (m, 3H), 7.03 – 6.93 (m, 2H).
δ 8.15 (dd, *J* = 9.0, 7.9 Hz, 1H), 8.03 (dd, *J* = 3.8, 1.3 Hz, 1H), 7.75 (dd, *J* = 5.0, 1.3 Hz, 1H), 7.25 – 7.19 (m, 3H), 6.98 (dd, *J* = 7.8, 5.8 Hz, 2H).
**$^{13}$C NMR (126 MHz, $CDCl_3$)** δ 163.88, 161.78, 161.28, 159.32, 155.95, 155.86, 152.14, 137.31, 135.56, 134.59, 133.51, 131.61, 128.35, 117.94, 117.91, 114.62, 114.55, 111.48, 111.27, 107.28, 107.23, 107.13, 107.08.

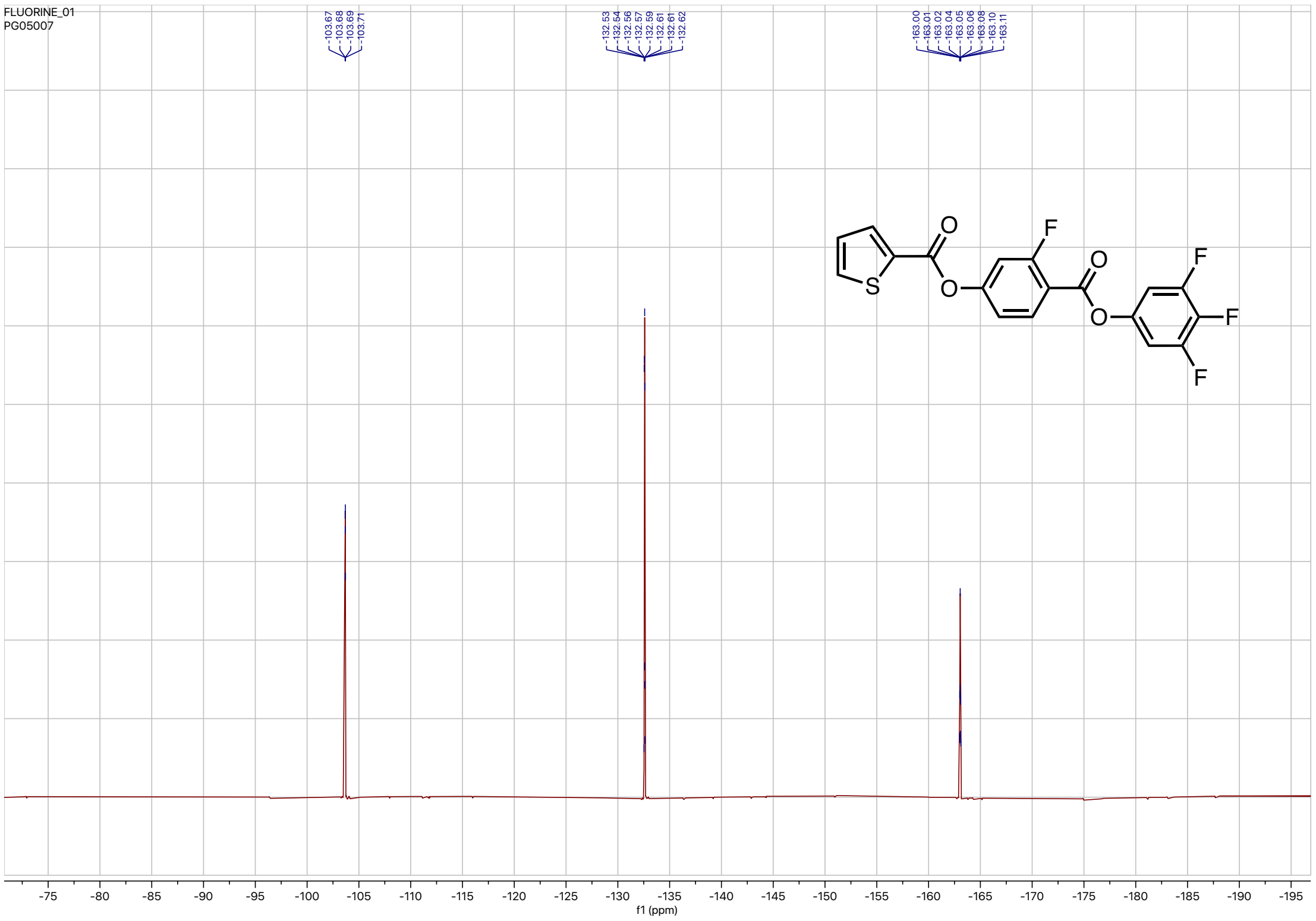

**Figure S32**. *[19]F-NMR of 3-fluoro-4-[(3,4,5-trifluorophenoxy) carbonyl] phenyl-2-thiophenecarboxylate*

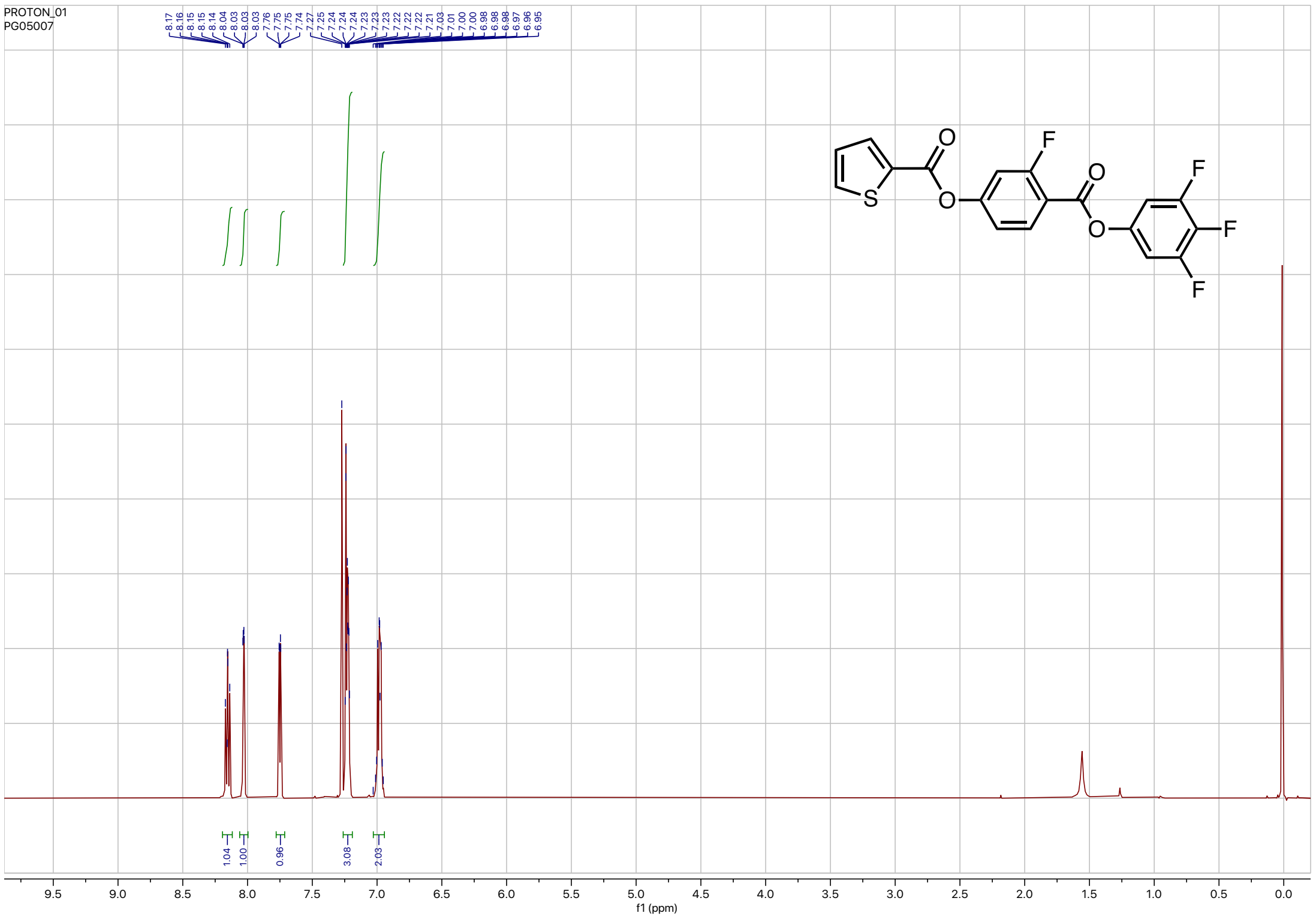

**Figure S33**. *[1]H-NMR of 3-fluoro-4-[(3,4,5-trifluorophenoxy) carbonyl] phenyl-2-thiophenecarboxylate*

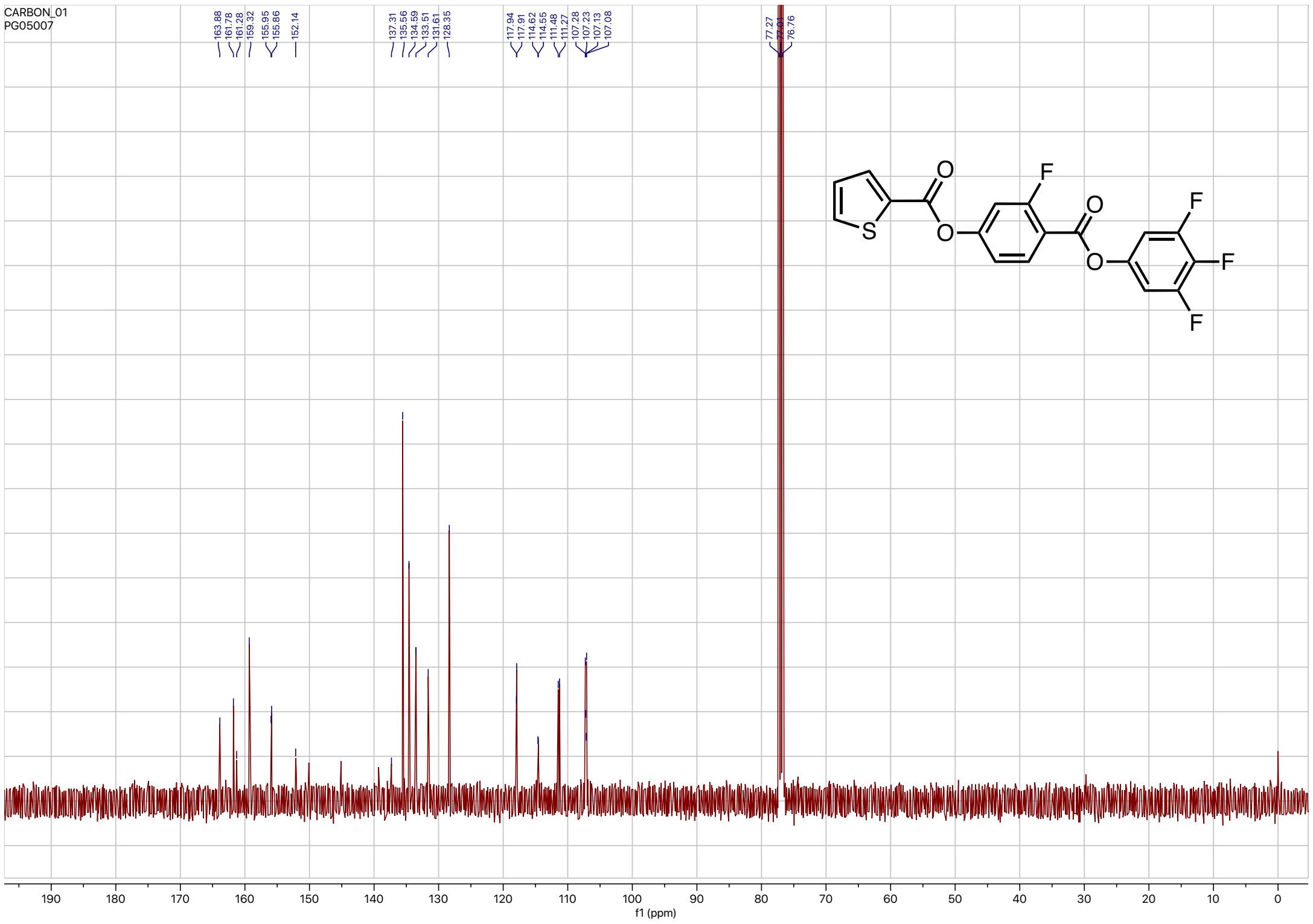

**Figure S34**. *[13]C-NMR of 3-fluoro-4-[(3,4,5-trifluorophenoxy) carbonyl] phenyl-2-thiophenecarboxylate*

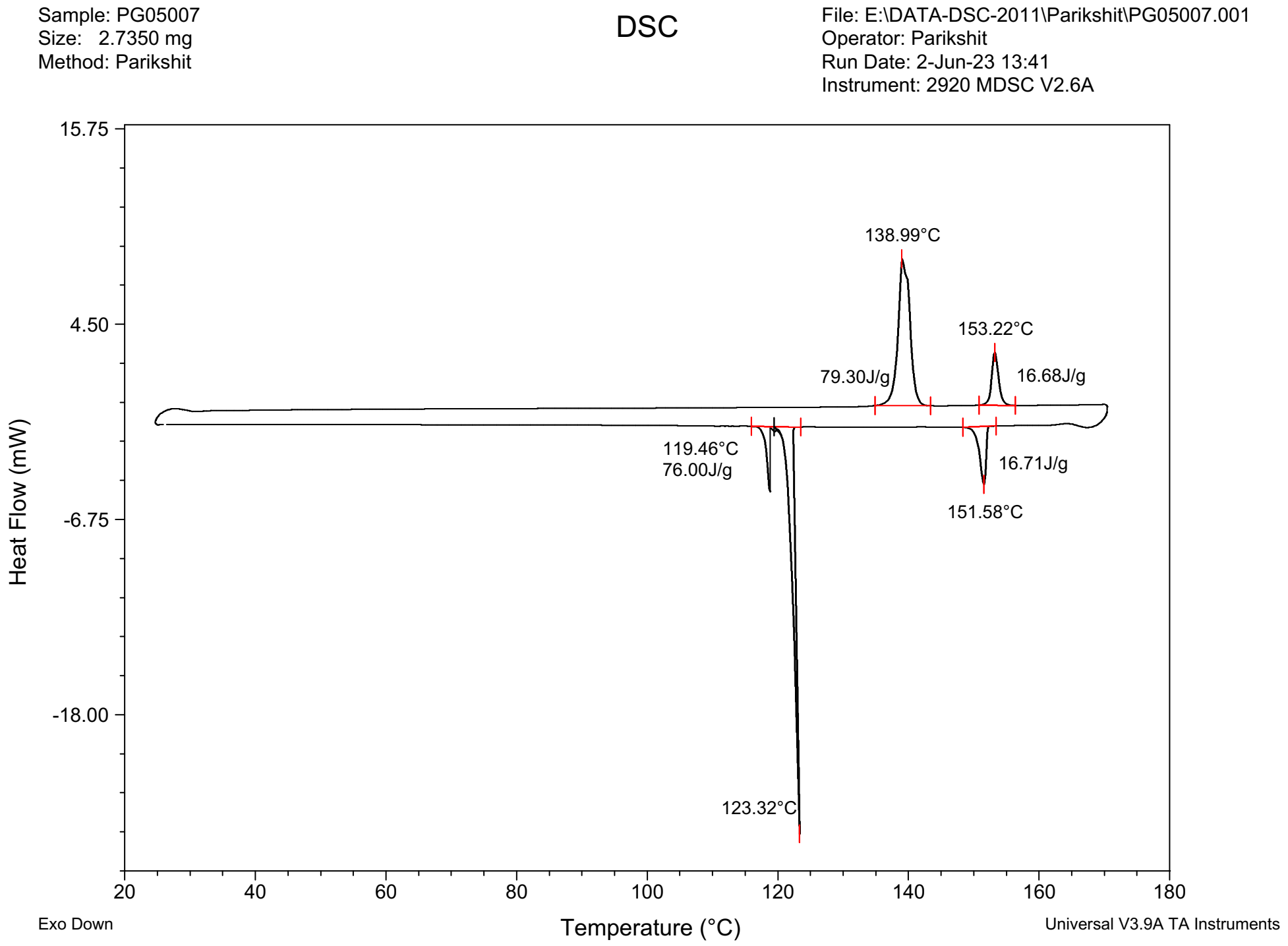

**Figure S35**. *DSC scan at 5°C/min rate of 3-fluoro-4-[(3,4,5-trifluorophenoxy) carbonyl] phenyl-2-thiophenecarboxylate*

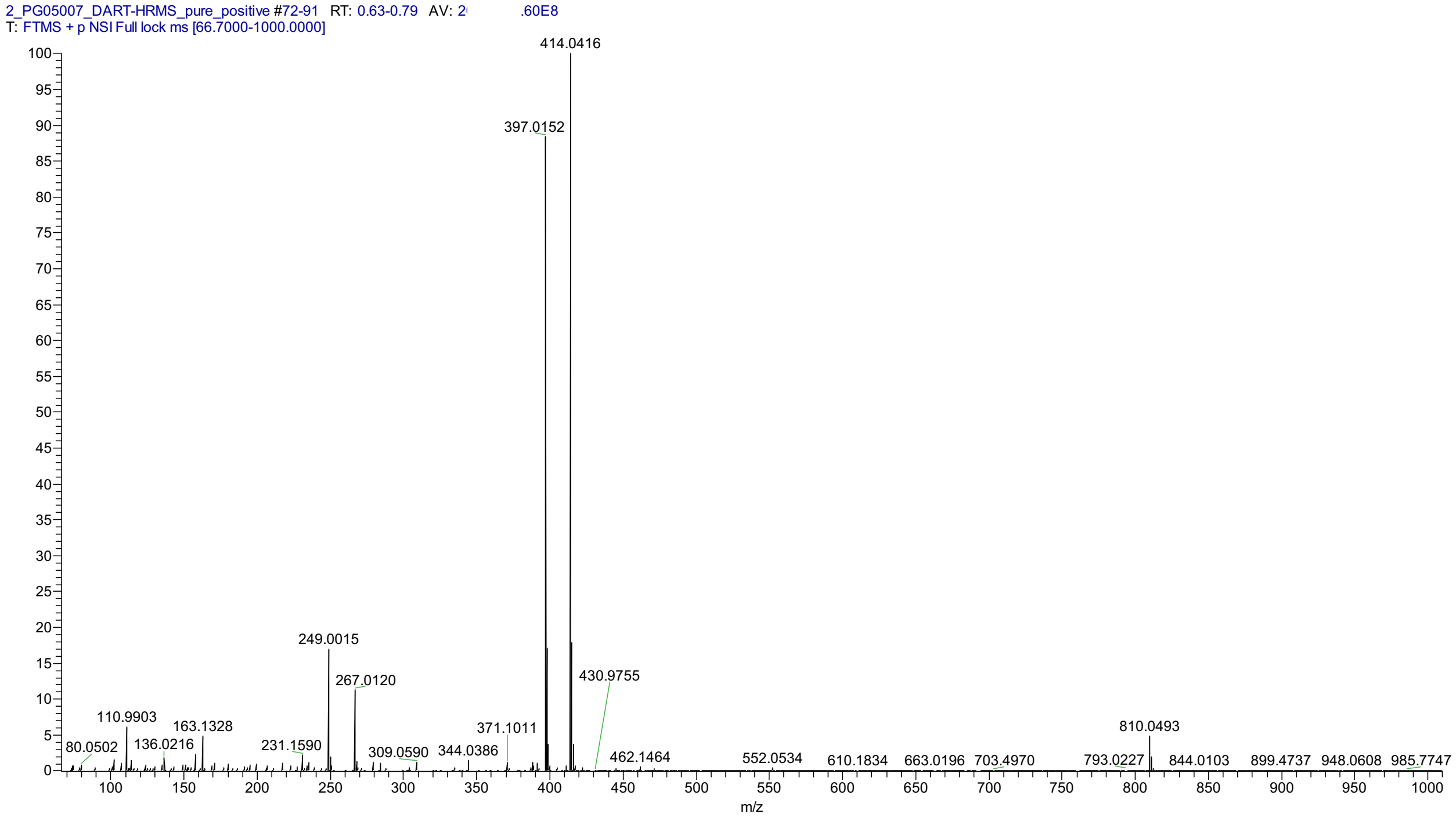

**Figure S36**. *HRMS data for target compound **5**. The second strongest signal at 397.0152 is for* $[C_{18}H_8F_4O_4S+H]^+$ *with base peak at 414.0416 for* $[C_{18}H_8F_4O_4S+NH_4]^+$.

## C. Preparation and characterization of the Reference Compounds

*4-[(4-Nitrophenoxy)carbonyl]phenyl 2,4-dimethoxybenzoate (**RM734**)* CAS 2196195-87-4

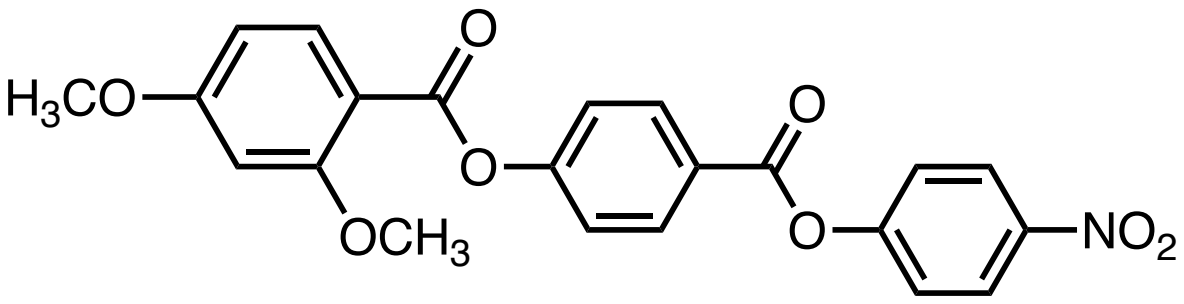

Since the original reports of RM734 in 2017 at least forty references with a substance role as product have appeared in CAS. The chemical characterization properties of the RM734 we have prepared are in accordance with the original reports. [SI 1, SI 2]

*4-[(4-Nitrophenoxy) carbonyl] phenyl 2-thiophenecarboxylate (**PN03111**)*

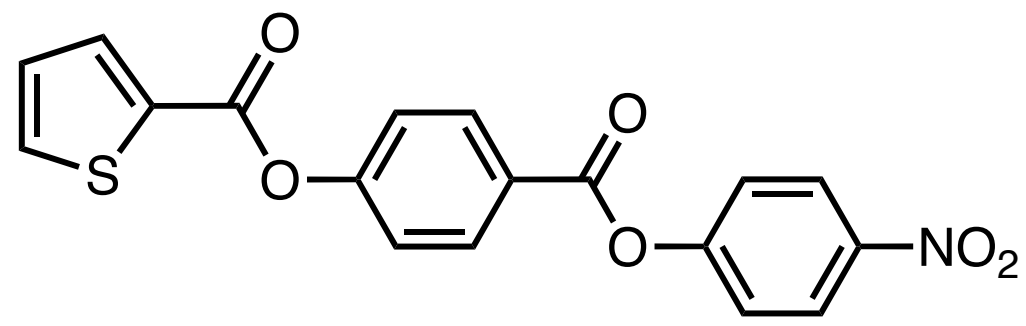

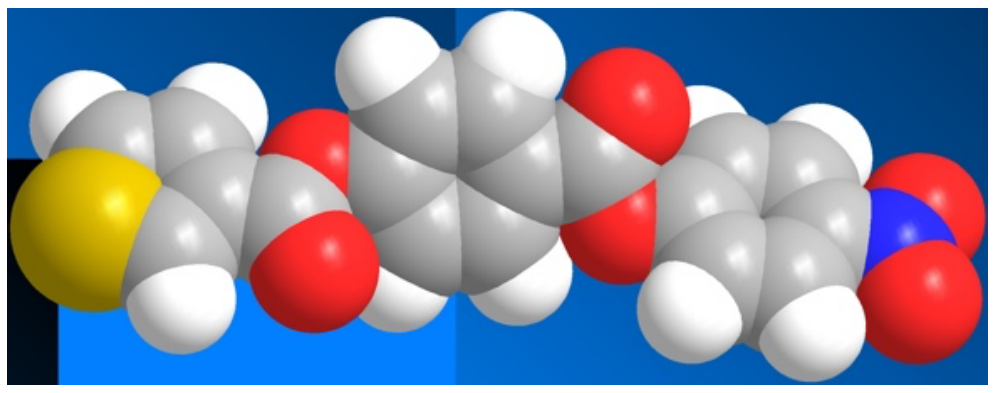

In a 200 ml round bottom flask with stirbar was placed 4-nitrophenyl 4-hydroxybenzoate (0.778 g, 3.0 mmol), and pyridine (20 ml). The mixture was placed in an ice bath and 2-thiophenecarbonyl chloride (0.439 g, 3.0 mmol) was added. After an hour a precipitate began to form. The resulting mixture was slowly warmed to room temperature and stirred overnight. After this time, TLC indicated complete consumption of the starting materials to give a single less polar product. The reaction was quenched by dropwise addition of cold water (100 ml). The solid precipitate obtained was isolated by suction filtration, air dried and recrystallized from acetonitrile (yield = 0.683 g, 62%).

**$^1$H NMR (400 MHz, $CDCl_3$)** δ 8.40 – 8.33 (m, 2H), 8.32 – 8.26 (m, 2H), 8.04 (dd, *J* = 3.8, 1.3 Hz, 1H), 7.74 (dd, *J* = 5.0, 1.3 Hz, 1H), 7.48 – 7.39 (m, 4H), 7.23 (dd, *J* = 5.0, 3.8 Hz, 1H).
**$^{13}$C NMR (101 MHz, $CDCl_3$)** δ 163.48, 159.88, 155.62, 155.23, 145.49, 135.31, 134.26, 132.07, 132.02, 128.28, 126.14, 125.35, 122.66, 122.31, 122.23, 122.17.

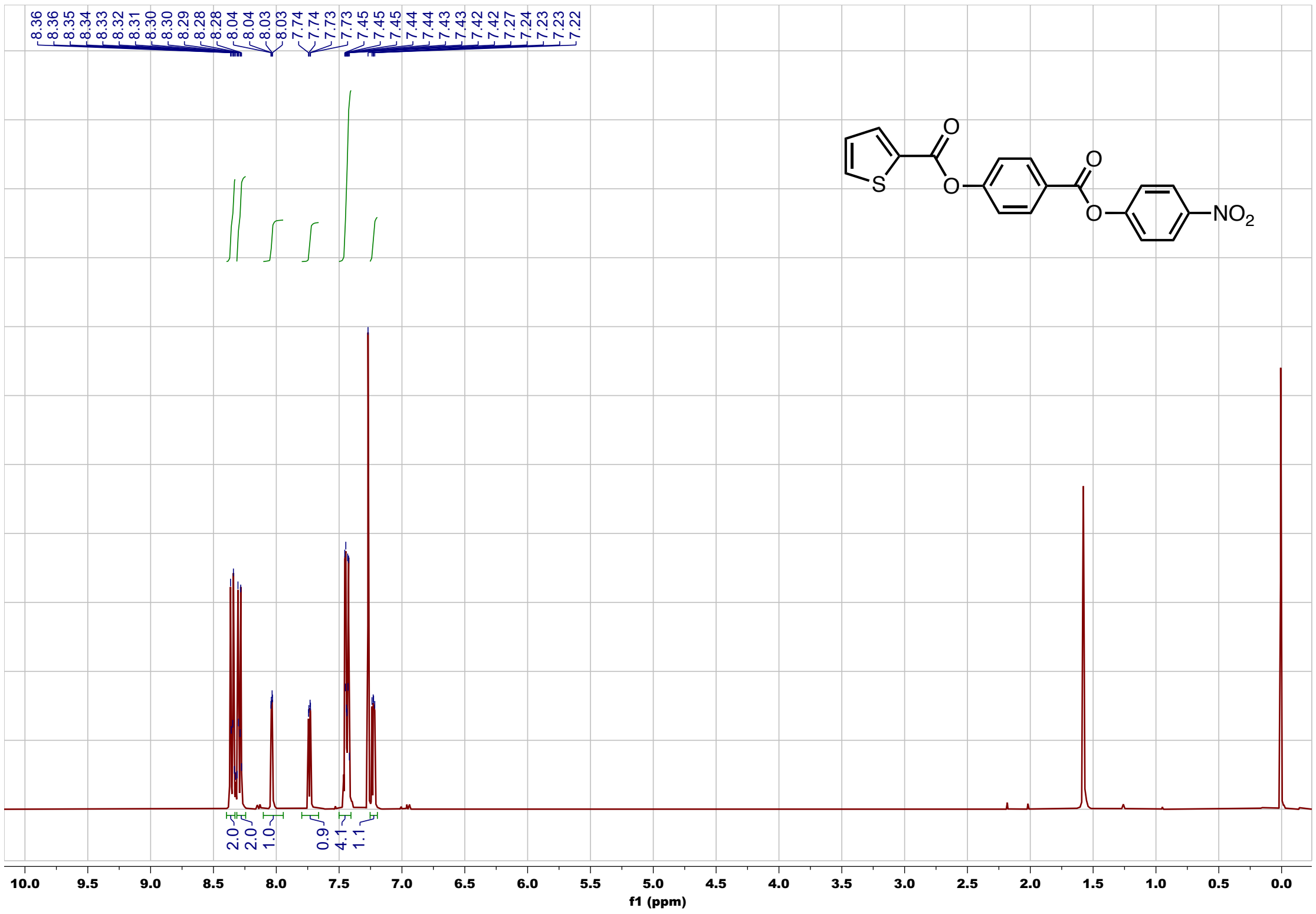

**Figure S37**. *$^1$H-NMR of 4-[(4-Nitrophenoxy) carbonyl] phenyl 2-thiophenecarboxylate*

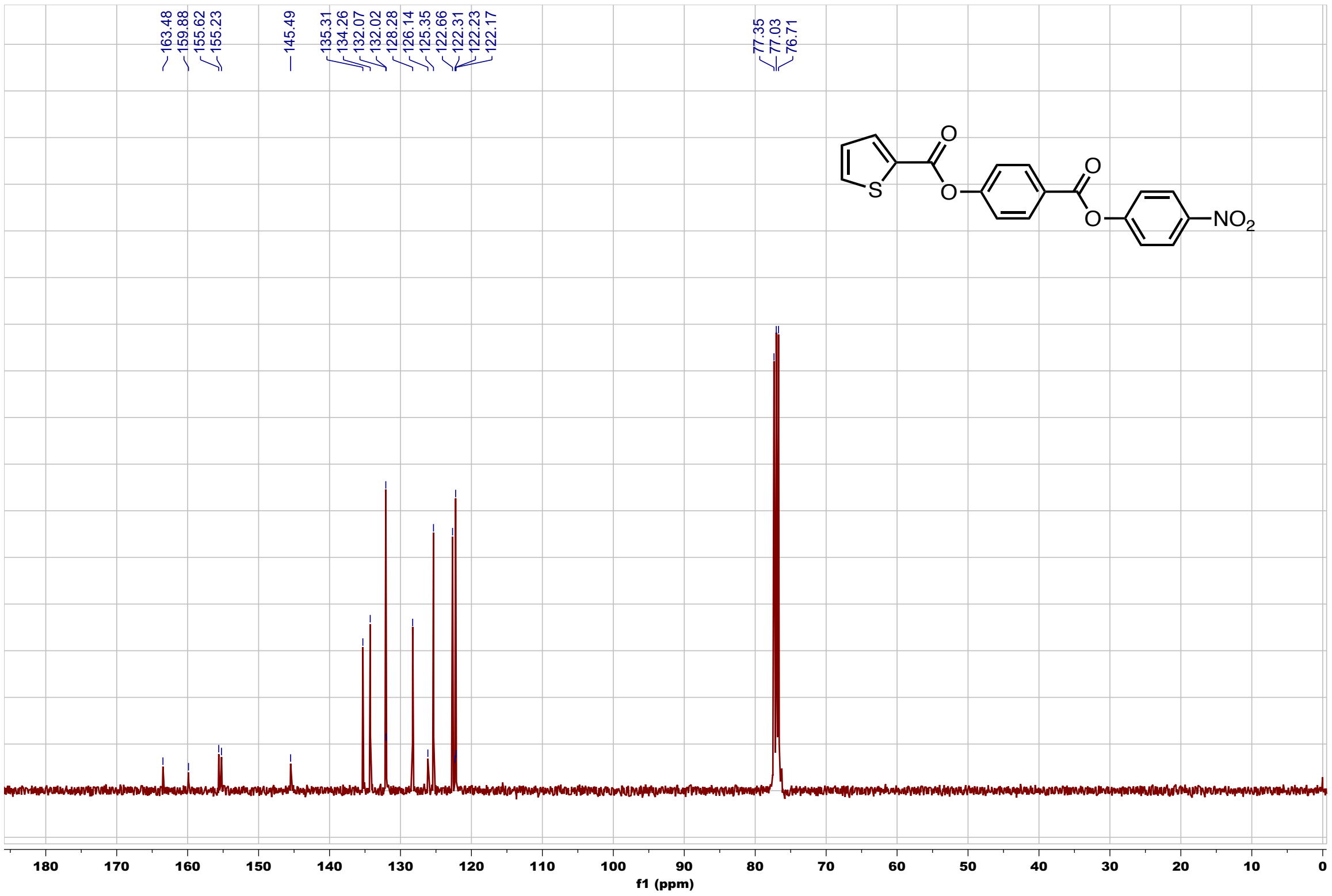

**Figure S38**. *[13]C-NMR of 4-[(4-Nitrophenoxy) carbonyl] phenyl 2-thiophenecarboxylate*

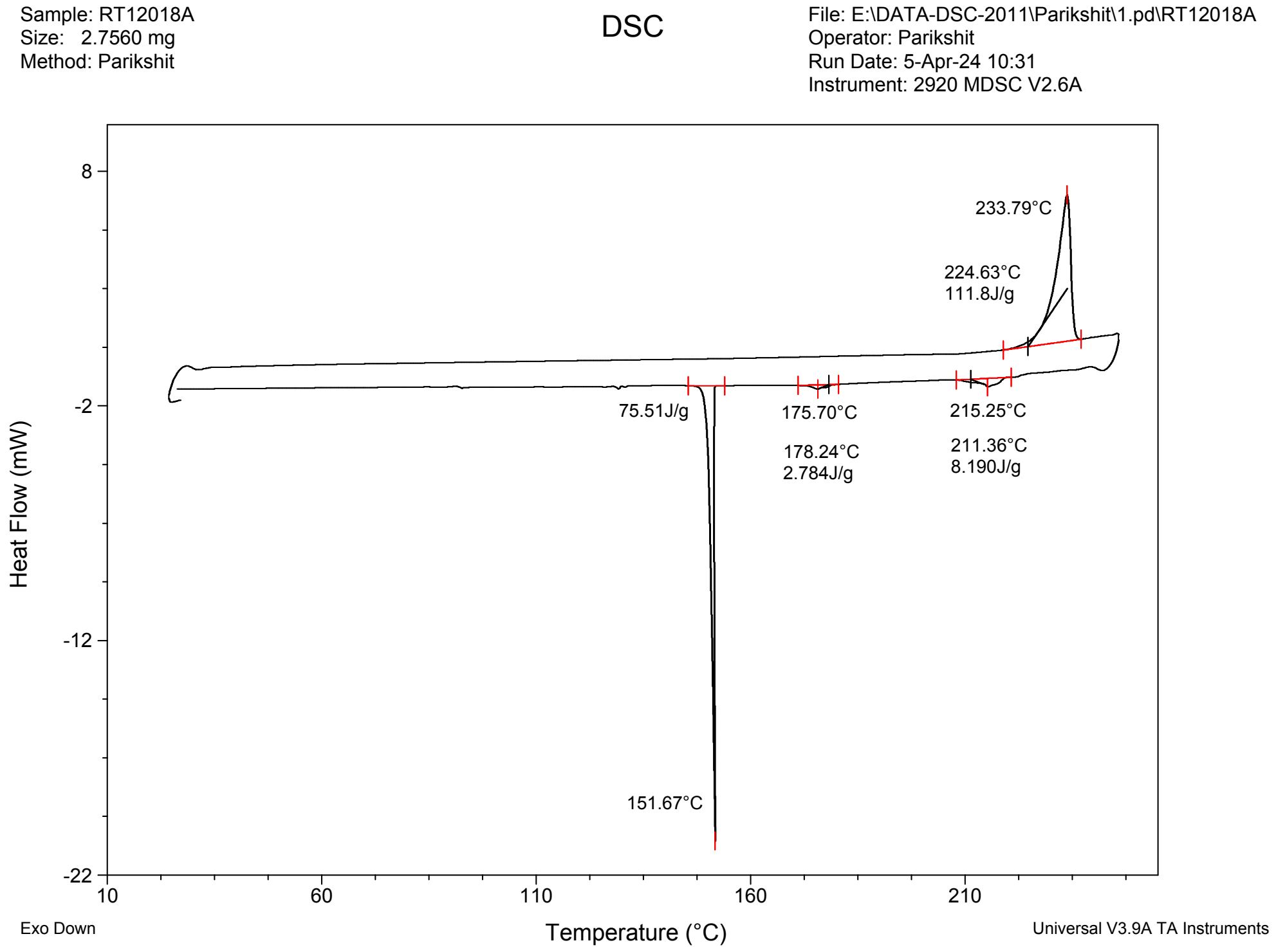

**Figure S39**. *A DSC scan at 5°C/min rate of of 4-[(4-Nitrophenoxy) carbonyl] phenyl 2-thiophenecarboxylate*

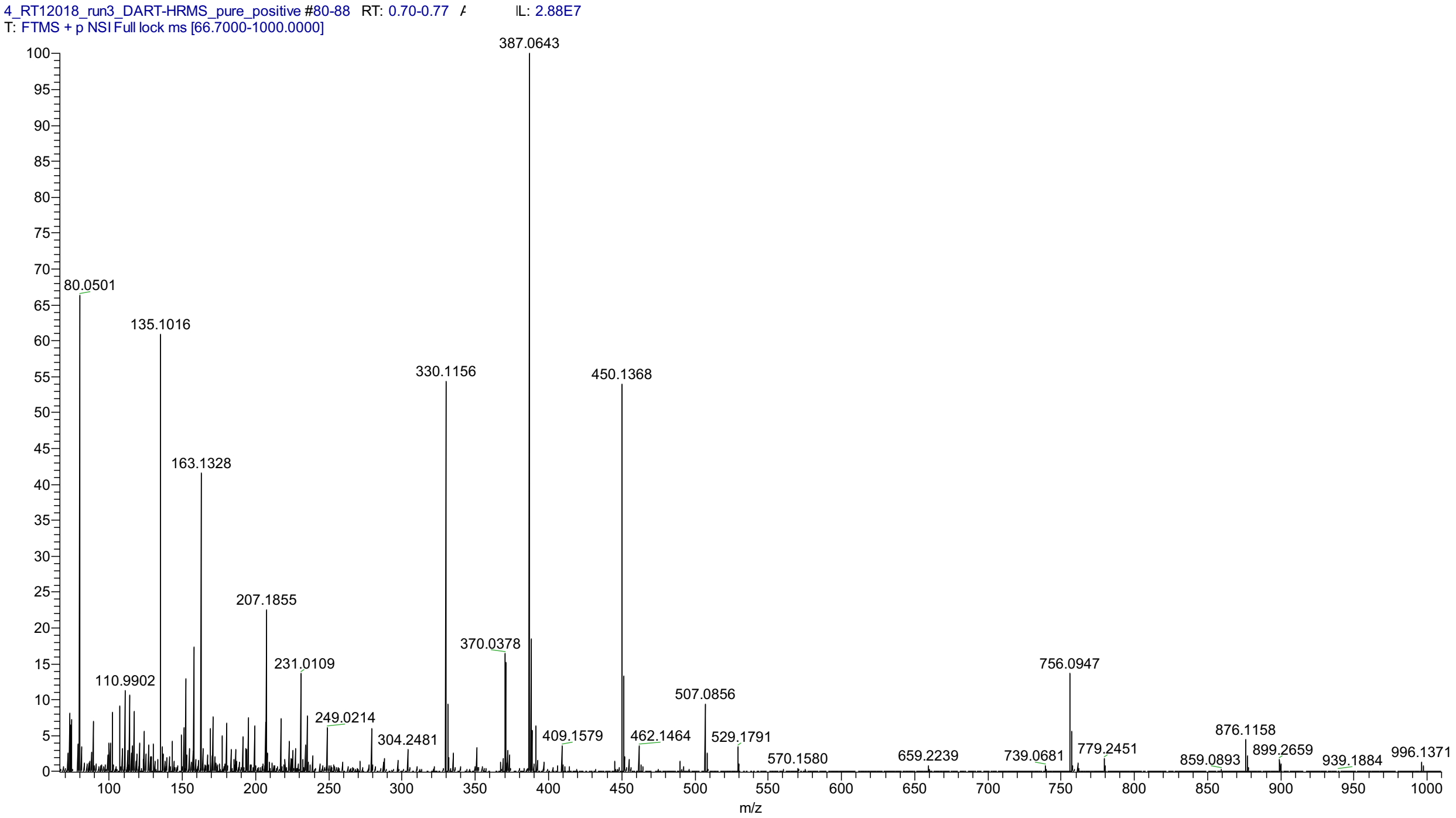

**Figure S40**. *HRMS data for PN03111. The small signal at 370.0378 is for* $[C_{18}H_{11}NO_6S+H]^+$ *with base peak at 387.0643 for* $[C_{18}H_{11}NO_6S+NH_4]^+$.

The physical properties of *PN03111* provided here are in accordance with the literature [SI 3].

**SI references**